\documentclass{amsproc}

\usepackage{graphicx}
\usepackage{hyperref}
\usepackage{appendix}

\theoremstyle{definition}

\theoremstyle{remark}

\numberwithin{equation}{section}

\newlength{\myhght}
\newlength{\mydpth}
\newcommand{\mybox}[1]{%
  \settoheight{\myhght}{#1}%
  \settodepth{\mydpth}{#1}%
  \addtolength{\mydpth}{3pt}%
  \addtolength{\myhght}{3pt}%
  \addtolength{\myhght}{\mydpth}%
  \rule[-\mydpth]{0pt}{\myhght}#1}

\begin{document}

\title[The Nikiforov--Uvarov Method]{On Potentials Integrated by the Nikiforov--Uvarov method}

\author{Lina Ellis}
\address{Department of Mathematics and Statistics, Northern Arizona University,
  P.~O.\ Box 5717, Flag\-staff, AZ 86011, U.S.A.}
\email{linabellis@gmail.com}

\author{Ikumi Ellis}
\address{School of Mathematical and Statistical Sciences, Arizona State University,
  P.~O.\ Box 871804, Tempe, AZ 85287-1804, U.S.A.}
\email{irellis@asu.edu}

\author{Christoph Koutschan}
\address{Johann Radon Institute for Computational and Applied Mathematics,
  Austrian Academy of Sciences, Altenberger Stra{\ss}e 69, 4040 Linz, Austria}
\email{christoph.koutschan@ricam.oeaw.ac.at}
\thanks{C.~K.~was supported by the Austrian Science Fund (FWF): I6130-N}

\author{Sergei K. Suslov}
\address{School of Mathematical and Statistical Sciences, Arizona State University,
  P.~O.\ Box 871804, Tempe, AZ 85287-1804, U.S.A.}
\email{sergei@asu.edu}



\date{April 3, 2023}

\begin{abstract}
  We discuss basic potentials of the nonrelativistic and relativistic quantum
  mechanics that can be integrated in the Nikiforov and Uvarov paradigm with
  the aid of a computer algebra system. This approach may help the readers to
  study modern analytical methods of quantum physics.
\end{abstract}

\maketitle

{\scriptsize{
Building on ideas of {\scshape{de Broglie}} and {\scshape{Einstein}}, I tried to show that the
ordinary differential equations of mechanics, which attempt to define the
co-ordinates of a mechanical system as functions of the time, are no longer
applicable for \textquotedblleft small\textquotedblright\ systems; instead
there must be introduced a certain \textit{partial} differential equation,
which defines a variable $\psi $ (\textquotedblleft wave
function\textquotedblright ) as a function of the co-ordinates and the time.}

\begin{flushright}
\it{Erwin~Schr\"{o}dinger} \cite{SchrodingerCohrent}
\end{flushright}
}

\section{Introduction}

Discovery of the relativistic and nonrelativistic Schr\"{o}dinger equations \cite{Schrodinger2010}, \cite{SchroedingerOscillator}, \cite{SchrodingerParabolic},
\cite{SchrodingerCohrent}, \cite{SchrodingerTimeDependent}, \cite{SchrodingerReview} is discussed in \cite{Barleyetal2021} (see also the references therein).
Invented about a century ago, the stationary Schr\"{o}dinger equation turns out to have an enormously wide range of applications, from the quantum
theory of atoms and molecules to solid state physics, quantum crystals, superfluidity, and superconductivity.
Finding of the energy levels and the corresponding normalized wave functions of various systems is one of the basic problems of quantum physics.
Only in a few elementary cases the exact solutions are known. They are usually investigated by different techniques.
Nonetheless, those completely integrable problems are important in creation of mathematical models for complex quantum systems. Moreover,
they may provide a useful testing ground for verification of numerical methods.
A true story of calculations of the energy levels for the two-electron atoms \cite{Pek58} is presented in \cite{KZ10}.
We assemble analytical solutions for a range of potentials in the nonrelativistic and relativistic quantum mechanics that are available in the literature.
Data for most of the potentials that can be studied, in a unified way, by the so-called Nikiforov--Uvarov method \cite{Ni:Uv} are collected, independently verified, and completed with the help of the Mathematica computer algebra system. Only bound states are discussed.
On the contrary, in a traditional approach, for each of those problems one has to identify and factor out the singularities of the corresponding square integrable wave functions
and find the remaining terminating power series expansions or use algebraic methods (see, for example, \cite{Bethe:Sal}, \cite{Blokh}, \cite{Dav}, \cite{DelSol}, \cite{Flugge}, \cite{Fock1978}, \cite{Greiner}, \cite{Karp70}, \cite{La:Lif}, \cite{Schiff}, \cite{Sommerfeld1951}).
As a result, each of such problems has to be treated separately, which is not suitable for a unified computer algebra approach.
Our review article is organized as follows. In the next section, we introduce the basics of the Nikiforov--Uvarov approach and then, successively, apply it to the main problems of introductory quantum mechanics, such as harmonic oscillators, Bessel functions, Coulomb problems, P\"{o}schl--Teller potential holes, Kratzer's molecular potential, Hulth\'{e}n potentials, and Morse potentials, in the forthcoming sections. All calculations are verified in a complementary Mathematica notebook that can serve both educational and research purposes.
Appendices~\ref{app:A} and~\ref{app:B} contain, for the reader's convenience, the data for classical orthogonal polynomials and a useful integral evaluation, respectively,
in order to make our presentation as self-contained as possible. Appendix~\ref{app:c} describes the Mathematica notebook.
This review is written for those who study quantum mechanics and would like to see more details than in the classical textbooks by utilizing the advanced computer algebra system, Mathematica.
It is motivated by an introductory course in mathematics of quantum mechanics which one of the authors (SKS) has been teaching at Arizona State University
for more than two decades.
\section{Summary of the Nikiforov--Uvarov approach\/}

The generalized equation of the hypergeometric type
\begin{equation}
u^{\prime \prime }+\frac{\widetilde{\tau }(x)}{\sigma (x)}u^{\prime }+\frac{%
\widetilde{\sigma }(x)}{\sigma ^{2}(x)}u=0  \label{A1}
\end{equation}
($\sigma (x) ,$ $\widetilde{\sigma }(x)$ are polynomials of degrees at most $2$ %
and $\widetilde{\tau }(x)$ is a polynomial degree at most one) by the substitution%
\begin{equation}
u=\varphi (x)y(x)  \label{A2}
\end{equation}%
can be reduced to the form%
\begin{equation}
\sigma (x)y^{\prime \prime }+\tau (x)y^{\prime }+\lambda y=0  \label{A3}
\end{equation}%
if:%
\begin{equation}
\frac{\varphi ^{\prime }}{\varphi }=\frac{\pi (x)}{\sigma (x)},\qquad \pi
(x)=\frac{1}{2}\left( \tau (x)-\widetilde{\tau }(x)\right)  \label{A4}
\end{equation}%
(or, $\tau (x)=\widetilde{\tau }+2\pi ,$ for later),%
\begin{equation}
k=\lambda -\pi ^{\prime }(x)\qquad (\text{or, }\lambda =k+\pi ^{\prime }),
\label{A5}
\end{equation}%
and%
\begin{equation}
\pi (x)=\frac{\sigma ^{\prime }-\widetilde{\tau }}{2}\pm \sqrt{\left( \frac{%
\sigma ^{\prime }-\widetilde{\tau }}{2}\right) ^{2}-\widetilde{\sigma }%
+k\sigma }  \label{A6}
\end{equation}%
is a linear function. (Use the choice of the constant~$k$ to complete the square
under the radical sign; see \cite{Ni:Uv} and our argument below for more details.)

In Nikiforov--Uvarov's method, the energy levels can be obtained from the
quantization rule:%
\begin{equation}
\lambda +n\tau ^{\prime }+\frac{1}{2}n(n-1) \sigma ^{\prime
\prime }=0\qquad (n=0,1,2,\dots)  \label{A7}
\end{equation}%
and the corresponding square-integrable solutions are classical orthogonal polynomials, up to a factor.
They can be found by the \textit{Rodrigues-type formula\/} \cite{Ni:Uv}:
\begin{equation}
y_n(x) = \dfrac{B_n}{\rho(x)} \left(\sigma^n(x) \rho(x)\right)^{(n)}, \qquad (\sigma \rho)'=\tau \rho , \label{A8}
\end{equation}%
where $B_n$ is a constant (see also \cite{Suslov2020} and Table~\ref{tab:cop}).
(The corresponding data for basic nonrelativistic and relativistic problems are
presented in the Tables~\ref{tab:StatSchr}--\ref{tab:gMp} below.)

Let us try to transform the differential equation (\ref{A1}) to the simplest form
by the change of unknown function $u=\varphi(x) \, y$ with the
help of some special choice of function $\varphi(x) $.

Substituting $u=\varphi(x) \;y$ in (\ref{A1}) one gets
\begin{equation}
y^{\prime \prime }+\left( \frac{\widetilde{\tau }}{\sigma }+2\frac{\varphi
^{\prime }}{\varphi }\right) y^{\prime }+\left( \frac{\widetilde{\sigma }}{%
\sigma ^{2}}+\frac{\widetilde{\tau }}{\sigma }\frac{\varphi ^{\prime }}{%
\varphi }+\frac{\varphi ^{\prime \prime }}{\varphi }\right) y=0. \label{e21}
\end{equation}
Equation (\ref{e21}) should not be more complicated than our original equation
(\ref{A1}). Thus, it is natural to assume that the coefficient in front of $%
y^{\prime }$ has the form $\tau(x) /\sigma(x) $,
where $\tau(x) $ is a polynomial of degree at most one. This
implies the following first-order differential equation
\begin{equation}
\frac{\varphi ^{\prime }}{\varphi }=\frac{\pi(x) }{\sigma(x) } \label{e22}
\end{equation}
for the function $\varphi(x) $, where
\begin{equation}
\pi(x) =\frac{1}{2}\left( \tau(x) -\widetilde{\tau
}(x) \right) \label{e23}
\end{equation}
is a polynomial of degree at most one. As a result, equation (\ref{e21}) takes
the form
\begin{equation}
y^{\prime \prime }+\frac{\tau(x) }{\sigma(x) }%
\;u^{\prime }+\frac{\overline{\sigma }(x) }{\sigma ^{2}(x) }\;u=0, \label{e24}
\end{equation}
where
\begin{equation}
\overline{\sigma }(x) =\widetilde{\sigma }(x) +\pi
^{2}(x) +\pi(x) \left[ \widetilde{\tau }(x) -\sigma ^{\prime }(x) \right] +\pi ^{\prime }(x) \sigma(x).
\end{equation}
The functions $\tau(x) $ and $\overline{\sigma }(x) $
are polynomials of degrees at most one and two in $x$, respectively.
Therefore, equation (\ref{e24}) is an equation of the same type as our original
equation (\ref{A1}).

By using a special choice of the polynomial $\pi(x) $ we can
reduce (\ref{e24}) to the simplest form assuming that
\begin{equation}
\overline{\sigma }(x) =\lambda \;\sigma(x) , \label{e26}
\end{equation}
where $\lambda $ is some constant. Then equation (\ref{e24}) takes the form (\ref{A3}).
We call equation (\ref{A3}) a \textit{differential equation of hypergeometric
type} and its solutions \textit{functions of hypergeometric type}. In
this context, it is natural to call equation (\ref{A1}) a \textit{generalized
differential equation of hypergeometric type\/} \cite{Ni:Uv}.

The condition (\ref{e26}) can be rewritten as
\begin{equation}
\pi ^{2}+\left( \widetilde{\tau }-\sigma ^{\prime }\right) \;\pi +\widetilde{%
\sigma }-k\sigma =0, \label{eq:quad}
\end{equation}
where
\begin{equation}
k=\lambda -\pi ^{\prime }(z) \label{e29}
\end{equation}
is a constant. Assuming that this constant is known, we can find $\pi(x)$
as a solution (\ref{A6}) of the quadratic equation~\eqref{eq:quad}.
But $\pi(x) $ is a polynomial, therefore the second degree
polynomial
\begin{equation}
p(x) =\left( \frac{\sigma ^{\prime }(x) -\widetilde{%
\tau }(x) }{2}\right) ^{2}-\widetilde{\sigma }(x)
+k\sigma(x)
\end{equation}
under the radical should be a square of a linear function and the
discriminant of $p(x) $ should be zero. This condition gives an
equation for the constant $k$, which is, generally, a quadratic equation.
Given $k$ as a solution of this equation, we find $\pi(x) $ by the quadratic formula
(\ref{A6}), then $\tau(x) $ and $\lambda $ by (\ref{e23}) and (\ref{e29}).
Finally, we find the function $\varphi(x) $ as a solution of (\ref{e22}).
It is clear that the reduction of equation (\ref{A1}) to the simplest form (\ref{A3})
can be accomplished by a few different ways in accordance with different
choices of the constant $k$ and different signs in (\ref{A6}) for $\pi(x)$.

A closed form for the constant $k$ can be obtained as follows \cite{Barleyetal2021}.
Let
\begin{equation}
p(x)={\left( \frac{%
\sigma ^{\prime }-\widetilde{\tau }}{2}\right) ^{2}-\widetilde{\sigma }%
+k\sigma } =q(x) + k\sigma(x)\/,  \label{A9}
\end{equation}
where
\begin{equation}
q(x)=\left(\frac{\sigma ^{\prime }-\widetilde{\tau }}{2}\right)^2 -\widetilde{\sigma } \/. \label{A10}
\end{equation}
Completing the square, one gets
\begin{equation}
p(x)=\dfrac{p^{\prime \prime }}{2}\left(x+\frac{p^{\prime}(0)}{p^{\prime \prime }}\right)^2 -\dfrac{\left(p^{\prime}(0)\right)^2-2p^{\prime \prime }p(0)}{2p^{\prime \prime }} \/, \label{A11}
\end{equation}
where the last term must be eliminated:%
\begin{equation}
{\left(p^{\prime}(0)\right)^2-2p^{\prime \prime }p(0)} = 0 \/. \label{A12}
\end{equation}
Therefore,
\begin{equation}
\left(q^{\prime}(0)+k\sigma^{\prime} (0)\right)^2-2\left(q^{\prime \prime } + k \sigma ^{\prime \prime } \right) \left(q(0) + k\sigma(0) \right) = 0 \/, \label{A13}
\end{equation}
which results in the following quadratic equation:
\begin{equation}
ak^2+2bk+c=0. \label{A14}
\end{equation}
Here,
\begin{align}
&a = \left(\sigma^{\prime}(0) \right)^2  - 2\sigma^{\prime \prime} \sigma(0) \/, \\ 
&b = q^{\prime}(0) \sigma^{\prime}(0) - \sigma^{\prime \prime} q(0) -\sigma(0) q^{\prime \prime} \/,   \\ 
&c =   \left(q^{\prime}(0) \right)^2  - 2q^{\prime \prime} q(0) \/. 
\end{align}
Solutions are
\begin{equation}
k_0 = -\dfrac{c}{2b}, \qquad \text{if} \quad a=0 \label{A18}
\end{equation}
and
\begin{equation}
k_{1,2}=\dfrac{-b \pm \sqrt{d}}{a}, \qquad \text{if} \quad a\ne 0 \/. \label{A19}
\end{equation}
Here,
\begin{equation}
d=b^2-ac=\left(\sigma(0)q^{\prime \prime} - \sigma^{\prime \prime} q(0) \right)^2 - 2\left(\sigma^{\prime}(0)q^{\prime \prime} - \sigma^{\prime \prime} q^{\prime}(0) \right)
\left(\sigma(0)q^{\prime }(0) - \sigma^{\prime }(0)q(0) \right) \/. \label{A20}
\end{equation}
As a result,
\begin{equation}
p(x)=\left[\dfrac{p^{\prime \prime }}{2}\left(x+\frac{p^{\prime}(0)}{p^{\prime \prime }}\right)^2 \right]_{k=k_{0, 1, 2}} \/, \label{A21}
\end{equation}
which allows evaluating the linear function $\pi(x)\/$ in the Nikiforov--Uvarov technique.

Examples, for the most integrable cases that are available in the literature, are presented
in the Tables~\ref{tab:StatSchr}--\ref{tab:gMp} below, where
all these analytical arguments are implemented in a supplementary Mathematica notebook
(Appendix~\ref{app:c} ).

%

%
\section{Harmonic Oscillator\/}

Let us consider the one-dimensional stationary Schr\"{o}dinger equation for the
harmonic oscillator:%
\begin{equation}
-\frac{\hbar ^{2}}{2m}\frac{d^{2}\psi }{dx^{2}}+\frac{1}{2}m
\omega^{2}x^{2}\psi =E\psi   \label{1DOscillator}
\end{equation}%
with the orthonormal real-valued wave function%
\begin{equation}
\int_{-\infty }^{\infty }\psi ^{2}(x) \ dx=1.
\label{1DOscillatorNorm}
\end{equation}%
Introducing dimensionless variables%
\begin{equation}
\psi(x) =u(\xi) ,\qquad x=\xi \sqrt{\frac{\hbar }{%
m\omega }},\quad E=\hbar \omega \varepsilon   \label{1DOscillatorVariables}
\end{equation}%
one gets%
\begin{equation}
u^{\prime \prime }+\left( 2\varepsilon -\xi ^{2}\right) u=0.
\label{1DOscillatorGeneralized}
\end{equation}%
Here, $\sigma(\xi) =1,$ $\widetilde{\tau }(\xi)
=0,$ and $\widetilde{\sigma }(\xi) =2\varepsilon -\xi ^{2}.$
Therefore,%
\begin{equation}
\pi(\xi) =\pm \sqrt{k-2\varepsilon +\xi ^{2}}=\pm \xi ,\qquad
k=2\varepsilon .\label{1DOscillatorPi}
\end{equation}%
We pick $\pi =-\xi ,$ which gives a negative derivative for%
\begin{equation}
\tau(\xi) =\widetilde{\tau }(\xi) +2\pi(\xi) =-2\xi .\label{1DOscillatorTau}
\end{equation}%
Then%
\begin{equation}
\frac{\varphi ^{\prime }}{\varphi }=\frac{\pi(\xi) }{\sigma
(\xi) }=-\xi ,\qquad \varphi(\xi) =e^{-\xi ^{2}/2}%
\label{1DOscillatorPhi}
\end{equation}%
and $\lambda =2\varepsilon -1,$ $\rho(\xi) =e^{-\xi ^{2}}.$
The energy levels are $\varepsilon =\varepsilon _{n}=n+1/2,$ $(n=0,1,2,\dots)$ from (\ref{A7}).
The eigenfunctions,%
\begin{equation}
y_{n}(\xi) =B_{n}e^{\xi ^{2}}\frac{d^{n}}{d\xi ^{n}}\left(
e^{-\xi ^{2}}\right) ,\label{1DOscillatorY}
\end{equation}%
are, up to a normalization, the Hermite polynomials (Table~\ref{tab:cop}).

\begin{figure}
\centering
\includegraphics[width=0.8\textwidth]{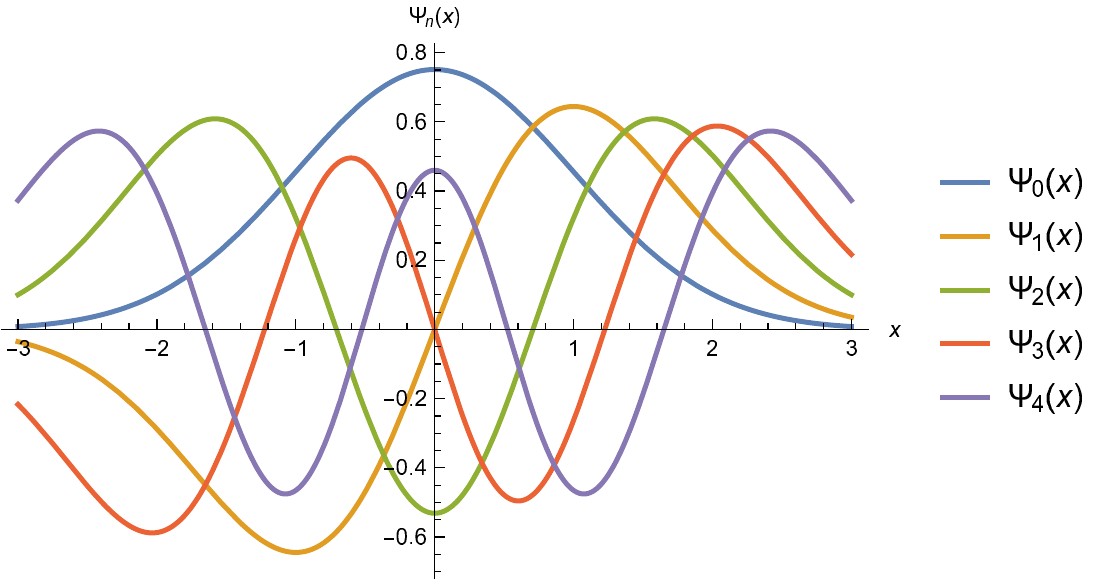}
\caption{\textquotedblleft The first five proper vibrations of the Planck oscillator
according to undulatory mechanics. Outside the region $-3\le x\le 3$,
represented here, all five functions approach $x$-axis in monotonic fashion.\textquotedblright
--- generated by Mathematica following the original Schr\"{o}dinger
article~\cite{SchrodingerCohrent}.}
\label{Figure1}
\end{figure}

As a result, the orthonormal wave functions are given by \cite{SchroedingerOscillator}, \cite{SchrodingerCohrent}, \cite{La:Lif}%
\begin{equation}
\psi(x) =\left( \frac{m\omega }{\pi \hbar }\right) ^{1/4}\frac{1%
}{\sqrt{2^{n}n!}}\exp \left( -\frac{m\omega }{2\hbar }x^{2}\right)
H_{n}\left( x\sqrt{\frac{m\omega }{\hbar }}\right) ,\label{1DOscillatorPsi}
\end{equation}%
corresponding to the discrete energy levels%
\begin{equation}
E_{n}=\hbar \omega \left( n+\frac{1}{2}\right) \qquad (n=0,1,2,\dots),\label%
{1DOscillatorE}
\end{equation}%
in Gaussian units; see also Figure~\ref{Figure1} for graphs of the first five wave functions.
(More general, ``missing'', solutions of the time-dependent Schr\"{o}\-dinger
equation are discussed in \cite{Ko:Sua:Su}, \cite{Kryuch:Sus:Vega12}, and \cite{Lop:Sus:VegaHarm}.)
%


%
\def\arraystretch{1.3}
\begin{table}
\caption[]{Stationary Schr\"{o}dinger equation for the harmonic potential
  $U(x)=\frac{1}{2} m \omega^{2} x^{2}.$}
\label{tab:StatSchr}
\begin{center}
\begin{tabular}{|l|l|}
\hline
$\sigma (\xi)$ & $1$ \\ \hline
$\widetilde{\sigma }(\xi)$ & $2\varepsilon - \xi^{2} $
\\ \hline
$\widetilde{\tau }(\xi)$ & $0$ \\ \hline
$k$ & $ 2\varepsilon$ \\ \hline
$\pi(\xi) $ & $\pm \xi$ \\ \hline
$\tau(\xi) =\widetilde{\tau }+2\pi \qquad $ & $ -2 \xi $ \\ \hline
$\lambda =k+\pi ^{\prime }$ & $2\varepsilon - 1 $ \\ \hline
$\varphi(\xi) $ & $e^{-\xi^2/2}$ \\ \hline
$\rho (\xi)$ & $e^{-\xi^2} $ \\ \hline
$y_{n}(\xi)$ & $C_{n}H_{n}(\xi) $ \\ \hline
$C_{n}^{2}$ & \mybox{$\dfrac{1}{\sqrt{\pi} \, 2^n n! }$} \\ \hline
\end{tabular}
\end{center}
\end{table}
%


%
\section{Bessel Functions\/}

Let us also mention some solutions of the {\it Bessel equation\/}:
\begin{equation}
z^2u^{\prime\prime}+zu^\prime +\left( z^2-\nu^2\right) u=0. \label{eqn1213}
\end{equation}
\noindent
With the aid of the change of the function $u=\varphi(z)y$
when $\varphi(z)=z^\nu e^{iz}$ this equation can be reduced to the
hypergeometric form
\begin{equation}
zy^{\prime\prime}+(2iz+2\nu +1)y^\prime +i(2\nu +1)y=0 \label{eqn1214}
\end{equation}
\noindent
(Table~\ref{tab:Bessel}) and one can obtain the {\it Poisson integral representations\/} for the Bessel functions of the first kind, $J_\nu(z)$,
and the Hankel functions of the first and second kind,
$H_\nu^{(1)}(z)$ and $H_\nu^{(2)}(z)$:
\begin{align}
J_{\nu}(z)&=\frac{(z/2)^\nu}{\sqrt{\pi}\,\Gamma(\nu +1/2)}\,
\int_{-1}^1\,\left( 1-t^2\right) ^{\nu -1/2}\cos(zt)\, dt, \label{eqn1215}\\
\nonumber \\
H_{\nu}^{(1,2)}(z)&=\sqrt{\frac2{\pi z}}\;
\frac{e^{\pm i\left( z-\frac\pi 2\nu -\frac\pi 4\right) }}
{\Gamma(\nu +1/2)}\,\int_0^\infty\,
e^{-t}t^{\nu -1/2}\left( 1\pm\frac{it}{2z}\right) ^{\nu -1/2} dt, \label{eqn1216}
\end{align}
\noindent
where $\text{Re}\,\nu >-1/2$. It is then possible to deduce from
these integral representations all the remaining properties of
these functions. (For details, see \cite{Ni:Uv} and \cite{Suslov2020} or \cite{Watson} and \cite{Whi:Wat}).
%


%
\begin{table}
\caption[]{Bessel's equation.}
\label{tab:Bessel}
\begin{center}
\begin{tabular}{|l|l|}
\hline
$\sigma (z)$ & $z$ \\ \hline
$\widetilde{\sigma }(z)$ & $z^2 - \nu^{2} $
\\ \hline
$\widetilde{\tau }(z)$ & $1$ \\ \hline
$k$ & $ \pm 2i\nu$ \\ \hline
$\pi(z) $ & $\pm \nu \pm iz$ \\ \hline
$\tau(z) =\widetilde{\tau }+2\pi \qquad $ & $ 1+2\nu+2iz $ \\ \hline
$\lambda =k+\pi ^{\prime }$ & $i (2\nu+1) $ \\ \hline
$\varphi(z) $ & $z^{\pm \nu}e^{\pm iz}$ \\ \hline
$\rho (z)$ & $z^{2\nu}e^{2iz} $ \\ \hline
\end{tabular}
\end{center}
\end{table}
%


%
\section{Central Field: Spherical Harmonics\/}

The stationary Schr\"{o}dinger equation in the central
field with the potential energy $U(r) $ is given by%
\begin{equation}
\Delta \psi +\frac{2m}{\hbar ^{2}}\left( E-U(r) \right) \psi =0.
\label{s6}
\end{equation}%
The Laplace operator in the spherical coordinates $r,$ $\theta ,$ $\varphi $
has the form \cite{Ni:Su:Uv}, \cite{Ni:Uv}\/:%
\begin{equation}
\Delta =\Delta _{r}+\frac{1}{r^{2}}\Delta _{\omega }  \label{s7}
\end{equation}%
with%
\begin{equation}
\Delta _{r}=\frac{1}{r^{2}}\frac{\partial }{\partial r}\left( r^{2}\frac{%
\partial }{\partial r}\right) ,\quad \Delta _{\omega }=\frac{1}{\sin \theta }%
\frac{\partial }{\partial \theta }\left( \sin \theta \frac{\partial }{%
\partial \theta }\right) +\frac{1}{\sin ^{2}\theta }\frac{\partial ^{2}}{%
\partial \varphi ^{2}}.  \label{s8}
\end{equation}%
and separation of the variables $\psi =R(r) Y(\theta,\varphi) $ gives%
\begin{equation}
\Delta _{\omega }Y(\theta,\varphi) +\mu Y(\theta,\varphi) =0,  \label{s11}
\end{equation}%
\begin{equation}
\frac{1}{r^{2}}\frac{d}{dr}\left( r^{2}\frac{dR(r) }{dr}\right)
+\left( \frac{2m}{\hbar ^{2}}\left( E-U(r) \right) -\frac{%
\mu }{r^{2}}\right) R(r) =0.  \label{s12}
\end{equation}%
Bounded single-valued solutions of equation (\ref{s11}) on the sphere $S^{2}$
exist only when $\mu =l(l+1) $ with $l=0,1,2,\dots\;.$ They
are the spherical harmonics $Y=Y_{lm}(\theta,\varphi)$.
\smallskip

Looking for solutions in the form $Y=e^{im\varphi}f(\theta)\/$ with $m= 0, \pm 1, \pm 2,\dots$ one gets
\begin{equation}
\dfrac{1}{\sin \theta}\dfrac{d}{d\theta}\left(\sin \theta \,\dfrac{dg}{d\theta}\right)-\dfrac{m^2}{\sin^2 \theta} \,g + \mu g=0 . \label{s12f}
\end{equation}
%
%
\noindent
The following change of variables $\xi = \cos\theta$ and $F(\xi)=f(\theta)$ results in the generalized equation of hypergeometric type:
\begin{equation}
\left(1-\xi^2 \right) F^{\prime \prime}- 2\xi F^{\prime} + \left(\mu - \dfrac{m^2}{1-\xi^2 } \right) F =0 , \label{s12F}
\end{equation}
with $\sigma(\xi)=1-\xi^2, \, \widetilde{\tau }(\xi)=-2\xi\/$ and $\widetilde{\sigma }(\xi)=\mu \left(1-\xi^2\right) - m^{2}\/,$ which can be reduced to
the simpler form by the standard substitution $F=\varphi\, y\/:$
\begin{equation}
\left(1-\xi^2\right) y^{\prime \prime } - 2 \left(|m|+1\right) \xi \, y^{\prime }
+\left(\mu - |m|\left(|m|+1\right) \right) y=0  \label{A3F}
\end{equation}%
Indeed, by (\ref{A6})
\begin{equation}
\pi(\xi) = \pm \sqrt{(\mu - k) \xi^2 + k + m^2 - \mu} ,
\end{equation}
or%
\begin{equation}
\pi (\xi) = \left\{
\begin{array}{c}
\pm |m|, \quad k=\mu \\
\pm |m| \, \xi,\quad k=\mu - m^2%
\end{array}%
\right.
\end{equation}%
where we should choose the case when the linear function $\tau =\widetilde{\tau }%
+2\pi $ will have a negative derivative and a zero on the interval $(-1,1)$.

Then
\begin{equation}
\dfrac{\varphi^\prime}{\varphi} = \dfrac{- |m| \xi}{1-\xi^2}, \qquad \ln \varphi = - |m| \int \frac{\xi \, d\xi}{1-\xi^2}=\dfrac{1}{2} \ln\left(1-\xi^2\right)
\end{equation}
and
\begin{equation}
\varphi(\xi) = \left(1-\xi^2\right)^{|m|/2}=\left(\sin \theta\right)^{|m|} .
\end{equation}
(see our complementary Mathematica notebook and Table~\ref{tab:SphHarm} for further details of calculations).
\smallskip

The final result is given by
\begin{equation}
Y_{l\,m}(\theta, \varphi)= A_m \, \dfrac{e^{im\varphi}}{2^{|m|} \, l!} {\sqrt{\dfrac{2l+1}{4\pi}\, (l-m)!(l+m)!} } \;
\left(\sin \theta\right)^{|m|}\, P_{l-|m|}^{(|m|, |m|)}(\cos \theta) . \label{A3Y}
\end{equation}
Here, $P_n^{(\alpha, \, \alpha)}(\xi)$ are the Jacobi polynomials (Table~\ref{tab:cop}); $A_m=(-1)^m,\, m\ge0\/$ and $A_m=1,\, m<0\/.$
(See \cite{Ni:Su:Uv}, \cite{Ni:Uv}, \cite{Varshalovich1988} for more details.)
\smallskip
%

%
\begin{table}
\caption[]{Equation for spherical harmonics. }
\label{tab:SphHarm}
\begin{center}
\begin{tabular}{|l|l|}
\hline
$\sigma (\xi)$ & $1-\xi^2, \qquad \xi= \cos \theta , \quad 0 \le \theta \le \pi $ \\ \hline
$\widetilde{\sigma }(\xi)$ & $\mu (1-\xi^2) - m^{2} $ \\ \hline
$\widetilde{\tau }(\xi)$ & $-2\xi$ \\ \hline
$k$ & $ \mu -m^2 $ \\ \hline
$\pi(\xi) $ & $ -|m| \xi$ \\ \hline
$\tau(\xi) =\widetilde{\tau }+2\pi \qquad $ & $ -2 \left(|m|+1\right) \xi $ \\ \hline
$\lambda =k+\pi ^{\prime }$ & $\mu - |m|\left(|m|+1\right) $ \\ \hline
$\varphi(\xi) $ & $(1-\xi^2)^{|m|/2} = (\sin \theta)^{|m|}$ \\ \hline
$\rho (\xi)$ & $ (1-\xi^2)^{|m|} $ \\ \hline
$y_{n}(\xi)$ & \mybox{$N_{l\, m}\  P_{l-|m|}^{(|m|, \, |m|)}\left( \cos \theta \right), \quad n=l-|m| $} \\ \hline
$N_{l\, m}$ & \mybox{$\dfrac{1}{2^{|m|} \, l!} {\sqrt{\dfrac{2l+1}{2}\, (l-m)!(l+m)!} }$} \\ \hline
\end{tabular}
\end{center}
\end{table}
\smallskip


%
\section{Nonrelativistic Coulomb Problem\/}

In view of identity%
\begin{equation}
\frac{1}{r^{2}}\frac{d}{dr}\left( r^{2}\frac{dR}{dr}\right) =\frac{1}{r}%
\frac{d^{2}}{dr^{2}}\left( rR\right) , \label{IdentityR}
\end{equation}%
the substitution $F(r) =rR(r) $ into (\ref{s12}) results in the
standard radial equation%
\begin{equation}
F^{\prime \prime }+\left[ \frac{2m_{e}}{\hbar ^{2}}\left( E-U(r)
\right) -\frac{l(l+1)}{r^{2}}\right] F=0,\quad U(r)
=-\frac{Ze^{2}}{r}  \label{RadialCoulomb}
\end{equation}%
for the nonrelativistic Coulomb problem in spherical coordinates. In
dimensionless  units\/,%
\begin{equation}
F(r) =u(x) ,\quad x=\frac{r}{a_{0}},\quad
\varepsilon_0 =\frac{E}{E_{0}}\quad \left( a_{0}=\frac{\hbar ^{2}}{m_{e}e^{2}}%
,\quad E_{0}=\frac{e^{2}}{a_{0}}\right) \label{RadialUnitsCoulomb}
\end{equation}%
the radial equation is a generalized equation of hypergeometric type,%
\begin{equation}
u^{\prime \prime }+\left[ 2\left( \varepsilon_0 +\frac{Z}{x}\right) -\frac{%
l(l+1)}{x^{2}}\right] u=0,  \label{RadialXCoulomb}
\end{equation}%
where%
\begin{equation}
\sigma(x) =x,\qquad \widetilde{\tau }(x) =0,\qquad
\widetilde{\sigma }(x) =2\varepsilon_0 x^{2}+2Zx-l(l+1).  \label{RadialXCoefficients}
\end{equation}%
Therefore, one can utilize Nikiforov and Uvarov's approach in order to determine
the corresponding wave functions and discrete energy levels. \smallskip

We transform (\ref{RadialXCoulomb}) to the equation of hypergeometric type (\ref{A3}).
The linear function $\pi(x) $ takes the form%
\begin{equation}
\pi(x) =\frac{1}{2}\pm \sqrt{\frac{1}{4}-2\varepsilon_0
x^{2}-2x+l(l+1)+kx},\label{PiCoulomb}
\end{equation}%
or%
\begin{equation}
\pi(x) =\frac{1}{2}\pm \left\{
\begin{array}{c}
\sqrt{-2\varepsilon_0 }\ x+l+1/2,\quad k=2Z+(2l+1) \sqrt{%
-2\varepsilon_0 } \\
\sqrt{-2\varepsilon_0 }\ x-l-1/2,\quad k=2Z-(2l+1) \sqrt{%
-2\varepsilon_0 }%
\end{array}%
\right.
\end{equation}%
where we should choose the case when the linear function $\tau =\widetilde{\tau }%
+2\pi $ will have a negative derivative and a zero on $(0,+\infty)$:%
\begin{equation*}
\tau(x) =2\left( l+1-x\sqrt{-2\varepsilon_0 }\right) .
\end{equation*}%
This choice corresponds to%
\begin{equation*}
\pi(x) =l+1-x\sqrt{-2\varepsilon_0 },\quad \varphi(x) =
x^{l+1}\exp \left( -x\sqrt{-2\varepsilon_0 }\right)
\end{equation*}%
and%
\begin{equation*}
\lambda =k+\pi ^{\prime }=2\left[ Z-(l+1) \sqrt{-2\varepsilon_0 }%
\ \right] .
\end{equation*}

The energy values are given by (\ref{A7}):
\begin{equation}
\varepsilon_0 =\frac{E}{E_{0}}=-\frac{Z^{2}}{2\left( n_{r}+l+1\right) ^{2}}%
,\qquad E_{0}=\frac{e^{2}}{a_{0}}.\label{EnergyCoulomb}
\end{equation}%
Here, $n=n_{r}+l+1$ is known as the principal quantum number.\smallskip

In order to use the Rodrigues formula, one finds%
\begin{equation*}
\frac{\rho ^{\prime }}{\rho }=\frac{\tau -\sigma ^{\prime }}{\sigma }=\frac{%
2l+1}{x}-\frac{2Z}{n},
\end{equation*}%
or%
\begin{equation*}
\rho(x) =x^{2l+1}\exp \left( -\frac{2Z}{n}x\right) ,\qquad x=%
\frac{r}{a_{0}}.
\end{equation*}%
Therefore,%
\begin{equation}
y_{n_{r}}(x) =\frac{B_{n_{r}}}{x^{2l+1}e^{-\eta }}\frac{d^{n_{r}}%
}{dx^{n_{r}}}\left( x^{n_{r}+2l+1}e^{-\eta }\right) =L_{n_{r}}^{2l+1}(\eta),\label{WavefunctionsLaguerreCoulomb}
\end{equation}%
where%
\begin{equation*}
\eta =\frac{2Z}{n}x=\frac{2Z}{n}\left( \frac{r}{a_{0}}\right) =2x\sqrt{-2\varepsilon_0} ,
\end{equation*}%
and, up to a constant,%
\begin{equation}
F(r) =rR(r) =C_{nl}\ \eta ^{l+1}e^{-\eta
/2}L_{n_{r}}^{2l+1}(\eta).\label%
{WavefunctionWeightLaguerreCoulomb}
\end{equation}%
In view of the normalization condition%
\begin{equation*}
1=\int_{0}^{\infty }F^{2}\ dr=C_{nl}^{2}\left( \frac{na_{0}}{2Z}\right)
\int_{0}^{\infty }\eta ^{2l+2} e^{-\eta} \left[ L_{n_{r}}^{2l+1}(\eta) %
\right] ^{2}\ d\eta ,
\end{equation*}%
the three-term recurrence relation%
\begin{equation*}
\eta L_{n}^{\alpha }=-(n+1) L_{n+1}^{\alpha }+(\alpha+2n+1) L_{n}^{\alpha }-(\alpha+n) L_{n-1}^{\alpha },
\end{equation*}%
and the orthogonality property of the Laguerre polynomials (Table~\ref{tab:cop}\/), one gets%
\begin{equation}
C_{nl}^{2}=\frac{Z}{a_{0}n^{2}}\frac{(n-l-1)!}{(n+l)!}.\label{ConstantLaguerreCoulomb}
\end{equation}
More details can be found in \cite{Ni:Uv} and \cite{Suslov2020}.
(See also Appendices~\ref{app:A} and~\ref{app:B}.)
\smallskip

As a result, the nonrelativistic Coulomb wave functions obtained by the method of
separation of the variables in spherical coordinates, see above, are%
\begin{equation}
\psi =\psi _{nlm}({\bf r}) =R_{nl}(r) \ Y_{lm}(\theta,\varphi),  \label{nrc1}
\end{equation}%
where $Y_{lm}(\theta,\varphi) $ are the spherical harmonics,
the radial functions $R_{nl}(r) $ are given in terms of the
Laguerre polynomials (Table~\ref{tab:cop}) \cite{Bethe:Sal}, \cite{Flugge}, \cite{La:Lif}, \cite{Ni:Uv}, \cite%
{Schiff} :%
\begin{equation}
R(r) =R_{nl}(r) =\frac{2}{n^{2}}\left( \frac{Z}{a_{0}%
}\right) ^{3/2}\sqrt{\frac{(n-l-1)!}{(n+l)!}}\
e^{-\eta /2}\eta ^{l}\ L_{n-l-1}^{2l+1}(\eta)  \label{nrc2}
\end{equation}%
with%
\begin{equation}
\eta =\frac{2Z}{n}\left( \frac{r}{a_{0}}\right) ,\qquad a_{0}=\dfrac{\hbar
^{2}}{{m_e}e^{2}}  \label{nrc2a}
\end{equation}%
and the normalization is%
\begin{equation}
\int_{0}^{\infty }R_{nl}^{2}(r) r^{2}\ dr=1.  \label{nrc2b}
\end{equation}%
Here $n=1,2,3,\dots$ is the principal quantum number of the hydrogen-like
atom in the nonrelativistic Schr\"{o}dinger theory; $l=0,1,\dots,n-1$ and
$m=-l,-l+1,\dots,l-1,l$ are the quantum numbers of the angular momentum and
its projection on the $z$-axis, respectively. The corresponding discrete
energy levels in the cgs units are given by Bohr's formula \cite{Schrodinger2010}\/:%
\begin{equation}
E=E_{n}=-\frac{{m_e}Z^{2}e^{4}}{2\hbar ^{2}n^{2}},  \label{nrc2c}
\end{equation}%
where $n=1,2,3,\dots$ is the principal quantum number; they do not depend
on the quantum number of the orbital angular momenta $l\/.$
%


%
\begin{table}
\caption[]{The Schr\"{o}dinger equation for the Coulomb potential $U(r)=-Ze^{2}/r$.
Dimensionless quantities: $r=a_{0}x$,
$a_{0}=\hbar^{2}/(m_{e}e^{2})\simeq 0.5\cdot 10^{-8}\;\text{cm}$,
$E_{0}=e^{2}/a_{0}$, $R(r) =F(x) =u(x)/x$.}
\label{tab:Coulomb}
\begin{center}
\begin{tabular}{|l|l|}
\hline
$\sigma (x)$ & $x$ \\ \hline
$\widetilde{\sigma }(x)$ & $2\varepsilon_0 \;x^{2}+2Z\;x-l(l+1)\/, \quad {\varepsilon_0 =E/E_0 } $
\\ \hline
$\widetilde{\tau }(x)$ & $0$ \\ \hline
$k$ & $2Z-(2l+1) \sqrt{-2\varepsilon_0\/}$ \\ \hline
$\pi(x) $ & $l+1-\sqrt{-2\varepsilon_0 }\;x$ \\ \hline
$\tau(x) =\widetilde{\tau }+2\pi \qquad $ & $2\left( l+1-\sqrt{%
-2\varepsilon_0 }\;x\right) $ \\ \hline
$\lambda =k+\pi ^{\prime }$ & $2\left( Z-(l+1)\sqrt{-2\varepsilon_0 }\right) $
\\ \hline
$\varphi(x) $ & $x^{l+1}e^{-x\sqrt{-2\varepsilon_0 }}$ \\ \hline
$\rho (x)$ & $x^{2l+1}e^{-(2Z\ x)/n},\quad x=r/a_{0}$ \\ \hline
$y_{n_{r}}(x)$ & \mybox{$C_{n_{r}}L_{n_{r}}^{2l+1}(\eta) ,\quad \eta =
\dfrac{2Z}{n}x=\dfrac{2Z}{n}\left( \dfrac{r}{a_{0}}\right)$} \\ \hline
$C_{n_{r}}^{2}$ & \mybox{$\dfrac{Z}{a_{0}n^{2}}\dfrac{(n-l-1)!}{(n+l)!},\quad n_{r}=n-l-1$} \\ \hline
\end{tabular}
\end{center}
\end{table}
%


{\bf{Remark\/.}} In the original Nikiforov--Uvarov approach, the variable coefficients $\sigma(x)$ and $\tau(x)$ in (\ref{A3}) should not depend on the eigenvalue $\lambda$.
Here, we obtain
\begin{equation}
xy^{\prime\prime}+2\left(l+1-x\sqrt{-2\varepsilon_0}\right)y^{\prime}+2\left(Z-(l+1)\sqrt{-2\varepsilon_0}\right)y=0\/.
\end{equation}
Nonetheless, the change of variables $y(x)=Y(\eta)$ with $\eta=2x\sqrt{-2\varepsilon_0}$ results in
\begin{equation}
\eta Y^{\prime\prime}+\left(2l+2-\eta\right)Y^{\prime}+\left(\dfrac{Z}{\sqrt{-2\varepsilon_0}} - l-1\right)Y=0\/.
\end{equation}
Thus the Nikiforov--Uvarov method can be applied and the uniqueness of square integrable solutions holds.
\smallskip

In a similar fashion, one can consider solution of the Kepler problem in the so-called parabolic coordinates,
which is important in the theory of Stark effect  \cite{SchrodingerParabolic}, \cite{La:Lif}.

\section{Relativistic Schr\"{o}dinger Equation\/}

The stationary relativistic Schr\"{o}dinger equation has the form \cite{Barleyetal2021}, \cite{Dav}, \cite{Greiner}, \cite{Schiff}:
\begin{equation}
\left( E+\frac{Ze^{2}}{r}\right) ^{2}\chi =\left( -\hbar ^{2}c^{2}\Delta
+m^{2}c^{4}\right) \chi .  \label{rel6}
\end{equation}%
%
We separate the variables in spherical coordinates, $\chi(r,\theta,\varphi)
=R(r)Y_{lm}(\theta ,\varphi ),$ where $Y_{lm}(\theta
,\varphi )$ are the spherical harmonics with familiar properties \cite%
{Varshalovich1988}.\ As a result,%
\begin{equation}
\frac{1}{r^{2}}\frac{d}{dr}\left( r^{2}\frac{dR}{dr}\right) +\left[ \frac{%
\left( E+Ze^{2}/r\right) ^{2}-m^{2}c^{4}}{\hbar ^{2}c^{2}}-\frac{l(l+1)}{r^{2}}\right]
R=0\qquad (l=0,1,2,\dots). \label{sol0}
\end{equation}%
In the dimensionless quantities,%
\begin{equation}
\varepsilon =\frac{E}{mc^{2}},\qquad x=\beta r=\frac{mc}{\hbar }r,\qquad \mu
=\frac{Ze^{2}}{\hbar c},  \label{sol1}
\end{equation}%
for the new radial function,%
\begin{equation}
R(r)=F(x)=\frac{u(x)}{x},  \label{sol2}
\end{equation}%
one gets%
\begin{equation}
\frac{1}{x^{2}}\frac{d}{dx}\left( x^{2}\frac{dF}{dx}\right) +\left[ \left(
\varepsilon +\frac{\mu }{x}\right) ^{2}-1-\frac{l(l+1)}{x^{2}}%
\right] F=0.  \label{sol3}
\end{equation}%
Given the identity $(x^2F')'=x(xF)''$, we obtain%
\begin{equation}
u^{\prime \prime }+\left[ \left( \varepsilon +\frac{\mu }{x}\right) ^{2}-1-%
\frac{l(l+1)}{x^{2}}\right] u=0.  \label{sol5}
\end{equation}%
This is the generalized equation of hypergeometric form (\ref{A1}),
when%
\begin{equation}
\sigma(x) =x,\qquad \widetilde{\tau }(x)\equiv 0,\qquad
\widetilde{\sigma }(x)=\left( \varepsilon ^{2}-1\right) x^{2}+2\mu
\varepsilon x+\mu ^{2}-l(l+1)  \label{sol6}
\end{equation}%
The normalization condition takes the form:
\begin{equation}
\int_{0}^{\infty }R^{2}(r)r^{2}\ dr=1,\qquad \text{or\quad }\int_{0}^{\infty
}u^{2}(x)\;dx=\beta ^{3},\quad \beta =\frac{mc}{\hbar }.  \label{sol7}
\end{equation}%
Here, $u=\varphi \, y\/.$ For further computational details, see our supplementary Mathematica notebook,
as well as Refs.~\cite{Barleyetal2021} and \cite{Ni:Uv}.
Final results are presented in Table~\ref{tab:RelSchr}. \smallskip

In particular, one gets
Schr\"{o}dinger's fine structure formula (for a charged
spin-zero particle in the Coulomb field):
\begin{equation}
E=E_{n_{r}}=\frac{mc^{2}}{\sqrt{1+\left( \dfrac{\mu }{n_{r}+\nu +1}\right)
^{2}}}\qquad \left( n=n_{r}=0,1,2,\dots\right) .  \label{sol16}
\end{equation}%
Here,%
\begin{equation}
\mu =\frac{Ze^{2}}{\hbar c},\qquad \nu =-\frac{1}{2}+\sqrt{\left( l+\frac{1}{%
2}\right) ^{2}-\mu ^{2}}.  \label{sol17}
\end{equation}

The corresponding eigenfunctions are given by the Rodrigues-type formula%
\begin{equation}
y_{n}(x) =\frac{B_{n}}{x^{2\nu +1}e^{-2ax}}\frac{d^{n}}{dx^{n}}%
\left( x^{n+2\nu +1}e^{-2ax}\right) =C_{n}L_{n}^{2\nu +1}(2ax) .
\label{sol18}
\end{equation}%
Up to a constant, they are Laguerre polynomials (Table~\ref{tab:cop}\/). In view of the normalization condition (\ref{sol7}):
\begin{align}
\beta ^{3} &=\int_{0}^{\infty }u^{2}(x)\;dx=C_{n}^{2}\int_{0}^{\infty }
\left[ \varphi(x) L_{n}^{2\nu +1}(2ax) \right]
^{2}\ dx  \notag \\
&=\frac{C_{n}^{2}}{(2a) ^{2\nu +3}}\int_{0}^{\infty } e^{-\xi} \, \xi ^{2\nu
+2}\left( L_{n}^{2\nu +1}(\xi) \right) ^{2}\ d\xi ,\quad \xi
=2ax.  \label{sol19}
\end{align}%
The corresponding integral is given by (see \cite{Sus:Trey}, \cite{Suslov2020},
and Appendix~\ref{app:B}):%
\begin{equation}
I_{1}= J^{\alpha \alpha}_{nn 1} =\int_{0}^{\infty }e^{-x}x^{\alpha +1}\left( L_{n}^{\alpha }(x)
\right) ^{2}\ dx=(\alpha +2n+1) \frac{\Gamma(\alpha +n+1)}{n!}.  \label{sol20}
\end{equation}%
As a result,%
\begin{equation}
C_{n}=2(a\beta) ^{3/2}(2a) ^{\nu }\sqrt{\frac{n!}{\Gamma(2\nu+n+2)}}.  \label{sol21}
\end{equation}%
The normalized radial eigenfunctions, corresponding to the relativistic energy levels (%
\ref{sol16}), are explicitly given by%
\begin{equation}
R(r) =R_{n_{r}}(r) =2(a\beta)^{3/2}%
\sqrt{\frac{n_{r}!}{\Gamma(2\nu+n_{r}+2)}}\;\xi ^{\nu}e^{-\xi /2}\;L_{n_{r}}^{2\nu +1}(\xi),
\label{sol22}
\end{equation}%
where%
\begin{equation}
\xi =2ax=2a\beta r=2\sqrt{1-\varepsilon ^{2}}\,\frac{mc}{\hbar}\,r.
\label{sol23}
\end{equation}%
(More details can be found in \cite{Barleyetal2021}, \cite{Dav}, \cite{Schiff}.)
\smallskip

Let us analyze a nonrelativistic limit of Schr\"{o}dinger's fine structure
formula (\ref{sol16})--(\ref{sol17}):%
\begin{align}
\varepsilon_{\text{Schr\"{o}dinger}} = \frac{E_{n_{r},\, l}}{mc^{2}} &=\frac{1}{\sqrt{1+\dfrac{\mu ^{2}}{\left(
n_{r}+\frac{1}{2}+\sqrt{\left( l+\frac{1}{2}\right) ^{2}-\mu ^{2}}\right)^2 }}}
\notag \\
&=1-\frac{\mu ^{2}}{2n^{2}}-\frac{\mu ^{4}}{2n^{4}}\left( \frac{n}{l+1/2}-%
\frac{3}{4}\right) +\text{O}(\mu^{6}) ,\quad \mu \rightarrow 0,
\label{lim2}
\end{align}
which can be derived by a direct Taylor expansion and/or verified by a
computer algebra system (see our supplementary Mathematica notebook). Here, $n=n_{r}+l+1$ is the corresponding
nonrelativistic principal quantum number. The first term in this expansion
is simply the rest mass energy $E_{0}=mc^{2}$ of the charged spin-zero
particle, the second term coincides with the energy eigenvalue in the
nonrelativistic Schr\"{o}dinger theory and the third term gives the
so-called fine structure of the energy levels, which removes the degeneracy between
states of the same $n$ and different $l\/.$


\begin{table}
\caption[]{Relativistic Schr\"{o}dinger equation for Coulomb potential
$U(r)=-Ze^{2}/r$.
Dimensionless quantities: $\varepsilon = E/(mc^{2})$, $x=\beta r=(mc/\hbar)r$,
$\mu = Ze^{2}/(\hbar c)$, $R(r) =F(x) =u(x)/x$.}
\label{tab:RelSchr}
\begin{center}
\begin{tabular}{|l|l|}
\hline
$\sigma (x)$ & $x$ \\ \hline
$\widetilde{\sigma }(x)$ & $\left( \varepsilon ^{2}-1\right) \;x^{2}+2\mu
\varepsilon \;x+\mu ^{2}-l(l+1) $ \\ \hline
$\widetilde{\tau }(x)$ & $0$ \\ \hline
$k$ & \mybox{$2\mu \varepsilon -(2\nu+1) \sqrt{1-\varepsilon ^{2}}%
,\quad \nu =-\dfrac{1}{2}+\sqrt{\left( l+\dfrac{1}{2}\right) ^{2}-\mu ^{2}}$}
\\ \hline
$\pi(x) $ & $\nu +1-a\;x,\qquad a=\sqrt{1-\varepsilon ^{2}}$ \\
\hline
$\tau(x) =\widetilde{\tau }+2\pi $ & $2\left( \nu
+1-a\;x\right), \quad \tau^\prime < 0 $ \\ \hline
$\lambda =k+\pi ^{\prime }$ & $2\left( \mu \varepsilon -(\nu +1)a\right) $
\\ \hline
$\varphi(x) $ & $x^{\nu +1}e^{-a\ x}$ \\ \hline
$\rho (x)$ & $x^{2\nu +1}e^{-2a\ x}$ \\ \hline
$y_{n_{r}}(x)$ & \mybox{$C_{n_{r}}L_{n_{r}}^{2\nu +1}(\xi) ,\quad \xi
=2ax=2a\beta r=2\sqrt{1-\varepsilon ^{2}}\dfrac{mc}{\hbar }r$} \\ \hline
$C_{n_{r}}$ & \mybox{$2(a\beta)^{3/2}(2a)^{\nu }\sqrt{%
\dfrac{n_{r}!}{(\nu +n_{r}+1)\Gamma( 2\nu +n_{r}+2)}}$} \\ \hline
\end{tabular}
\end{center}
\end{table}
%


%
Once again, our equation,
\begin{equation}
x y^{\prime\prime} + 2(\nu+1-a x)y^{\prime} +2\left(\mu\varepsilon - (\nu+1) a\right) y=0\/,
\end{equation}
by the change of variables $y(x)=Y(\xi)$ with $\xi=2ax$ can be transformed into the required form:
\begin{equation}
\xi Y^{\prime\prime} + (2\nu+2-\xi) Y^{\prime}  \left(\dfrac{\mu\varepsilon}{\sqrt{1-\varepsilon^2}}-\nu-1\right) Y=0\/.
\end{equation}
Therefore, the set of the square integrable solutions above is unique.

\section{Relativistic Coulomb Problem: Dirac Equation\/}
\subsection{System of radial equations}
The radial Dirac equations are derived in Refs.~\cite{Ni:Uv}, \cite{Sus:Trey}, and \cite{Suslov2020} by separation of variables in spherical coordinates
(see also \cite{Bethe:Sal}, \cite{Dav}, \cite{Flugge}, \cite{Fock1978}, and \cite{Greiner}).
Then the radial functions $F(r)$ and $G(r) $ satisfy the system of two first-order
ordinary differential equations%
\begin{align}
\dfrac{dF}{dr}+\frac{1+\kappa }{r}\ F &=\dfrac{mc^{2}+E-U(r) }{%
\hbar c}\ G,  \label{ds12} \\
\dfrac{dG}{dr}+\frac{1-\kappa }{r}\ G &=\dfrac{mc^{2}-E+U(r) }{%
\hbar c}\ F,  \label{ds13}
\end{align}
where $\kappa =\kappa _{\pm }=\pm(j+1/2) =\pm 1,\pm 2,\pm 3,\dots$ respectively.
For the relativistic Coulomb problem, when $U=-Ze^{2}/r,$ we introduce the
dimensionless quantities%
\begin{equation}
\varepsilon =\frac{E}{mc^{2}},\qquad x=\beta r=\frac{mc}{\hbar }r,\qquad \mu
=\frac{Ze^{2}}{\hbar c}  \label{rrc1}
\end{equation}%
and change the variable in radial functions%
\begin{equation}
f(x) =F(r) ,\qquad g(x) =G(r) .  \label{rrc2}
\end{equation}%
The Dirac radial system 
becomes%
\begin{align}
\dfrac{df}{dx}+\frac{1+\kappa }{x}\ f &=\left( 1+\varepsilon +\frac{\mu }{x}%
\right) g,  \label{rrc3} \\
\dfrac{dg}{dx}+\frac{1-\kappa }{x}\ g &=\left( 1-\varepsilon -\frac{\mu }{x}%
\right) f.  \label{rrc4}
\end{align}%
(One can show later that in the nonrelativistic limit, $c\rightarrow \infty \/,$ the
following estimate holds: $\left| f(x) \right| \gg {\left| g(x) \right|}\/;$ see,
for example, Refs.~\cite{Ni:Uv}, \cite{Sus:Trey}, and \cite{Suslov2020}
for more details.)
\smallskip
\subsection{Decoupling of the radial system}
We follow \cite{Ni:Uv} with somewhat different details. Let us rewrite the
system (\ref{rrc3})--(\ref{rrc4}) in a matrix form. If%
\begin{equation}
u=\begin{pmatrix}
u_{1}\medskip \\
u_{2}%
\end{pmatrix} =
\begin{pmatrix}
xf(x) \medskip \\
xg(x)%
\end{pmatrix},\qquad
u^{\prime }=
\begin{pmatrix}
u_{1}^{\prime }\medskip \\
u_{2}^{\prime }%
\end{pmatrix}.  \label{rrc5}
\end{equation}%
Then%
\begin{equation}
u^{\prime }=Au,  \label{rrc6}
\end{equation}%
where%
\begin{equation}
A=\begin{pmatrix}
a_{11} & a_{12}\medskip \\
a_{21} & a_{22}%
\end{pmatrix} =
\begin{pmatrix}
-\dfrac{\kappa }{x} & 1+\varepsilon +\dfrac{\mu }{x}\; \\[\medskipamount]
1-\varepsilon -\dfrac{\mu }{x} & \dfrac{\kappa }{x}
\end{pmatrix}.  \label{rrc7}
\end{equation}%
To find $u_{1}(x)$, we eliminate $u_{2}(x) $ from
the system (\ref{rrc6}), obtaining a second-order differential equation%
\begin{align}
&u_{1}^{\prime \prime }-\left( a_{11}+a_{22}+\dfrac{a_{12}^{\prime }}{a_{12}%
}\right) u_{1}^{\prime }  \label{rrc8} \\
&\qquad +\left( a_{11}a_{22}-a_{12}a_{21}-a_{11}^{\prime }+\dfrac{%
a_{12}^{\prime }}{a_{12}}\;a_{11}\right) u_{1}=0.  \notag
\end{align}%
Similarly, eliminating $u_{1}(x)$, one gets an equation for $u_{2}(x)$:
\begin{align}
&u_{2}^{\prime \prime }-\left( a_{11}+a_{22}+\dfrac{a_{21}^{\prime }}{a_{21}%
}\right) u_{2}^{\prime }  \label{rrc9} \\
&\qquad +\left( a_{11}a_{22}-a_{12}a_{21}-a_{22}^{\prime }+\dfrac{%
a_{21}^{\prime }}{a_{21}}\;a_{22}\right) u_{2}=0.  \notag
\end{align}

The components of the matrix $A$ have the following generic form%
\begin{equation}
a_{ik}=b_{ik}+c_{ik}/x,  \label{me1}
\end{equation}%
where $b_{ik}$ and $c_{ik}$ are constants. Equations (\ref{rrc8}) and (\ref%
{rrc9}) are not generalized equations of hypergeometric type (\ref{A1}).
Indeed,%
\begin{equation*}
\dfrac{a_{12}^{\prime }}{a_{12}}=-\frac{c_{12}}{c_{12}x+b_{12}x^{2}},
\end{equation*}%
and the coefficients of $u_{1}^{\prime }(x) $ and $u_{1}(x) $ in (\ref{rrc8}) are%
\begin{align*}
&a_{11}+a_{22}+\dfrac{a_{12}^{\prime }}{a_{12}}=\frac{p_{1}(x)
}{x}-\frac{c_{12}}{c_{12}x+b_{12}x^{2}}, \\
&a_{11}a_{22}-a_{12}a_{21}-a_{11}^{\prime }+\dfrac{a_{12}^{\prime }}{a_{12}}%
a_{11}=\frac{p_{2}(x) }{x^{2}}-\frac{c_{12}(c_{11}+b_{11}x)}{(c_{12}+b_{12}x) x^{2}},
\end{align*}%
where $p_{1}(x) $ and $p_{2}(x) $ are polynomials of
degrees at most one and two, respectively (see supplementary Mathematica notebook for their explicit forms). Equation (\ref{rrc8}) will become
a generalized equation of hypergeometric type (\ref{A1}) with $\sigma
(x) =x$ if either $b_{12}=0$ or $c_{12}=0\/.$
\smallskip
\subsection{Similarity transformation\/}
The following
consideration helps. By a linear transformation%
\begin{equation}
\begin{pmatrix}
v_{1}\medskip \\
v_{2}%
\end{pmatrix} =C
\begin{pmatrix}
u_{1}\medskip \\
u_{2}%
\end{pmatrix} \label{rrc10}
\end{equation}%
with a nonsingular matrix $C$ that is independent of $x$, we transform the
original system (\ref{rrc6}) to a similar one%
\begin{equation}
v^{\prime }=\widetilde{A}v,  \label{rrc11}
\end{equation}%
where%
\begin{equation*}
v=\begin{pmatrix}
v_{1}\medskip \\
v_{2}%
\end{pmatrix} ,\qquad \widetilde{A}=CAC^{-1}=
\begin{pmatrix}
\widetilde{a}_{11} & \widetilde{a}_{12}\medskip \\
\widetilde{a}_{21} & \widetilde{a}_{22}%
\end{pmatrix}.
\end{equation*}%
The new coefficients $\widetilde{a}_{ik}$ are linear combinations of the
original ones $a_{ik}.$ Hence they have a similar form%
\begin{equation}
\widetilde{a}_{ik}=\widetilde{b}_{ik}+\widetilde{c}_{ik}/x,  \label{me2}
\end{equation}%
where $\widetilde{b}_{ik}$ and $\widetilde{c}_{ik}$ are constants.
\smallskip

The equations for $v_{1}(x) $ and $v_{2}(x) $ are
similar to (\ref{rrc8}) and (\ref{rrc9}):%
\begin{align}
&v_{1}^{\prime \prime }-\left( \widetilde{a}_{11}+\widetilde{a}_{22}+\dfrac{%
\widetilde{a}_{12}^{\prime }}{\widetilde{a}_{12}}\right) v_{1}^{\prime }
\label{rrc12} \\
&\qquad +\left( \widetilde{a}_{11}\widetilde{a}_{22}-\widetilde{a}_{12}%
\widetilde{a}_{21}-\widetilde{a}_{11}^{\prime }+\dfrac{\widetilde{a}%
_{12}^{\prime }}{\widetilde{a}_{12}}\;\widetilde{a}_{11}\right) v_{1}=0,
\notag
\end{align}%
\begin{align}
&v_{2}^{\prime \prime }-\left( \widetilde{a}_{11}+\widetilde{a}_{22}+\dfrac{%
\widetilde{a}_{21}^{\prime }}{\widetilde{a}_{21}}\right) v_{2}^{\prime }
\label{rrc13} \\
&\qquad +\left( \widetilde{a}_{11}\widetilde{a}_{22}-\widetilde{a}_{12}%
\widetilde{a}_{21}-\widetilde{a}_{22}^{\prime }+\dfrac{\widetilde{a}%
_{21}^{\prime }}{\widetilde{a}_{21}}\;\widetilde{a}_{22}\right) v_{2}=0.
\notag
\end{align}%
The calculation of the coefficients in (\ref{rrc12}) and (\ref{rrc13}) is
facilitated by a similarity of the matrices $A$ and $\widetilde{A}:$%
\begin{equation*}
\widetilde{a}_{11}+\widetilde{a}_{22}=a_{11}+a_{22},\qquad \widetilde{a}_{11}%
\widetilde{a}_{22}-\widetilde{a}_{12}\widetilde{a}%
_{21}=a_{11}a_{22}-a_{12}a_{21}.
\end{equation*}%
By a previous consideration, in order for (\ref{rrc12}) to be an equation of
hypergeometric type, it is sufficient to choose either $\widetilde{b}_{12}=0$
or $\widetilde{c}_{12}=0.$ Similarly, for (\ref{rrc13}): either $\widetilde{b%
}_{21}=0$ or $\widetilde{c}_{21}=0.$ These conditions impose certain
restrictions on our choice of the transformation matrix $C.$ Let%
\begin{equation}
C=\begin{pmatrix}
\alpha & \beta \smallskip \\
\gamma & \delta%
\end{pmatrix}.  \label{rrc14}
\end{equation}%
Then%
\begin{equation*}
C^{-1}=\dfrac{1}{\Delta}
\begin{pmatrix}
\delta & -\beta \smallskip \\
-\gamma & \alpha%
\end{pmatrix},\qquad \Delta =\det C=\alpha \delta -\beta \gamma ,
\end{equation*}%
and%
 \begin{flalign} \label{rrc15}
& \widetilde{A}=CAC^{-1}  \\ \notag
& =\dfrac{1}{\Delta}
\begin{pmatrix}
a_{11} \alpha \delta -a_{12} \alpha \gamma + a_{21}\beta \delta -a_{22} \beta \gamma
\; & a_{12} \alpha ^{2}- a_{21} \beta^{2}+(a_{22}-a_{11}) \alpha \beta \medskip \\
a_{21}\delta ^{2}-a_{12}\gamma ^{2}+(a_{11}-a_{22}) \gamma \delta
& a_{12}\alpha \gamma -a_{11}\beta \gamma +a_{22}\alpha \delta -a_{21}\beta
\delta%
\end{pmatrix}.  \phantom{pppp} 
 \end{flalign}%
(Here, we have corrected typos in Eqs.~(3.74) of \cite{Suslov2020}; see also \cite{Ni:Uv} and the supplementary Mathematica notebook.)
For the Dirac system (\ref{rrc6})--(\ref{rrc7}):%
\begin{alignat*}{2}
a_{11} &= -\dfrac{\kappa }{x},&\qquad a_{12}\medskip &= 1+\varepsilon +\dfrac{\mu}{x}, \\
a_{21} &= 1-\varepsilon -\dfrac{\mu }{x},& a_{22} &= \dfrac{\kappa }{x}
\end{alignat*}%
and%
\begin{align}
\Delta \ \widetilde{a}_{12} &=\alpha ^{2}-\beta ^{2}+\left( \alpha
^{2}+\beta ^{2}\right) \varepsilon +\dfrac{\left( \alpha ^{2}+\beta
^{2}\right) \mu +2\alpha \beta \kappa }{x},  \label{rrc16} \\
\Delta \ \widetilde{a}_{21} &=\delta ^{2}-\gamma ^{2}-\left( \delta
^{2}+\gamma ^{2}\right) \varepsilon -\dfrac{\left( \delta ^{2}+\gamma
^{2}\right) \mu +2\gamma \delta \kappa }{x}.  \label{rrc17}
\end{align}%
\begin{equation*}
\begin{array}{ccccc}
\text{The} & \text{condition} & \widetilde{b}_{12}=0 & \text{yields} &
(1+\varepsilon) \alpha ^{2}-(1-\varepsilon) \beta
^{2}=0, \\
" & " & \widetilde{c}_{12}=0 & " & \left( \alpha ^{2}+\beta ^{2}\right) \mu
+2\alpha \beta \kappa =0, \\
" & " & \widetilde{b}_{21}=0 & " & (1+\varepsilon) \gamma
^{2}-(1-\varepsilon) \delta ^{2}=0, \\
" & " & \widetilde{c}_{21}=0 & " & \left( \delta ^{2}+\gamma ^{2}\right) \mu
+2\gamma \delta \kappa =0.%
\end{array}%
\end{equation*}%
We see that there are several possibilities to choose the elements
$\alpha ,$ $\beta ,$ $\gamma ,$ $\delta $ of the transition matrix~$C.$ All
quantum mechanics textbooks use the original one, namely, $\widetilde{b}%
_{12}=0$ and $\widetilde{b}_{21}=0,$ due to Darwin \cite{Dar}
and Gordon \cite{Gor}.
Nikiforov and
Uvarov \cite{Ni:Uv} take another path, they choose $\widetilde{c}_{12}=0$
and $\widetilde{c}_{21}=0$ and show that it is more convenient for taking
the nonrelativistic limit $c\rightarrow \infty$. These conditions are
satisfied if%
\begin{equation}
C=\begin{pmatrix}
\mu & \smallskip \nu -\kappa \\
\smallskip \nu -\kappa & \mu%
\end{pmatrix},  \label{rrc18}
\end{equation}%
where $\nu =\sqrt{\kappa ^{2}-\mu ^{2}},$ and we finally arrive at the
following system of first-order equations for $v_{1}(x)$ and $v_{2}(x)$:
\begin{align}
v_{1}^{\prime } &=\left( \dfrac{\varepsilon \mu }{\nu }-\dfrac{\nu }{x}%
\right) v_{1}+\left( 1+\dfrac{\varepsilon \kappa }{\nu }\right) v_{2},
\label{rrc19} \\
v_{2}^{\prime } &=\left( 1-\dfrac{\varepsilon \kappa }{\nu }\right)
v_{1}+\left( \dfrac{\nu }{x}-\dfrac{\varepsilon \mu }{\nu }\right) v_{2}.
\label{rrc20}
\end{align}%
Here%
\begin{equation}
\text{Tr}\ \widetilde{A}=\widetilde{a}_{11}+\widetilde{a}_{22}=0,\quad \det
\widetilde{A}=\varepsilon ^{2}-1+\dfrac{2\varepsilon \mu }{x}-\dfrac{\nu ^{2}%
}{x^{2}},\quad \nu ^{2}=\kappa ^{2}-\mu ^{2},  \label{rrc21}
\end{equation}%
which is simpler than the original choice in 
\cite{Ni:Uv}. The corresponding
second-order differential equations (\ref{rrc12})--(\ref{rrc13}) become%
\begin{align}
v_{1}^{\prime \prime }+\dfrac{\left( \varepsilon ^{2}-1\right)
x^{2}+2\varepsilon \mu x-\nu(\nu+1)}{x^{2}}\;v_{1} &=0,
\label{rrc22} \\
v_{2}^{\prime \prime }+\dfrac{\left( \varepsilon ^{2}-1\right)
x^{2}+2\varepsilon \mu x-\nu(\nu-1)}{x^{2}}\;v_{2} &=0.
\label{rrc23}
\end{align}%
They are generalized equations of hypergeometric type (\ref{A1}) of the
simplest form $\widetilde{\tau }=0,$ thus resembling the one-dimensional
Schr\"{o}dinger equation; the second equation can be obtained from the first one
by replacing $\nu \rightarrow -\nu $
%
(see also Eqs.~(3.81)--(3.82) in Ref.~\cite{Suslov2020}).
\smallskip
\subsection{Nikiforov--Uvarov paradigm\/}
All details of the calculations 
are presented in Table~\ref{tab:Dirac}
(see also our supplementary Mathematica notebook and Refs.~\cite{Ni:Uv}, \cite{Sus:Trey}, and \cite{Suslov2020} for more details).
Then the corresponding energy levels $\varepsilon
=\varepsilon _{n}$ are determined by%
\begin{equation}
\varepsilon \mu =a(\nu+n+1) ,  \label{rrc31}
\end{equation}%
and the eigenfunctions are given by the Rodrigues--type formula%
\begin{equation}
y_{n}(x) =\frac{C_{n}}{\rho(x) }\left( \sigma
^{n}(x) \rho(x) \right) ^{(n)}=C_{n}\;x^{-2\nu -1}e^{2ax}\frac{d^{n}}{dx^{n}}\left( x^{2\nu
+n+1}e^{-2ax}\right) .  \label{rrc32}
\end{equation}%
These functions are, up to certain constants, 
Laguerre polynomials $L_{n}^{2\nu +1}(\xi) $ (Table~\ref{tab:cop}) with $\xi =2ax.$ %
%
The corresponding eigenfunctions have the form%
\begin{equation}
v_{1}(x) =\left\{
\begin{array}{l}
0,\qquad n=0,\medskip \\
A_{n}\xi ^{\nu +1}e^{-\xi /2}L_{n-1}^{2\nu +1}(\xi) ,\qquad
n=1,2,3,\dots\;.%
\end{array}%
\right.  \label{rrc34}
\end{equation}%
They are square integrable functions on $(0,\infty)$. The
counterparts are%
\begin{equation}
v_{2}(x) =B_{n}\xi ^{\nu }e^{-\xi /2}L_{n}^{2\nu -1}(\xi) ,\qquad n=0,1,2,\dots\;.  \label{rrc35}
\end{equation}%
It is easily seen that the solution $\varepsilon =-\nu /\kappa $
is included in this formula when $n=0.$
\smallskip

As a result,%
\begin{align}
xf(x) &= \frac{B_{n}}{2\nu(\kappa-\nu)}\xi ^{\nu
}e^{-\xi /2}\left( f_{1}\xi L_{n-1}^{2\nu +1}(\xi)
+f_{2}L_{n}^{2\nu -1}(\xi) \right) ,  \label{rrc40} \\
xg(x) &= \frac{B_{n}}{2\nu(\kappa-\nu)}\xi ^{\nu
}e^{-\xi /2}\left( g_{1}\xi L_{n-1}^{2\nu +1}(\xi)
+g_{2}L_{n}^{2\nu -1}(\xi) \right) ,  \label{rrc41}
\end{align}%
where%
\begin{equation}
f_{1}=\frac{a\mu }{\varepsilon \kappa -\nu },\quad f_{2}=\kappa -\nu ,\quad
g_{1}=\frac{a(\kappa-\nu)}{\varepsilon \kappa -\nu },\quad
g_{2}=\mu  . \label{rrc42}
\end{equation}%
%
(These formulas remain valid for $n=0;$ in this case the terms containing $%
L_{-1}^{2\nu +1}(\xi) $ have to be taken to be zero.) Thus we
derive the representation for the radial functions 
up to the
constant $B_{n}$ in terms of Laguerre polynomials (Table~\ref{tab:cop}\/). The normalization condition
\begin{equation}
\int_{{\mathbb{R}}^{3}}\psi^{\dagger }\ \psi \ dv=\int_{0}^{\infty
}r^{2}\left( F^{2}(r) +G^{2}(r) \right) \, dr=1
\label{ds15}
\end{equation}%
gives the value of this constant as follows \cite{Ni:Uv}:
\begin{equation}
B_{n}=a\beta ^{3/2}\sqrt{\frac{(\kappa-\nu)(\varepsilon\kappa-\nu)n!}{\mu \Gamma(n+2\nu)}}.
\label{rrc43}
\end{equation}%
(This is verified in section~5.4 of Ref.~\cite{Suslov2020}. Observe that Eq.~(\ref{rrc43}%
) applies when $n=0.$)
\smallskip
\subsection{Summary: wave functions and energy levels\/}
The end results, namely, the complete wave functions and the corresponding discrete energy levels, are given by Eqs.\ (3.11)--(3.17) of Ref.~\cite{Suslov2020}.
The WKB, or semiclassical, approximation for the Dirac equation with Coulomb potential is discussed in \cite{Barleyetal2021}.
\medskip

The relativistic energy levels of an
electron in the central Coulomb field are given by%
\begin{equation}
E=E_{n_{r},\, j}=\frac{mc^{2}}{\sqrt{1+\mu ^{2}/(n_{r}+\nu)^{2}}}%
,\quad \mu =\frac{Ze^{2}}{\hbar c}\quad (n_{r}=0,1,2,\dots) . \label{lim0}
\end{equation}%
In Dirac's theory,%
\begin{equation}
\nu =\nu_{\text{Dirac}} =\sqrt{(j+1/2)^{2}-\mu^{2}},  \label{lim1}
\end{equation}%
where $j=1/2,3/2,5/2,\dots$ is the total angular momentum including the
spin of the relativistic electron. More details on the solution of this problem,
including the nonrelativistic limit, can be found in \cite{{Ni:Uv}}, \cite{Sus:Trey}, \cite%
{Suslov2020} (following Nikiforov--Uvarov's paradigm), or in classical sources \cite{Bethe:Sal},
\cite{Dav}, \cite{Dar}, \cite{Fock1978}, \cite{Gor}, \cite{Schiff}.
\medskip

In Dirac's theory of the relativistic electron, the corresponding limit has the
form \cite{Bethe:Sal}, \cite{Dav},  \cite{Schiff}, \cite{Suslov2020}:%
\begin{equation}
\varepsilon_{\text{Dirac}} = \frac{E_{n_{r},\, j}}{mc^{2}}= 1-\frac{\mu ^{2}}{2n^{2}}-\frac{\mu ^{4}}{%
2n^{4}}\left( \frac{n}{j+1/2}-\frac{3}{4}\right) +\text{O}(\mu^{6}) ,\quad \mu \rightarrow 0,  \label{lim3}
\end{equation}%
where $n=n_{r}+j+1/2$ is the principal quantum number of the nonrelativistic
hydrogenlike atom. Once again, the first term in this expansion is the rest
mass energy of the relativistic electron, the second term coincides with the
energy eigenvalue in the nonrelativistic Schr\"{o}dinger theory and the
third term gives the so-called fine structure of the energy levels --- the
correction obtained for the energy in the Pauli approximation which includes
the interaction of the spin of the electron with its orbital angular momentum.
(See our supplementary Mathematica notebook for a computer algebra proof.)

%

\begin{table}
\caption[]{Dirac equation for the Coulomb potential $U(r)=-Ze^{2}/r$.
Dimensionless quantities: $\varepsilon =E/(mc^{2})$, $x=\beta r=(mc/\hbar)r$,
$\mu = Ze^{2}/(\hbar c)$, $F(r) =f(x) =u_1(x)/x$, $G(r) =g(x) =u_2(x)/x$.}
\label{tab:Dirac}
\begin{center}
\begin{tabular}{|l|l|}
\hline
$\sigma (x)$ & $x$ \\ \hline
$\widetilde{\sigma }(x)$ & \mybox{$(\varepsilon^{2}-1)x^{2}+2\mu
\varepsilon \,x - \nu(\nu+1); \nu =\sqrt{\kappa^2 - \mu^2},
\kappa = \pm \left(j+\dfrac{1}{2}\right)$} \\ \hline
$\widetilde{\tau }(x)$ & $0$ \\ \hline
$k$ & $2\mu \varepsilon \pm \sqrt{1-\varepsilon ^{2}} (2\nu+1)$
\\ \hline
$\pi(x) $ & \mybox{$ \dfrac{1}{2} \pm \left({\sqrt{1-\varepsilon^2}} \, x \pm \left(\nu + \dfrac{1}{2}\right)\right) $} \\
\hline
$\tau(x) =\widetilde{\tau }+2\pi $ & $2(\nu+1-a\;x), \quad a= \sqrt{1-\varepsilon^2}; \quad \tau^\prime < 0 $ \\ \hline
$\lambda =k+\pi ^{\prime }$ & $2\left( \mu \varepsilon -(\nu +1)a\right) $ \\ \hline
$\varphi(x) $ & $x^{\nu +1}e^{-a x}$ \\ \hline
$\rho (x)$ & $x^{2\nu +1}e^{-2a x}$ \\ \hline
$y_{n}(x)$ & \mybox{$A_{n}L_{n}^{2\nu +1}(\xi) ; \; \, \xi
=2ax=2a\beta r=2\sqrt{1-\varepsilon ^{2}}\,\dfrac{mc}{\hbar }\,r$} \\
& $n=n_{r}=0,1,\dots$ \\ \hline
$A_{n}=\dfrac{a}{\kappa \varepsilon -\nu}\, B_{n}$ & \mybox{$ B_{n}=a {\beta^{3/2}}\sqrt{%
\dfrac{(\kappa - \nu)(\kappa \varepsilon -\nu)\,n!}{\mu \Gamma(2\nu+n)}} $} \\ \hline
\end{tabular}
\end{center}
\end{table}
%



%
\section{A Model of the 3D-confinement Potential}

Looking for solutions of the Schr\"{o}dinger equation (\ref{s6}) in spherical coordinates,
\begin{equation}
\psi = \dfrac{1}{r}\, R(r) Y_{lm}(\theta, \varphi) , \label{Conf1}
\end{equation}
with the following model central field potential,
\begin{equation}
U(r) = V_0 \left(\dfrac{r}{a}-\dfrac{a}{r}\right)^2 , \label{Conf2}
\end{equation}
one gets the radial equation of the form
\begin{equation}
R''+\dfrac{2m}{\hbar^2} \left[(E+2V_0) -V_0\left(\dfrac{a^2}{r^2}+\dfrac{r^2}{a^2}\right) - \dfrac{\hbar^2 l(l+1)}{2m r^2} \right] R=0
\quad \; (l=0,1,2,\dots).
\end{equation}

This is not a generalized equation of hypergeometric type and, therefore, cannot be treated right away by the Nikiforov--Uvarov method. By using the substitution
\begin{equation}
R(r) = u(\xi), \qquad \xi = \alpha r^2 , \qquad \alpha^2 =  \dfrac{2m V_0}{\hbar^2 a^2},
\end{equation}
we finally obtain equation (\ref{A1}) with the following coefficients:
\begin{equation}
\sigma (\xi) = \xi, \quad  \widetilde{\tau } =\dfrac{1}{2} , \quad
\widetilde{\sigma }(\xi)=\dfrac{1}{4}  \left[\dfrac{2m }{\alpha \hbar^2}(E+2V_0)\xi - {\alpha}^2 a^4 -l(l+1) -\xi^2 \right] . \label{Conf4}
\end{equation}
In the Nikiforov--Uvarov method, the energy levels and the corresponding radial wave functions can be obtained by (\ref{A7}) and (\ref{A8}).
As a result, they are given by
\begin{equation}
E_{n,l}=\hbar \sqrt{\dfrac{8V_0}{ma^2}} \left[n+\dfrac{1}{2} + \dfrac{1}{2}\left(\sqrt{\dfrac{2m{V_0}a^2}{\hbar^2}+\left(l+\dfrac{1}{2}\right)^2} - \sqrt{\dfrac{2m{V_0}a^2}{\hbar^2}} \right) \right]   \label{Conf5}
\end{equation}
and
\begin{equation}
R(r)=R_{n,l}(r)=C_n \xi^{(\beta-1/2)/2} \exp(-\xi/2) L^{\beta}_n(\xi),  \label{Conf6}
\end{equation}
provided
\begin{equation}
\int_0^\infty R^2(r)\, dr=1 ,  \label{Conf7}
\end{equation}
respectively. Here
\begin{equation}
C^2_n =\dfrac{2 n! \sqrt{\alpha} }{\Gamma(\beta +n +1)}, \qquad \alpha=\dfrac{\sqrt{2m V_0}}{\hbar a} ,   \qquad
\beta =\sqrt{\dfrac{2m{V_0}a^2}{\hbar^2}+\left(l+\dfrac{1}{2}\right)^2} . \label{Conf8}
\end{equation}
Details of the calculations are presented in Table~\ref{tab:conf3}. (The case $l=0$ corresponds to a one-dimensional problem from \cite{Gol:Kriv}.)

In this case, the spectrum is linear, as for the harmonic oscillator. There is no continuous spectrum thus resembling the confinement property in quantum chromodynamics.
%
%

\begin{table}
\caption[]{A model of the $3D$-confinement potential $U(r)$ given in~\eqref{Conf2}.
Dimensionless quantities: $\xi=\alpha r^2$, $\alpha=\sqrt{2m V_0}/(\hbar a)$,
$R(r) =u(\xi)$.}
\label{tab:conf3}
\begin{center}
\begin{tabular}{|l|l|}
\hline
$\sigma (\xi)$ & $\xi$ \\ \hline
$\widetilde{\sigma }(\xi)$ & \mybox{$ \dfrac{1}{4}  \left[\dfrac{2m }{\alpha \hbar^2}(E+2V_0)\xi - {\alpha}^2 a^4 -l(l+1) -\xi^2 \right]$} \\ \hline
$\widetilde{\tau }(\xi)$ & $1/2$ \\ \hline
$k$ & \mybox{$\dfrac{1}{2}\left[\dfrac{m}{\alpha \hbar^2} (E+2V_0) \pm \sqrt{\alpha^2 a^4 +(l+1/2)^2} \right]$} \\ \hline
$\pi (\xi) $ & \mybox{$ \dfrac{1}{4} \pm \dfrac{\xi \pm \sqrt{\alpha^2 a^4 + (l+1/2)^2}}{2}$} \\
\hline
$\tau (\xi) =\widetilde{\tau }+2\pi $ & \mybox{$ 1+\beta - \xi , \quad \beta =\sqrt{\dfrac{2m{V_0}a^2}{\hbar^2}+\left(l+\dfrac{1}{2}\right)^2}$} \\ \hline
$\lambda =k+\pi ^{\prime }$ & \mybox{$ \dfrac{1}{2}\left[\dfrac{m}{\alpha \hbar^2} (E+2V_0) - \sqrt{\alpha^2 a^4 +(l+1/2)^2} \right] -\dfrac{1}{2}$} \\ \hline
$\varphi (\xi) $ & \mybox{$\xi^{\nu/2}\exp{(-\xi/2)} , \; \nu =  \beta + \dfrac{1}{2} = \dfrac{1}{2} + \sqrt{\alpha^2 a^4 +\left(l+ \dfrac{1}{2}\right)^2} $} \\ \hline
$\rho (\xi)$ & $\xi^{\beta} \exp{(-\xi)}$ \\ \hline
$y_{n}(\xi)$ & $C_{n}L_{n}^{\beta}(\xi) , \quad \xi
=\alpha r^2 \/, \quad n=0,1,\dots$ \\ \hline
$C^2_{n}$ & \mybox{$\dfrac{2 n! \sqrt{\alpha} }{\Gamma(\beta +n +1)} , \quad \alpha=\dfrac{\sqrt{2m V_0}}{\hbar a}$} \\ \hline
\end{tabular}
\end{center}
\end{table}
%


%
\section{3D-Spherical Oscillator}

Looking for solutions of the Schr\"{o}dinger equation (\ref{s6}) with harmonic potential
\begin{equation}
U(r)=\dfrac{1}{2}m\omega^2 r^2  \label{3DHO1}
\end{equation}
in spherical coordinates (\ref{Conf1}), one gets the following radial equation:
\begin{equation}
R''+\left[\dfrac{2mE}{\hbar^2} - \dfrac{m^2\omega^2}{\hbar^2} r^2 - \dfrac{l(l+1)}{r^2} \right] R=0
\quad \; (l=0,1,2,\dots). \label{3DHO2}
\end{equation}
Using the abbreviations \cite{Flugge}
\begin{equation}
\dfrac{2mE}{\hbar^2}=\kappa^2, \quad \dfrac{m\omega}{\hbar}=\mu, \quad \dfrac{\kappa^2}{2\mu}=\dfrac{E}{\hbar \omega} = \varepsilon , \label{3DHO3}
\end{equation}
the radial equation can be rewritten in the standard form
\begin{equation}
\dfrac{d^2R}{dr^2}+\left[\kappa^2 -\mu^2 r^2 -\dfrac{l(l+1)}{r^2} \right]R=0. \label{3DHO4}
\end{equation}
Finally, the substitution $R(r)=u(\xi)$ with $\xi = \mu r^2 $
results in the generalized equation of hypergeometric type with
\begin{equation}
\sigma(\xi)= \xi , \quad \widetilde{\tau }(\xi) = \dfrac{1}{2} , \quad
\widetilde{\sigma }(\xi) = \dfrac{1}{4} \left[\dfrac{\kappa^2}{\mu}  \xi  - \xi^2 -l(l+1) \right]\/ .
\label{3DHO5}
\end{equation}
Therefore,
\begin{equation}
2k-\varepsilon = \pm \left(l+ \dfrac{1}{2} \right) , \quad \pi(\xi) = \dfrac{1}{4} \pm \dfrac{1}{2} \left[\xi \pm \left(l+ \dfrac{1}{2} \right)  \right] .
\end{equation}
Further details of calculation are presented in Table~\ref{tab:3DHO} and in the corresponding Mathematica file.

As a result, the energy levels are given by
\begin{equation}
E_0=\hbar \omega \left(2n+l+ \dfrac{3}{2}\right) , \quad n=0,1,2,\dots   \label{3DHO6}
\end{equation}
and the corresponding radial wave functions are related to the Laguerre polynomials (Table~\ref{tab:cop}\/):
\begin{equation}
R_n(r)= C_n \xi^{l+1} \exp\left(-\dfrac{\xi}{2}\right) \ L^{l+1/2}_n(\xi) , \quad n=0,1,2,\dots\;.
\label{3DHO7}
\end{equation}
Here,
\begin{equation}
\int_0^\infty R_n^2(r) \, dr =1 .  \label{3DHO8}
\end{equation}
Extension to the case of $n$-dimensions is discusses in \cite{Ni:Su:Uv}.

\begin{table}
\caption[]{Data for the $3D$-spherical harmonic oscillator potential
$U(x)= \frac{1}{2}m\omega^2 r^2$.
Substitution: $\xi=\mu r^2$, $\mu = m\omega/\hbar$, and $R(r) =u(\xi)$.}
\label{tab:3DHO}
\begin{center}
\begin{tabular}{|l|l|}
\hline
$\sigma (\xi)$ & $\xi$ \\ \hline
$\widetilde{\sigma }(\xi)$ & \mybox{$\dfrac{1}{4} \left[2\varepsilon \  \xi  - \xi^2 -l(l+1) \right] , \quad  \varepsilon = \dfrac{E}{\hbar \omega}$} \\ \hline
$\widetilde{\tau }(\xi)$ & \mybox{$\dfrac{1}{2}$} \\ \hline
$k$ & \mybox{$-\dfrac{1}{2}\left( l+\dfrac{1}{2}-\varepsilon \right)$} \\ \hline
$\pi (\xi) $ & \mybox{$\dfrac{1}{2} (l+1-\xi)$} \\ \hline
$\tau (\xi) =\widetilde{\tau }+2\pi $ & \mybox{$l+ \dfrac{3}{2} - \xi$} \\ \hline
$\lambda =k+\pi ^{\prime }$ & \mybox{$-\dfrac{1}{2}\left(l+\dfrac{3}{2} - \varepsilon \right)$} \\ \hline
$\varphi(\xi)$ & \mybox{$\xi^{(l+1)/2}\exp{(-\xi/2)}$} \\ \hline
$\rho(\xi)$ & \mybox{$\xi^{l+1/2} \exp{(-\xi)}$} \\ \hline
$y_{n}(\xi)$ & \mybox{$C_{n}L_{n}^{l+1/2}(\xi), \quad n=0,1,\dots$} \\ \hline
$C^2_{n}$ & \mybox{$\dfrac{2 n! \sqrt{\mu} }{\Gamma(l +n +3/2)} , \quad \mu = \dfrac{m\omega}{\hbar}  $} \\ \hline
\end{tabular}
\end{center}
\end{table}
%


%
\section{P\"{o}schl--Teller Potential Hole\/}

Let us consider the one-dimensional stationary Schr\"{o}dinger equation:%
\begin{equation}
-\frac{\hbar ^{2}}{2m}\frac{d^{2}\psi }{dx^{2}}+U(x) \psi =E\psi, \label{PT1}
\end{equation}%
where
\begin{equation}
U(x)=\dfrac{1}{2} V_0 \left[\dfrac{a(a-1)}{\sin^2(\alpha x)}
+ \dfrac{b(b-1)}{\cos^2(\alpha x)} \right] , \quad V_0 = \dfrac{\hbar^2 \alpha^2}{m}
\label{PT2}
\end{equation}
with real-valued parameters $a>1, \, b>1$ in the finite region $0<x<\pi/(2\alpha) \/$ bounded by the singularities of $U(x)\/$
(see \cite{Flugge}, \cite{PT}, \cite{RM} for original references and applications).
Here, we are looking for orthonormal real-valued wave functions:%
\begin{equation}
\int_{ 0 }^{ \pi/(2\alpha) }\psi ^{2}(x) \, dx=1.
\label{PT3}
\end{equation}%
Introducing new quantities%
\begin{equation}
\psi(x) =u(\xi) ,\qquad \xi = \sin^2(\alpha x), \quad 1-\xi = \cos^2(\alpha x) \/,  \label{PT4}
\end{equation}%
one gets the following generalized equation of hypergeometric type:%
\begin{align}
&\xi (1-\xi) u^{\prime \prime } + \left(\dfrac{1}{2}-\xi \right) u^{\prime } \\
& \qquad  + \dfrac{1}{4} \left[\dfrac{c^2}{\alpha^2} - \dfrac{a(a-1)}{\xi} - \dfrac{b(b-1)}{1-\xi} \right] u =0 , \quad c^2=\dfrac{2mE}{\hbar^2} \/. \notag
\label{PT5}
\end{align}%
Here,
\begin{align}
&\sigma(\xi)= \xi (1 - \xi) , \qquad \widetilde{\tau }(\xi) = \left(\dfrac{1}{2}-\xi \right) \/ , \\
&\widetilde{\sigma }(\xi) = \dfrac{1}{4} \left[\dfrac{c^2}{\alpha^2}  \xi (1 - \xi) - a(a-1) (1-\xi) - b(b-1) \xi  \right]\/  \notag
\label{PT6}
\end{align}
and the boundary conditions take the form $u(0)=u(1)=0\/$ .

Therefore,
\begin{align}
&p(\xi)=\dfrac{1}{4}\left[(\kappa +1) \xi^2 -(\kappa+1+(a-b)(a+b-1)) \xi +a(a-1) + \dfrac{1}{4}\right]\/, \\
&\qquad \qquad \kappa=\dfrac{c^2}{\alpha^2} -4k \/ . \notag \label{PT7}
\end{align}
Equation (\ref{A12}) takes the form
\begin{equation}
\left[\kappa+1+(a-b)(a+b-1)\right]^2=(\kappa+1)\left[4a(a-1)+1\right]\/ .
\label{PT8}
\end{equation}
There are two solutions
\begin{equation}
\kappa_1=(a+b)(a+b-2), \quad \kappa_2=-(b-a-1)(b-a+1)
\label{PT9}
\end{equation}
If one chooses
\begin{equation}
\dfrac{c^2}{\alpha^2} - 4k = (a+b)(a+b-2) \/ ,
\label{PT10}
\end{equation}
then
\begin{equation}
\pi(\xi)=\dfrac{1}{2}\left(\dfrac{1}{2}-\xi\right) - \left(\dfrac{a+b-1}{2} \xi - \frac{(2a-1}{4} \right) =\dfrac{a}{2} -\dfrac{a+b}{2} \xi \/
\label{PT11}
\end{equation}
and
\begin{equation}
\tau(\xi)= a+\dfrac{1}{2} - (a+b+1) \xi \/  , \qquad \lambda = \dfrac{1}{4} \left[\dfrac{c^2}{\alpha^2} - (a+b)^2 \right] \/ .
\label{PT12}
\end{equation}
Further details of calculation are presented in Table~\ref{tab:PT} (see also the corresponding Mathematica file).

As a result, the energy levels are given by (\ref{A7}):
\begin{equation}
E_0=\dfrac{1}{2} V_0 (a+b+2n)^2, \quad n=0,1,2,\dots   \label{PT13}
\end{equation}
and the corresponding wave functions are related to the Jacobi polynomials:
\begin{equation}
\psi_n(x)= C_n \sin^a(\alpha x) \cos^b(\alpha x) \ P_n^{(a-1/2,\, b-1/2)}\left(\cos(2\alpha x)\right) \/ ,
\label{PT14}
\end{equation}
where $C_n$  is the normalization constant.

Indeed, by the Rodrigues-type formula (\ref{A8}):
\begin{equation}
y_n=\dfrac{B_n}{\xi^{a-1/2}(1-\xi)^{b-1/2}} \dfrac{d^n}{d\xi^n}\left[\xi^{n+a-1/2}(1-\xi)^{n+b-1/2} \right]
\label{PT15}
\end{equation}
and with the aid of the substitution $\eta=1-2\xi=\cos(2\alpha x)$ one gets
\begin{align}
y_n(\eta) &= \dfrac{(-1)^n}{2^n}\dfrac{B_n}{(1-\eta)^{a-1/2}(1+\eta)^{b-1/2}} \dfrac{d^n}{d\eta^n}\left[(1-\eta)^{n+a-1/2}(1+\eta)^{n+b-1/2} \right] \\
 &= C_n\, P_n^{(a-1/2,\,  b-1/2)}(\eta) \/ . \notag
\label{PT16}
\end{align}
Moreover, by the normalization condition:
\begin{align}
1&=\int_0^{\pi/(2\alpha)}\psi^2_n(x)\, dx = \dfrac{C_n^2}{2\alpha}\int_{-1}^{1}\left[ P_n^{(a-1/2,\,  b-1/2)}(\eta)  \right]^2 \dfrac{\varphi^2\ d\eta}{(1-\eta)^{1/2}(1+\eta)^{1/2}} \\
& = \dfrac{C_n^2}{(2\alpha)2^{a+b}}\int_{-1}^{1}\left[ P_n^{(a-1/2,\,  b-1/2)}(\eta) \right]^2 {(1-\eta)^{a-1/2}(1+\eta)^{b-1/2}}\ d\eta  \notag  \\ 
& = \dfrac{C_n^2}{2\alpha} \dfrac{\Gamma(a+n+1/2)\Gamma(b+n+1/2)}{n!(a+b+2n)\Gamma(a+b+n)} \/ .    \notag
\label{PT17}
\end{align}
(Here, the value of the squared norm $d^2_n$ for the Jacobi polynomials has been taken from Table~\ref{tab:cop}\/.)
As a result,
\begin{equation}
C_n^2=\dfrac{2\alpha n! (a+b+2n) \Gamma(a+b+n)}{\Gamma(a +n +1/2) \Gamma(b+n+1/2)} \/ .
\label{PT18}
\end{equation}
%

\begin{table}
\caption[]{The P\"{o}schl--Teller potential hole $U(x)$ is given in \eqref{PT2}.
Substitution: $\xi=\sin^2(\alpha x)$, $\psi(x) =u(\xi)$, $0<x<\pi/(2\alpha)$.}
\label{tab:PT}
\begin{center}
\begin{tabular}{|l|l|}
\hline
$\sigma (\xi)$ & $\xi (1-\xi) = \xi - \xi^2 $ \\ \hline
$\widetilde{\sigma }(\xi)$ & \mybox{$-\dfrac{1}{4}\left[\dfrac{c^2}{\alpha^2} \xi^2 - \left(\dfrac{c^2 }{\alpha^2} + (a-b)(a+b-1) \right) \xi + a(a-1) \right], c^2=\dfrac{2mE}{\hbar^2}\!$} \\ \hline
$\widetilde{\tau }(\xi)$ & \mybox{$\dfrac{1}{2} - \xi$} \\ \hline
$k$ & \mybox{$\dfrac{1}{4}\left[\dfrac{c^2}{\alpha^2} - (a+b)(a+b-2) \right]$} \\ \hline
$\pi (\xi) $ & \mybox{$\dfrac{1}{2}\left[a-(a+b)\, \xi \right]$} \\
\hline
$\tau (\xi) =\widetilde{\tau }+2\pi $ & $ (a+1/2) - (a+b+1) \xi  $ \\ \hline
$\lambda =k+\pi ^{\prime }$ & \mybox{$\dfrac{1}{4}\left[\dfrac{c^2}{\alpha^2} - (a+b)^2 \right]$} \\ \hline
$\varphi(\xi) $ & \mybox{$\xi^{a/2}(1-\xi)^{b/2} = \sin^a(\alpha x) \cos^b(\alpha x)$} \\ \hline
$\rho(\xi)$ & \mybox{$\xi^{a-1/2} (1-\xi)^{b-1/2}$} \\ \hline
$y_{n}(\xi)$ & \mybox{$C_{n} P_n^{(a-1/2,\, b-1/2)}\left((1-\xi)/2)\right)= C_{n} P_n^{(a-1/2,\, b-1/2)}\left(\cos(2\alpha x)\right)$} \\ \hline
$C^2_{n}$ & \mybox{$\dfrac{2\alpha n! (a+b+2n) \Gamma(a+b+n)}{\Gamma(a +n +1/2) \Gamma(b+n+1/2)}   \qquad  (n=0,1,2, \, \dots)$} \\ \hline
\end{tabular}
\end{center}
\end{table}
%


%
\section{Modified P\"{o}schl--Teller Potential Hole\/}

In order to solve the one-dimensional stationary Schr\"{o}dinger equation (\ref{PT1}) for the potential:
\begin{equation}
U(x)=-\dfrac{\hbar^2 \alpha^2}{2m}\dfrac{a(a-1)}{\cosh^2(\alpha x)}  \qquad (-\infty <x< \infty)
\label{MPT1}
\end{equation}
with $a>1\/$, one can use the following substitution $\psi(x)=u(\xi)\/ ,$ where $\xi=\cosh^2(\alpha x)\/ $ \cite{Flugge}, \cite{PT}.
As a result, we arrive at the generalized equation of hypergeometric type
\begin{equation}
\xi(1-\xi) u'' + \left(\dfrac{1}{2} -\xi\right) u' + \dfrac{1}{4} \left[\dfrac{c^2}{\alpha^2} - \dfrac{a(a-1)}{\xi} \right] u=0 \/ ,
\quad c^2=\dfrac{2m(-E)}{\hbar^2}\/ ,
\label{MPT2}
\end{equation}
where
\begin{align}
&\sigma(\xi)= \xi (1 - \xi) , \qquad \widetilde{\tau }(\xi) = \left(\dfrac{1}{2}-\xi \right) \/ , \\
&\widetilde{\sigma }(\xi) =  \dfrac{1}{4} \left[\dfrac{c^2}{\alpha^2}  \xi - a(a-1)  \right] (1-\xi)\/ .  \notag
\label{MPT3}
\end{align}
Using the standard substitution $u=\varphi(\xi)y$ with $\varphi(\xi)=\xi^{(1-a)/2}\/$, one gets the hypergeometric differential equation of the form:
\begin{equation}
\xi(1-\xi) y'' + \left[\dfrac{3}{2} -a + (a-2)\xi \right] y' + \dfrac{1}{4}\left[ \dfrac{c^2}{\alpha^2} - (a-1)^2  \right] y=0\/ .
\label{MPT4}
\end{equation}
Here, we concentrate only on the bounded states (continuous spectrum is discussed in \cite{Flugge}).

There is a finite number of negative discrete energy levels that are explicitly given by
\begin{equation}
E=E_n=-\dfrac{\hbar^2\alpha^2}{2m}(1-a+2n)^2\/ , \quad n=0,1,\ldots <(a-1)/2 \/.
\label{MPT5}
\end{equation}
The corresponding orthonormal wave functions are related to a set of Jacobi polynomials with a negative value of one parameter
that are orthogonal on an infinite interval $(1, +\infty)\/.$
They are given by the Rodrigues-type formula (\ref{A8}) or in terms of a terminating hypergeometric series:
\begin{align}
y_n(\xi) &= \dfrac{(1/2)_n}{n!} {}_2F_1\biggl(-n,\, 1-a + n;\, \dfrac{1}{2}; \, 1-\xi\biggr) \\
&= P_n^{(-1/2, \, 1/2 - a)}\left(\cosh(2\alpha x)\right)\/, \notag
\label{MPT6}
\end{align}
where $(1/2)_n = \Gamma(n+1/2)/\Gamma(1/2)\/.$ Cauchy's beta integral,
\begin{equation}
\int_0^\infty \dfrac{t^{A-1}}{(1+t)^{A+B}}\, dt = \dfrac{\Gamma(A)\,\Gamma(B)}{\Gamma(A+B)}\/ , \qquad \Re\, (A)>0, \; \Re\, (B)>0 \/ ,
\label{MPT7}
\end{equation}
should be used in order to find the value of the normalization constant (see \cite{Suslov2020}, Exercise~1.15 and \cite{NIST}, (5.12.3)).
Further details of calculation are presented in Table~\ref{tab:modPT} (see also the corresponding Mathematica file).
As a result, for the bound states (\ref{MPT5}), the normalized wave functions  are given by
\begin{align}
&\psi_n(x)=\sqrt{\alpha} \left[ \dfrac{ n! (a-1-2n)\, \Gamma(a-n-1/2)}{\Gamma(n +1/2)\,\Gamma(a-n)}  \right]^{1/2} \\
&\qquad\qquad \times \cosh^{1-a}(\alpha x)\, P_n^{(-1/2, \, 1/2-a)}\left(\cosh(2\alpha x)\right) \/ .   \notag
\label{MPT8}
\end{align}
%
%

\begin{table}
\caption[]{The modified P\"{o}schl--Teller potential hole $U(x)$ is defined in~\eqref{MPT1}.
Substitution: $\xi=\cosh^2(\alpha x)$, $\psi(x) = u(\xi)$.}
\label{tab:modPT}
\begin{center}
\begin{tabular}{|l|l|}
\hline
$\sigma (\xi)$ & $\xi (1-\xi)  $ \\ \hline
$\widetilde{\sigma }(\xi)$ & \mybox{$\dfrac{1}{4}\left[\dfrac{c^2}{\alpha^2} \xi - a(a-1) \right](1-\xi), \quad c^2=\dfrac{2m(-E)}{\hbar^2}$} \\ \hline
$\widetilde{\tau }(\xi)$ & \mybox{$\dfrac{1}{2} - \xi$} \\ \hline
$k$ & \mybox{$\dfrac{1}{4}\left(\dfrac{c^2}{\alpha^2} +1 - a^2 \right)$} \\ \hline
$\pi (\xi) $ & \mybox{$\dfrac{1-a}{2}(1-\xi)$} \\ \hline
$\tau (\xi) =\widetilde{\tau }+2\pi $ & \mybox{$\dfrac{3}{2} -a + (a-2) \xi$} \\ \hline
$\lambda =k+\pi ^{\prime }$ & \mybox{$\dfrac{1}{4}\left[\dfrac{c^2}{\alpha^2} - (a-1)^2 \right]$} \\ \hline
$\varphi(\xi) $ & \mybox{$\xi^{(1-a)/2}$} \\ \hline
$\rho(\xi)$ & \mybox{$\xi^{(1/2) -a} (\xi-1)^{-(1/2)}$} \\ \hline
$y_{n}(\xi)$ & \mybox{$C_{n} P_n^{(-1/2, \, 1/2 - a)}\left(\cosh(2\alpha x)\right)\/,  \;  n<(a-1)/2$} \\ \hline
$C^2_{n}$ & \mybox{$\alpha \dfrac{n! (a-1-2n) \Gamma(a-n-1/2)}{\Gamma(n +1/2) \Gamma(a-n)}$} \\ \hline
\end{tabular}
\end{center}
\end{table}
%


%
\section{Kratzer's Molecular Potential\/}

In order to investigate the rotation-vibration spectrum of a diatomic molecule, the potential
\begin{equation}
U(r)=-2D\left(\dfrac{a}{r}-\dfrac{1}{2}\dfrac{a^2}{r^2}\right)\/ \qquad D>0\/,
\label{KMP1}
\end{equation}
with a minimum $U(a)=-D\/$,
has been used \cite{Flugge}. Once again we are looking for solutions of
the Schr\"{o}dinger equation (\ref{s6}) in spherical coordinates (\ref{Conf1}) and introduce the dimensionless quantities:
\begin{equation}
x=\dfrac{r}{a}, \qquad \beta^2=-\dfrac{2ma^2}{\hbar^2}\ E, \qquad \gamma^2 =\dfrac{2ma^2}{\hbar^2}\ D
\label{KMP2}
\end{equation}
together with the standard substitution: $R(r)=u(x)\/.$

For bound states $E<0\/, \beta>0\/$ and the radial equation takes the form
\begin{equation}
u''+\left[-\beta^2 +\frac{2\gamma^2}{x}- \frac{\gamma^2 +l(l+1)}{x^2}\right] u =0\/.
\label{KMP3}
\end{equation}
Further computational details are presented in Table~\ref{tab:Kratzer} (see also the corresponding Mathematica file).
This case is somewhat similar to Coulomb and relativistic Coulomb problems.
As a result, the bound states are given by
\begin{equation}
E_n=-\dfrac{2m a^2 D^2}{\hbar^2 }\, \dfrac{1}{(\nu+n)^2}\/ ,
\label{KMP4}
\end{equation}
where
\begin{equation}
\nu = \dfrac{1}{2} + \sqrt{\gamma^2 + \left(l+\dfrac{1}{2}\right)^2} , \qquad \gamma^2 = \dfrac{2ma^2}{\hbar^2}\ D\/ .
\label{KMP5}
\end{equation}
We can obtain the same exact result with the aid of the Bohr--Sommerfeld quantization rule in the semiclassical approximation (the WKB-method \cite{Barleyetal2021}, \cite{Langer1937}, \cite{Ni:Uv}, \cite{Sommerfeld1951}).

The normalized radial wave functions,
\begin{equation}
\int_0^\infty R^2(r)\, dr =1\/ ,
\label{KMP6}
\end{equation}
are related to the Laguerre polynomials (Table~\ref{tab:cop}\/):
\begin{equation}
R_n(r)=C_n \, x^{\nu} \exp(-\beta x) L_n^{2\nu-1}(2\beta x)\/  \qquad \left(x=\dfrac{r}{a}\right) \/ ,
\label{KMP7}
\end{equation}
where
\begin{equation}
C^2_n=\dfrac{(2\beta)^{2\nu+1} n!}{a (2\nu+2n) \Gamma(2\nu+n)} \/ .
\label{KMP8}
\end{equation}
Once again, we have used the integral (\ref{sol20}). (See also \cite{AiOthman}, \cite{Berk2006},  \cite{Flugge}, \cite{Fues1926}, \cite{Gol:Kriv}, \cite{Karp70}
for some applications of Kratzer's molecular potential and further references.)
%


%
\begin{table}
\caption[]{Kratzer's molecular potential~\eqref{KMP1}.
Substitution: $x=r/a$, $R(r) = u(x)$; see also (\ref{KMP2}).}
\label{tab:Kratzer}
\begin{center}
\begin{tabular}{|l|l|}
\hline
$\sigma (x)$ & $x $ \\ \hline
$\widetilde{\sigma }(x)$ & $ 2 \gamma^2 x - \beta^2 x^2 -\gamma^2 - l(l+1)  $ \\ \hline
$\widetilde{\tau }(x)$ & $ 0 $ \\ \hline
$k$ & \mybox{$2\gamma^2 \pm \beta (2\nu - 1), \quad \nu = \dfrac{1}{2} + \sqrt{\gamma^2 + \left(l+\dfrac{1}{2}\right)^2}$} \\ \hline
$\pi(x) $ & $ \nu - \beta x $ \\ \hline
$\tau(x) =\widetilde{\tau }+2\pi \qquad $ & $ 2 (\nu - \beta x) $ \\ \hline
$\lambda =k+\pi ^{\prime }$ & $2( \gamma^2 - \nu \beta) $ \\ \hline
$\varphi(x) $ & $x^{\nu} \exp(-\beta x)$ \\ \hline
$\rho (x)$ & $ x^{2\nu - 1} \exp{{(-2\beta x)}}$ \\ \hline
$y_{n}(x)$ & $C_{n}\  L_n^{2\nu -1} (2\beta x)$ \\ \hline
$C_{n}^2$ & \mybox{$\dfrac{(2\beta)^{2\nu+1} n!}{a (2\nu+2n) \Gamma(2\nu+n)}$} \\ \hline
\end{tabular}
\end{center}
\end{table}
\section{Hulth\'{e}n Potential\/}
We are looking for solutions of the Schr\"{o}dinger equation (\ref{s6}) in spherical coordinates (\ref{Conf1})
for the following central field potential,
\begin{equation}
U(r) =-V_{0}\frac{e^{-r/a}}{1-e^{-r/a}} \qquad  (0\leq r<\infty ),
\label{hp1}
\end{equation}
when $l=0\/.$ (See, for example, (\ref{RadialCoulomb}) with $F=R$ and use these data for an explicit form of the corresponding radial equation.)
With the aid of the substitution%
\begin{equation}
R(r) =u(\xi) ,\qquad \xi =e^{-r/a},  \label{hp2}
\end{equation}%
when%
\begin{equation}
\alpha ^{2}=-\frac{2mE}{\hbar ^{2}}a^{2}>0,\qquad \beta ^{2}=\frac{2mV_{0}}{%
\hbar ^{2}}a^{2}>0\label{hp2a}
\end{equation}%
with $\alpha >0$ and $\beta >0,$ one obtains the following generalized
equation of hypergeometric type,%
\begin{equation}
\xi ^{2}\frac{d^{2}u}{d\xi ^{2}}+\xi \frac{du}{d\xi }+\left( -\alpha
^{2}+\beta ^{2}\frac{\xi }{1-\xi }\right) u=0 , \/  \label{hp3}
\end{equation}%
where%
\begin{equation}
\sigma(\xi) =\xi (1-\xi) ,\qquad \widetilde{\tau }%
(\xi) =1-\xi ,  \label{hp4}
\end{equation}%
and%
\begin{equation}
\widetilde{\sigma }(\xi) =(1-\xi) \left( \left(
\alpha ^{2}+\beta ^{2}\right) \xi -\alpha ^{2}\right) .\label{hp4a}
\end{equation}%
The boundary conditions are%
\begin{equation}
R(0) =\lim_{r\rightarrow 0^{+}}R(r) =0,\qquad
R(\infty) =\lim_{r\rightarrow \infty }R(r) =0,
\label{hp5}
\end{equation}%
or%
\begin{equation}
u(0) =u(1) =0.  \label{hp6}
\end{equation}%
We have%
\begin{equation}
k=\beta ^{2}\mp \alpha ,\qquad \pi(\xi) =-\frac{\xi }{2}\pm
\left( -\alpha \pm \frac{\xi }{2}+\alpha \xi \right) \label{hp6a}
\end{equation}
for all four possible sign combinations.
The solutions can be found by putting \cite{Flugge}%
\begin{equation}
u= \varphi(\xi) y(\xi) = \xi ^{\alpha }(1-\xi) y(\xi) ,\label{hp7}
\end{equation}%
which results in%
\begin{equation}
\xi(1-\xi) y^{\prime \prime }+\left[ 2\alpha +1-(2\alpha+3) \xi \right] y^{\prime }
-\left( 2\alpha +1-\beta ^{2}\right) y=0%
\label{hp8}
\end{equation}%
(details of calculations are presented in Table~\ref{tab:Hulthen} and in a complementary
Mathematica file).
As one can see, a direct quantization in terms of the classical orthogonal
polynomials by Nikiforov and Uvarov's approach is not applicable here, right away,
because the second coefficient,%
\begin{equation}
\tau(\xi) =2\alpha +1-(2\alpha+3) \xi ,\label{hp9}
\end{equation}%
does depend on $\alpha $ and therefore on the energy~$E$. We have to utilize the
boundary conditions (\ref{hp6}) instead, in a somewhat similar way to the consideration of a familiar case
of an infinite well.
Equation (\ref{hp8}) is a special case of the hypergeometric equation \cite{AndAskRoy}, \cite%
{NIST},%
\begin{equation}
\xi (1-\xi) w^{\prime \prime }+\left[ C-(A+B+1) \xi %
\right] w^{\prime }-ABw=0,\label{gauss}
\end{equation}%
with
\begin{equation}
A=1+\alpha +\gamma ,\quad B=1+\alpha -\gamma ,\quad C=2\alpha +1;\qquad
\gamma =\sqrt{\alpha ^{2}+\beta ^{2}}.\label{gausspars}
\end{equation}%
The required solution, that is bounded at $\xi =0,$ has the form%
\begin{equation}
y= {}_{2}F_{1}\biggl(
\begin{array}{@{}c@{}}
A,\ B \\
C%
\end{array}%
;\xi \biggr) = {}_{2}F_{1}\biggl(
\begin{array}{@{}c@{}}
1+\alpha +\gamma ,\ 1+\alpha -\gamma  \\
2\alpha +1%
\end{array}%
;\xi \biggr) , \label{gaussol}
\end{equation}%
up to a constant, and the first boundary condition is satisfied $u(0)=0\/,$ when $\alpha > 0 \/.$
As is known \cite{NIST},%
\begin{equation}
\lim_{\xi \rightarrow 1^{-}}(1-\xi) ^{A+B-C} {}_2F_1\biggl(
\begin{array}{@{}c@{}}
A,\ B \\
C%
\end{array}%
;\ \xi \biggr) =\frac{\Gamma(C)\,\Gamma(A+B-C)}{\Gamma(A)\,\Gamma(B)}\/,\label{gausslim}
\end{equation}%
provided ${\rm{\Re}}\, (C-A-B)<0\/.$
Thus%
\begin{align}
u(1)  &=\lim_{\xi \rightarrow 1^{-}}(1-\xi) \ _{2}F_{1}\biggl(
\begin{array}{@{}c@{}}
1+\alpha +\gamma ,\ 1+\alpha -\gamma  \\
2\alpha +1%
\end{array}%
;\xi \biggr) \label{gausslimu} \\
&= \frac{\Gamma(2\alpha+1)\,\Gamma(1)}{\Gamma(1+\alpha+\gamma)\,\Gamma(1+\alpha-\gamma)}=0,
\notag
\end{align}%
provided that%
\begin{equation}
\alpha -\gamma =-n=-1,-2,-3,\ \dots, \label{hp10}
\end{equation}
\begin{equation}
\alpha ={\alpha}_n =\frac{\beta ^{2}-n^{2}}{2n} > 0 .\label{hp11}
\end{equation}
As a result, the discrete energy levels are given by%
\begin{equation}
E_{n}=-V_{0}\left( \frac{\beta ^{2}-n^{2}}{2\beta n}\right) ^{2},\qquad
n=1,2,3,\ \dots \quad \left( n^{2}<\beta ^{2}\right) . \label{hpEnergy}
\end{equation}
There exists a minimum size of potential hole before any energy eigenvalue at all can be obtained, viz. $\beta^2=1\/.$
Equation $1\le n^2 \le \beta^2 $ determines the finite number of eigenvalues in a potential hole of a given size \cite{Flugge}.

The radial wave functions take the form%
\begin{equation}
R_{n}(r) =C_{n}\ \xi ^{\alpha }(1-\xi) \ _{2}F_{1}\biggl(
\begin{array}{@{}c@{}}
1-n,\ 1+2\alpha +n \\
2\alpha +1%
\end{array}%
;\xi \biggr) ,\qquad \xi =e^{-r/a},\label{hpFunctions}
\end{equation}%
where the hypergeometric series terminates and $C_n$ is a constant to be determined.
Thus, the energy levels can be obtained by the condition (\ref{A7}) and the corresponding wave functions are derived with the help of
the Rodrigues-type formula (\ref{A8}) as follows:
\begin{equation}
\left( \xi ^{2\alpha +n-1}(1-\xi) ^{n}\right)^{(n-1)} =
\frac{\Gamma(2\alpha+n)}{\Gamma(2\alpha+1)}\xi ^{2\alpha }(1-\xi)\
_{2}F_{1}\biggl(
\begin{array}{@{}c@{}}
1-n,\ 2\alpha +n+1 \\
2\alpha +1%
\end{array}%
;\xi \biggr).
\label{RodriguesFormula}
\end{equation}
This result follows also, as a special case, from (15.5.9) of \cite{NIST}.

Once again, we can use (\ref{KMP6}) for normalization of the radial wave function.
Then
\begin{equation}
a C^2_{n}\ \int_{0}^{1}\xi ^{2\alpha -1}(1-\xi)^{2}y_{n}^{2}(\xi) \, d\xi =1\label{hp12} \/ ,
\end{equation}
where%
\begin{equation}
\xi ^{2\alpha }(1-\xi) y_{n}(\xi) =\frac{\Gamma(2\alpha+1)}{\Gamma(2\alpha+n)}\left[ \xi
^{2\alpha +n-1}(1-\xi)^{n}\right] ^{(n-1)} \label%
{hp12a}
\end{equation}
by (\ref{RodriguesFormula}).
Moreover,
\begin{equation}
(1-\xi) y_{n}(\xi) = {}_{2}F_{1}\biggl(
\begin{array}{@{}c@{}}
-n,\ 2\alpha +n \\
2\alpha +1%
\end{array}%
;\xi \biggr) ,\label{hp12b}
\end{equation}%
by the familiar transformation \cite{Ni:Uv}:%
\begin{equation}
{}_{2}F_{1}\biggl(
\begin{array}{@{}c@{}}
A,\ B \\
C%
\end{array}%
;\xi \biggr) =(1-\xi)^{C-A-B} \ _{2}F_{1}\biggl(
\begin{array}{@{}c@{}}
C-A,\ C-B \\
C%
\end{array}%
;\xi \biggr) \label{hp13}
\end{equation}%
with $A=-n,$ $B=2\alpha +n,$ $C=2\alpha +1.$
Therefore,%
\begin{align}
&\int_{0}^{1}\xi ^{2\alpha -1}(1-\xi)^{2}y_{n}^{2}(\xi) \, d\xi  =
  \frac{\Gamma(2\alpha+1)}{\Gamma(2\alpha+n)}\label{hp14} \\
& \quad \times \int_{0}^{1}\left[ \xi ^{-1}\ _{2}F_{1}\biggl(
\begin{array}{@{}c@{}}
-n,\ 2\alpha +n \\
2\alpha +1%
\end{array}%
;\xi \biggr) \right] \left[ \xi ^{2\alpha +n-1}(1-\xi)^{n} \right] ^{(n-1)}\, d\xi \/ ,  \notag
\end{align}%
and integrating by parts $n-1$ times, one gets%
\begin{align}
&\int_{0}^{1}\xi ^{-1}\ _{2}F_{1}\biggl(
\begin{array}{@{}c@{}}
-n,\ 2\alpha +n \\
2\alpha +1%
\end{array}%
;\xi \biggr) \left[ \xi ^{2\alpha +n-1}(1-\xi)^{n}\right]^{(n-1)}\ d\xi
\\
&=\left. \left( \xi ^{-1}\ _{2}F_{1}\biggl(
\begin{array}{@{}c@{}}
-n,\ 2\alpha +n \\
2\alpha +1%
\end{array}%
;\xi \biggr) \left[ \xi ^{2\alpha +n-1}(1-\xi)^{n}\right]
^{(n-2)}\right) \right\vert _{\xi =0}^{1}  \notag
\\
&\; \; \quad -\int_{0}^{1}\left[ \xi ^{-1}\ _{2}F_{1}\biggl(
\begin{array}{@{}c@{}}
-n,\ 2\alpha +n \\
2\alpha +1%
\end{array}%
;\xi \biggr) \right] ^{\prime }\left[ \xi ^{2\alpha +n-1}(1-\xi)^{n}\right]^{(n-2)}\ d\xi   \notag
\\
&= \dots = \notag
\\
& (-1) ^{k-1}\left. \left( \left[ \xi ^{-1} {}_{2}F_{1}\biggl(
\begin{array}{@{}c@{}} -n,\ 2\alpha +n \\ 2\alpha +1 \end{array}%
;\xi \biggr) \right] ^{(k-1)}\!\!\!\!\left[ \xi ^{2\alpha +n-1}(1-\xi)^{n}\right] ^{(n-k-1)}\right) \right\vert _{\xi
=0}^{1}  \notag
\\
& \; \; \quad \quad +(-1)^{k}\int_{0}^{1}\left[ \xi ^{-1}\ _{2}F_{1}\biggl(
\begin{array}{@{}c@{}}
-n,\ 2\alpha +n \\
2\alpha +1%
\end{array}%
;\xi \biggr) \right] ^{(k)}\left[ \xi ^{2\alpha +n-1}(1-\xi)^{n}\right] ^{(n-k-1)}\ d\xi   \notag
\\
&= \dots =(-1)^{n-1}\int_{0}^{1}\left[ \xi ^{-1}\ _{2}F_{1}\biggl(
\begin{array}{@{}c@{}}
-n,\ 2\alpha +n \\
2\alpha +1%
\end{array}%
;\xi \biggr) \right] ^{(n-1)}\xi ^{2\alpha +n-1}(1-\xi) ^{n}\ d\xi \/ ,  \notag
\end{align}%
in view of the boundary conditions (\ref{hp5})--(\ref{hp6}).
By the power series expansion,%
\begin{align}
\left[ \xi ^{-1}\ _{2}F_{1}\biggl(
\begin{array}{@{}c@{}}
-n,\ 2\alpha +n \\
2\alpha +1%
\end{array}%
;\xi \biggr) \right] ^{(n-1)} &= \left( \frac{1}{\xi }\right)
^{(n-1)}+ \frac{(-n)_{n}(2\alpha+n)_{n}}{(n!) (2\alpha +1)_{n}}(\xi^{n-1})^{(n-1)} \\
&= (-1)^{n-1}\frac{(n-1)!}{\xi ^{n}}+ \frac{%
(-n)_{n}(2\alpha+n)_{n}}{(n!)(2\alpha+1)_{n}}(n-1)!,  \notag
\end{align}%
and our integral evaluation can be completed with the aid of the following Euler beta integrals (\ref{eqnAppB2}):
\begin{align*}
\int_{0}^{1}\xi^{2\alpha -1}(1-\xi)^{(n+1)-1}d\xi  &= \dfrac{\Gamma(2\alpha)(n!)}{\Gamma(2\alpha +n+1)}, \\
\int_{0}^{1}\xi^{2\alpha +n-1}(1-\xi)^{(n+1)-1}d\xi &= \dfrac{\Gamma(2\alpha+n) (n!)}{\Gamma(2\alpha+2n+1)} \/ .
\end{align*}
The final result is given by
\begin{equation}
C_{n}=\frac{(2\alpha)_{n}}{n!}\sqrt{\frac{(\alpha+n) (2\alpha+n)}{(2\alpha) a}},\qquad
\alpha =\frac{\beta ^{2}-n^{2}}{2n} \quad (n=1, 2, 3, \  \dots)  \label{HulthenNorm}
\end{equation}
as a complementary normalization in (\ref{hpFunctions}) (Table~\ref{tab:Hulthen}).
We were not able to find the value of this constant in the available literature (see, for example,  \cite{Flugge}).

The Hulth\'{e}n potential at small values of $r$ behaves like a Coulomb potential $U_{C}=-V_{0}a/r\/,$
whereas for large values of $r$ it decreases exponentially. (See \cite{Flugge} for more details and a numerical example.)
Section~17 below contains an extension of this potential that is suitable for diatomic molecules.
%

\begin{table}
\caption[]{The Hulth\'{e}n potential \eqref{hp1}
in the spherically symmetric case $l=0.$
Substitution: $R(r) = u(\xi)$, $\xi=\exp(-r/a)$, $\alpha^2 = - (2m E/\hbar^2) a^2 >0$,
$\beta^2 = (2m V_0/\hbar^2) a^2 >0$.}
\label{tab:Hulthen}
\begin{center}
\begin{tabular}{|l|l|}
\hline
$\sigma (\xi)$ & $\xi (1-\xi)  $ \\ \hline
$\widetilde{\sigma }(\xi)$ & $ (1-\xi) \left[\left( {\alpha}^2 + {\beta}^2 \right)\xi - {\alpha}^2 \right] $ \\ \hline
$\widetilde{\tau }(\xi)$ & $ 1 - \xi$ \\ \hline
$k$ & $ {\beta^2} - \alpha %
$
\\ \hline
$\pi(\xi) $ & $ \alpha - (\alpha + 1) \xi  $ \\
\hline
$\tau (\xi) =\widetilde{\tau }+2\pi $ & $ 2\alpha + 1 - (2\alpha +3) \xi  $ \\ \hline
$\lambda =k+\pi ^{\prime }$ & $ \beta^2 - 2\alpha -1  $ \\ \hline
$\varphi (\xi) $ & $\xi^{\alpha} (1-\xi) $ \\ \hline
$\rho (\xi) $ & $\xi^{2\alpha} (1-\xi) $ \\ \hline
$ E_n $ & \mybox{$-V_0 \left( \dfrac{\beta^2 - n^2}{2\beta n} \right)^2, \qquad n=1,2,3,\ \dots \quad \left(n^2<\beta^2 \right)$} \\ \hline
$y_{n}(\xi)$ & $ \ _2F_1(1-n, \, 1+2\alpha +n ;\, 2\alpha +1 ;\,  \xi)$ \\ \hline
$C_{n}$ & \mybox{$\dfrac{(2\alpha)_{n}}{n!}\sqrt{\dfrac{(\alpha+n) (2\alpha+n)}{(2\alpha) a}},\quad
\alpha = {\alpha}_n =\dfrac{\beta ^{2}-n^{2}}{2n} >0$} \\ \hline
\end{tabular}
\end{center}
\end{table}
%


%

\section{Morse Potential}

The following central field potential:%
\begin{equation}
U(r) =D\left( e^{-2\alpha x}-2e^{-\alpha x}\right) ,\quad x=%
\frac{r-r_{0}}{r_{0}}  \quad \left( 0\leq r<\infty \right) \label{mp1}
\end{equation}%
is used for the study of vibrations of two-atomic molecules  \cite{AiOthman}, \cite{Flugge}, \cite{Morse29}.
The corresponding Schr\"{o}dinger equation (\ref{s6}) can be solved in
spherical coordinates (\ref{Conf1}) when $l=0.$ Introducing new parameters%
\begin{equation}
\beta ^{2}=-\frac{2mEr_{0}^{2}}{\hbar ^{2}}>0,\qquad \gamma ^{2}=\frac{%
2mDr_{0}^{2}}{\hbar ^{2}} \label{mp2}
\end{equation}%
($\beta, \gamma> 0\/$), with the help of the following substitution%
\begin{equation}
R(r) =u(\xi) ,\qquad \xi =\frac{2\gamma }{\alpha }%
e^{-\alpha x},\quad x=\frac{r-r_{0}}{r_{0}}\label{mp3}
\end{equation}%
one gets%
\begin{equation}
\xi ^{2}\frac{d^{2}u}{d\xi ^{2}}+\xi \frac{du}{d\xi }+\left( -\frac{\beta
^{2}}{\alpha ^{2}}+\frac{\gamma }{\alpha }\xi -\frac{1}{4}\xi ^{2}\right)
u=0.\label{mp4}
\end{equation}
This is the generalized equation of hypergeometric type with%
\begin{equation}
\sigma(\xi) =\xi ,\qquad \widetilde{\tau }(\xi)
=1,\label{mp5}
\end{equation}%
and%
\begin{equation}
\widetilde{\sigma }(\xi) =-\frac{\beta ^{2}}{\alpha ^{2}}+\frac{%
\gamma }{\alpha }\xi -\frac{1}{4}\xi ^{2}.\label{mp6}
\end{equation}%
The substitution%
\begin{equation}
u=\varphi(\xi) y(\xi) = \xi ^{\beta /\alpha }e^{-\xi /2}y(\xi) \label{mp7}
\end{equation}%
results in the confluent hypergeometric equations:%
\begin{equation}
y^{\prime \prime }+\left( \frac{2\beta }{\alpha }+1-\xi \right) y^{\prime
}+\left( \frac{\gamma -\beta }{\alpha }-\frac{1}{2}\right) y=0\label{mp8}
\end{equation}%
with the following values of parameters:%
\begin{equation}
c=2\frac{\beta }{\alpha }+1,\quad a=\frac{1}{2}c-\frac{\gamma }{\alpha }=%
\frac{1}{2}+\frac{\beta -\gamma }{\alpha }\label{mp9}
\end{equation}%
(see Table~\ref{tab:Morse} and the corresponding Mathematica file for more details).
The general solution of (\ref{mp8}) has the form \cite{{Ni:Uv}}, \cite{NIST}:%
\begin{equation}
y=A\ _{1}F_{1}\biggl(
\begin{array}{@{}c}
a \\
c%
\end{array}%
;\xi \biggr) +B\ \xi ^{1-c}\ _{1}F_{1}\biggl(
\begin{array}{@{}c}
1+a-c \\
2-c%
\end{array}%
;\xi \biggr) .\label{mp10}
\end{equation}%
Here, the second constant must vanish, $B=0,$ due to the boundary condition $%
\lim_{r\rightarrow \infty }R(r) =u(0) =0,$ because%
\begin{equation}
1-c+\frac{\beta }{\alpha }=-\frac{\beta }{\alpha }<0.\label{mp11}
\end{equation}%
The first constant $A$ has to be determined by the normalization.
The second boundary condition, namely, $\lim_{r\rightarrow 0}R(r) =u(\xi_{0}) =0$
with $\xi _{0}=(2\gamma /\alpha) e^{\alpha }$, states%
\begin{equation}
\ _{1}F_{1}\biggl(
\begin{array}{@{}c}
a \\
c%
\end{array}%
;\xi _{0}\biggr) =0,  \label{mp12}
\end{equation}%
where both coefficients depend on energy in view of (\ref{mp2}) and (\ref%
{mp9}). This transcendent equation for the discrete energy levels cannot be
solved explicitly but for all real diatomic molecules $\xi _{0}\gg 1$~\cite%
{Flugge}. This is why one can use the familiar asymptotic:%
\begin{align}
{}_{1}F_{1}\biggl(\begin{array}{@{}c} a \\ c \end{array};\xi \biggr)
&= \frac{\Gamma(c)}{\Gamma(c-a)}
(-\xi)^{-a}\left[ 1+\text{O}\left( \frac{1}{\xi }\right) \right]  \label{mp13} \\
&\quad +\frac{\Gamma(c)}{\Gamma(a)}\ e^{\xi }\ \xi^{a-c}
\left[ 1+\text{O}\left( \frac{1}{\xi }\right) \right] ,\quad \xi\to\infty \notag
\end{align}
for the confluent hypergeometric function \cite{Ni:Uv}, \cite{NIST}.

By eliminating the largest asymptotic term with $\Gamma(a)=\infty$, an approximate quantization rule states:
\footnote{
The values $v=0$ can be added because $\exp(-\xi_{0}/2) \ll 1$ and the upper bound is due to convergence of the normalization integral (\ref{mp19}). }%
\begin{equation}
a=-v;\quad v=0,1,2, \ \dots  ;\quad v< \dfrac{\gamma }{\alpha } - \dfrac{1}{2} %
 \label{mp14}
\end{equation}%
(more details can be found in  \cite{Flugge}). The corresponding
approxiation to the discrete energy levels is given by%
\begin{equation}
-\beta ^{2}=-\gamma ^{2}+2\gamma \alpha \left( v+\frac{1}{2}\right) -\alpha
^{2}\left( v+\frac{1}{2}\right) ^{2}.\label{mp15}
\end{equation}%
This result can also be obtained in the Nikiforov-Uvarov approach by (\ref{A7}).
Hence, the approximate energy levels in terms of the vibrational quantum
number $v$ are%
\begin{equation}
E_{v}=-D+\frac{\hbar ^{2}}{2mr_{0}^{2}}\left[ 2\gamma \alpha \left( v+\frac{1%
}{2}\right) -\alpha ^{2}\left( v+\frac{1}{2}\right) ^{2}\right] ,\label{mp16}
\end{equation}%
where the last term reflects the anharmonicity correction.
This formula can be rewritten as follows~\cite{Flugge}:%
\begin{equation}
E_{v}=-D+\hbar \omega \left[ \left( v+\frac{1}{2}\right) -\frac{\alpha }{%
2\gamma }\left( v+\frac{1}{2}\right) ^{2}\right] ,\qquad \hbar \omega ={\hbar^{2}}\frac{%
\alpha \gamma }{mr_{0}^{2}} = \hbar \dfrac{\alpha}{ r_0 } \sqrt{\dfrac{2 D}{m}} \/ .\label{mp17}
\end{equation}%
The first two terms in this formula are in complete agreement with the harmonic oscillator energy levels.
The last term reflects the anharmonicity correction,
which shows that the anharmonic term never exceeds the harmonic one \cite{Flugge}.
The corresponding radial wave functions are given in terms of the Laguerre
polynomials (Table~\ref{tab:cop}\/):
\begin{equation}
R(r) =R_{v}(r) =\sqrt{\frac{(2\beta )\ v!}{%
r_{0}\Gamma \left( 2\beta /\alpha +v+1\right) }} \, \xi ^{\beta /\alpha
}e^{-\xi /2}\ L_{v}^{2\beta /\alpha }(\xi) .  \label{mp18}
\end{equation}%
Here, we have used the following normalization:%
\begin{align}
\int_{r=0}^{\infty }R^{2}(r) \ dr &=\frac{r_{0}}{\alpha }%
\int_{\xi =0}^{\xi _{0}}\xi ^{2\beta /\alpha - 1}e^{-\xi }y^{2}(\xi) \ d\xi   \label{mp19} \\
&\approx C^2_v \dfrac{r_{0}}{\alpha }\int_{\xi =0}^{\infty }\xi ^{2\beta /\alpha
- 1}e^{-\xi }\left( L_{v}^{2\beta /\alpha }(\xi) \right) ^{2}\
d\xi =1  \notag
\end{align}%
and the following integral\/:%
\begin{equation}
I_{-1}= J^{\delta \delta}_{mm,-1}=\int_{0}^{\infty }e^{-\xi }\xi ^{\delta -1}\left( L_{m}^{\delta }(\xi
)\right) ^{2}\ d\xi =\frac{\Gamma (\delta +m+1)}{m!\delta } \/.\label{mp20}
\end{equation}
Here $\delta =2\beta/\alpha > 0$  (see \cite{Sus:Trey}, \cite{Suslov2020}
and Appendix~\ref{app:B}; further details are left to the reader).
%

%
%

\begin{table}
\caption[]{The Morse potential \eqref{mp1} in the spherically symmetric case $l=0$.
Substitution: $R(r) = u(\xi)$, $\xi=(2\gamma/\alpha)\exp(-\alpha x)$,
$\beta^{2}=-2mEr_{0}^{2}/\hbar ^{2}>0$, $\gamma^2 = 2m D r^2_0/\hbar^2$.}
\label{tab:Morse}
\begin{center}
\begin{tabular}{|l|l|}
\hline
$\sigma (\xi)$ & $\xi  $ \\ \hline
$\widetilde{\sigma }(\xi)$ & \mybox{$-\dfrac{\beta^2}{{\alpha}^2} + \dfrac{\gamma}{\alpha} \xi - \dfrac{1}{4} {\xi}^2$} \\ \hline
$\widetilde{\tau }(\xi)$ & $ 1 $ \\ \hline
$k$ & \mybox{$\dfrac{\gamma-\beta}{\alpha}$} \\ \hline
$\pi(\xi) $ & \mybox{$\dfrac{\beta}{\alpha} - \dfrac{\xi}{2}$} \\ \hline
$\tau (\xi) =\widetilde{\tau }+2\pi $ & \mybox{$1+ \dfrac{2\beta}{\alpha}  - \xi$} \\ \hline
$\lambda =k+\pi ^{\prime }$ & \mybox{$\dfrac{\gamma-\beta}{\alpha}  - \dfrac{1}{2}$} \\ \hline
$\varphi (\xi) $ & $\xi^{\beta/\alpha} e^{-\xi/2} $ \\ \hline
$\rho (\xi) $ & $\xi^{2\beta/\alpha} e^{-\xi} $ \\ \hline
$ E_{v} $ & \mybox{$-D+\dfrac{\hbar ^{2}}{2mr_{0}^{2}}\left[ 2\gamma \alpha \left( v+\dfrac{1%
}{2}\right) -\alpha ^{2}\left( v+\dfrac{1}{2}\right) ^{2}\right]$} \\ \hline
$y_{v}(\xi)$ & \mybox{$C_v \ L_v^{2\beta/\alpha}(\xi) = C_v \dfrac{{\Gamma({2\beta/\alpha}+v+1)}}{{v!\,\Gamma({2\beta/\alpha}+1)}}
 \ _1F_1(-v; 2{\beta/\alpha}+1 \  ; \xi)$} \\ \hline
$C^2_{v}$ & \mybox{${\dfrac{(2\beta) v!}{r_0\Gamma(2\beta/\alpha +v+1)}}$} \\ \hline
\end{tabular}
\end{center}
\end{table}
%
%

\section{Rotation Correction of Morse Potential}

The standard centrifugal term \cite{SchroedingerOscillator}\/:%
\begin{equation}
\dfrac{l ( l+1 ) }{r^{2}}=
\dfrac{l ( l+1 ) }{r_0^2}
\dfrac{1}{(1+x)^2} \/,
\quad x=\dfrac{r-r_{0}}{r_{0}} \label{rmp1}
\end{equation}%
can be approximated, in the neighborhood of the minimum of the Morse
potential $r=r_{0}$ (or $x=0$), as follows \cite{Flugge}:%
\begin{equation}
\dfrac{l ( l+1 ) }{r^{2}}
\approx \frac{l(l+1)}{r_{0}^{2}}\left(
C_{0}+C_{1}e^{-\alpha x}+C_{2}e^{-2\alpha x}\right) \/ ,
\label{rmp2}
\end{equation}%
where%
\begin{equation}
C_{0}=1-\frac{3}{\alpha }+\frac{3}{\alpha ^{2}},\quad C_{1}=\frac{4}{\alpha }%
-\frac{6}{\alpha ^{2}},\quad C_{2}=-\frac{1}{\alpha }+\frac{3}{\alpha ^{2}}%
.\quad   \label{rmp3}
\end{equation}%
Indeed,%
\begin{equation}
\frac{1}{(1+x)^{2}}-\left( C_{0}+C_{1}e^{-\alpha
x}+C_{2}e^{-2\alpha x}\right) =x^{3}\left( -\frac{2}{3}\alpha ^{2}+3\alpha
-4\right) +\text{O}(x^{4}) ,\quad x\rightarrow 0.  \label{rmp4}
\end{equation}%
This consideration allows one to introduce a rotation correction to the Morse
potential without changing the mathematical model much~\cite{Fluggetal67}.
%

%
In this approximation, the radial Schr\"{o}dinger equation (%
\ref{s6}) in spherical coordinates (\ref{Conf1})
\begin{equation}
R^{\prime \prime }(r)+\left[ \frac{2m}{\hbar ^{2}}\left( E-D\left( e^{-2\alpha
x}-2e^{-\alpha x}\right)\right)  - \dfrac{l ( l+1 ) }{r^{2}}
\right] R(r)=0,\label%
{rmp5}
\end{equation}%
with the new variables
\begin{equation}
R(r) =u(\xi) ,\qquad \xi =\frac{2\gamma _{2}}{%
\alpha }e^{-\alpha x},\quad x=\frac{r-r_{0}}{r_{0}}\label{rmp6}
\end{equation}
and with the modified parameters
\begin{alignat}{3}
\beta _{1}^{2} &= \beta ^{2}+l(l+1) C_{0}, & \quad \beta ^{2}
&=-\frac{2mEr_{0}^{2}}{\hbar ^{2}}>0,\label{rmp7a}
\\
\gamma _{1}^{2} &= \gamma ^{2}-\frac{1}{2}l(l+1) C_{1}, & \quad
\gamma _{2}^{2} &= \gamma ^{2}+l(l+1) C_{2}, \quad
\gamma ^{2}=\frac{2mDr_{0}^{2}}{\hbar ^{2}},
\label{rmp7b}
\end{alignat}%
becomes the following generalized equation of hypergeometric type:%
\begin{equation}
\xi ^{2}u^{\prime \prime }+\xi u^{\prime }+\left( -\frac{\beta _{1}^{2}}{%
\alpha ^{2}}+\frac{\gamma _{1}^{2}}{\alpha \gamma _{2}}\xi -\frac{1}{4}\xi
^{2}\right) u=0.\label{rmp8}
\end{equation}%
Here%
\begin{equation}
\sigma(\xi) =\xi ,\qquad \widetilde{\tau }(\xi)
=1,\label{rmp9a}
\end{equation}%
and%
\begin{equation}
\widetilde{\sigma }(\xi) =-\frac{\beta _{1}^{2}}{\alpha ^{2}}+%
\frac{\gamma _{1}^{2}}{\alpha \gamma _{2}}\xi -\frac{1}{4}\xi ^{2}.\label%
{rmp9b}
\end{equation}
%

%
The following substitution%
\begin{equation}
u=\xi ^{\beta _{1}/\alpha }e^{-\xi /2}y(\xi)
\end{equation}%
results, once again, in the confluent hypergeometric equation
\begin{equation}
y^{\prime \prime }+\left( \frac{2\beta _{1}}{\alpha }+1-\xi \right)
y^{\prime }+\left( \frac{\gamma _{1}^{2}}{\alpha \gamma _{2}}-\frac{\beta
_{1}}{\alpha }-\frac{1}{2}\right) y=0
\end{equation}%
with the new values of the parameters:%
\begin{equation}
c_{1}=2\frac{\beta _{1}}{\alpha }+1,\quad a_{1}=\frac{1}{2}c_{1}-\frac{%
\gamma _{1}^{2}}{\alpha \gamma _{2}}=\frac{1}{2}+\frac{\beta _{1}}{\alpha }-%
\frac{\gamma _{1}^{2}}{\alpha \gamma _{2}}
\end{equation}%
(see Table~\ref{tab:rmp} and the corresponding Mathematica file for more details).
%

%
An approximate quantization rule states:%
\begin{equation}
a_{1}=-v;\quad v=0,1,2,\ \ldots \ .\quad   \label{rmp13}
\end{equation}%
Therefore,%
\begin{equation}
-\beta _{1}^{2}=-\left[ \frac{\gamma _{1}^{2}}{\gamma _{2}}-\alpha \left( v+%
\frac{1}{2}\right) \right] ^{2},  \label{rmp14}
\end{equation}%
and, in the energy formula, one has to replace $\gamma $ by%
\begin{equation}
\frac{\gamma _{1}^{2}}{\gamma _{2}}\approx \gamma \left[ 1-l(l+1)
  \frac{C_{1}+C_{2}}{2\gamma ^{2}}\right] , \quad \gamma \gg 1 \/ .  \label{rmp15}
\end{equation}%
As a result, we arrive at the following vibration-rotation energy levels:%
\begin{align}
E = E_{vl} &= \left. \frac{\hbar ^{2}}{2mr_{0}^{2}}\left[ -\gamma ^{2}+2\gamma
\alpha \left( v+\frac{1}{2}\right) -\alpha ^{2}\left( v+\frac{1}{2}\right)
^{2}+l(l+1) \right. \right.   \notag \\
&\quad\;\left. -\frac{3(\alpha-1)}{\alpha \gamma }\left( v+\frac{1}{2%
}\right) l(l+1) -\frac{9(\alpha-1) ^{2}}{4\alpha
^{4}\gamma ^{2}}l^{2}(l+1)^{2}\right] .  \label{rmp16}
\end{align}%
This formula can be presented in the form%
\begin{align}
E_{vl} &= -D+\hbar \omega \left[ \left( v+\frac{1}{2}\right) -\frac{\alpha }{%
2\gamma }\left( v+\frac{1}{2}\right) ^{2}\right] +\frac{\hbar ^{2}l(l+1)}{2mr_{0}^{2}}
\label{rmp16a} \\
&\quad\; -\frac{3(\alpha-1)}{2\alpha ^{2}D}\hbar \omega \left(
v+\frac{1}{2}\right) \frac{\hbar ^{2}l(l+1)}{2mr_{0}^{2}}-\frac{%
9(\alpha-1)^{2}}{4\alpha ^{2}D}\left( \frac{\hbar ^{2}l(l+1)}{2mr_{0}^{2}}\right) ^{2},  \notag
\end{align}%
where
\begin{equation}
\hbar \omega =\dfrac{\hbar ^{2}\alpha \gamma }{mr_{0}^{2}} = \hbar \left( \dfrac{\alpha}{ r_0 } \sqrt{\dfrac{2 D}{m}} \right) \/ .\label{rmp16b}
\end{equation}%
The first three terms of this formula are exactly the same as those derived
in the previous case; see (\ref{mp17}). The fourth term can be interpreted
as the molecule rotational energy at fixed distance $r_{0}\/.$
The next term represents a coupling of the vibrations and rotations,
which is negative because at higher vibrational quantum numbers the average nuclear distance increases beyond $r_0$ in consequence of the anharmonicity \cite{Flugge}.
The last term can be thought of as a negative second-order correction to the rotation energy.

The corresponding radial wave functions are given in terms of the Laguerre
polynomials (Table~\ref{tab:cop}\/):
\begin{equation}
R(r) =R_{vl}(r) =\sqrt{\frac{(2\beta _{1})\ v!}{%
r_{0}\Gamma( 2\beta _{1}/\alpha +v+1)}} \, \xi ^{\beta _{1}/\alpha
}e^{-\xi /2}\ L_{v}^{2\beta _{1}/\alpha }(\xi) .  \label{rmp17}
\end{equation}%
Once again, we have used the normalization similar to (\ref{mp19}).
Further details on the rotational corrections of Morse formulas are discussed in \cite{Flugge}, \cite{Fluggetal67}.
\smallskip

%
\begin{table}
\caption[]{The rotation correction of the Morse potential \eqref{rmp2},
where the coefficients $C_0$, $C_1$, and $C_2$ are given by \eqref{rmp3}.}
\label{tab:rmp}
\begin{center}
\begin{tabular}{|l|l|}
\hline
$\sigma (\xi)$ & $\xi$ \\ \hline
$\widetilde{\sigma }(\xi)$ & \mybox{$-\dfrac{\beta^2_1}{{\alpha}^2} + \dfrac{\gamma_1^2}{\alpha \gamma_2} \xi - \dfrac{1}{4} {\xi}^2$} \\ \hline
$\widetilde{\tau }(\xi)$ & $ 1 $ \\ \hline
$k$ & \mybox{$\dfrac{\gamma_1^2}{\alpha \gamma_2}  - \dfrac{\beta_1}{\alpha}$} \\ \hline
$\pi(\xi) $ & \mybox{$\dfrac{\beta_1}{\alpha} - \dfrac{\xi}{2}$} \\ \hline
$\tau (\xi) =\widetilde{\tau }+2\pi $ & \mybox{$1+ \dfrac{2\beta_1}{\alpha}  - \xi$} \\ \hline
$\lambda =k+\pi ^{\prime }$ & \mybox{$\dfrac{\gamma_1^2}{\alpha \gamma_2} - \dfrac{\beta_1}{\alpha} - \dfrac{1}{2}$} \\ \hline
$\varphi (\xi) $ & \mybox{$\xi^{\beta_1/\alpha} e^{-\xi/2}$} \\ \hline
$\rho (\xi) $ & \mybox{$\xi^{2\beta_1/\alpha} e^{-\xi}$} \\ \hline
\end{tabular}
\end{center}
\end{table}
%
%

%
\section{Modified Hulth\'{e}n Potential}
Once again, we are looking for solutions of the stationary Schr\"{o}dinger equation (\ref{s6}) in spherical coordinates (\ref{Conf1})
for the following central field potential,
\begin{equation}
U(r) =-V_{0}\ \frac{e^{-r/a}\left(1-be^{-r/a}\right)}{\left(1-e^{-r/a}\right)^2} \qquad  (0\leq r<\infty ) \/,
\label{mhp1}
\end{equation}
when $b>1\/$ and $l=0\/.$ (The special case $b=1$ corresponds to the original Hulth\'{e}n  potential (\ref{hp1}) above.)
By using substitution (\ref{hp2}) with the same positive parameters (\ref{hp2a}),
one obtains the following generalized equation of hypergeometric type,%
\begin{equation}
\xi ^{2}\frac{d^{2}u}{d\xi ^{2}}+\xi \frac{du}{d\xi }+\left[ -\alpha
^{2}+\beta ^{2}\ \frac{\xi \left(1-b \xi\right)}{\left(1-\xi\right)^2 }\right] u=0 , \/  \label{mhp2}
\end{equation}%
where%
\begin{equation}
\sigma(\xi) =\xi (1-\xi) ,\qquad \widetilde{\tau }%
(\xi) =1-\xi ,  \label{mhp3}
\end{equation}%
and%
\begin{equation}
\widetilde{\sigma }(\xi) = - \left(\alpha ^{2}+b\beta ^{2}\right) {\xi^2} +
\left(2{\alpha}^{2} + {\beta}^2\right)\xi - {\alpha}^{2} \/ .\label{mhp4}
\end{equation}%
For the modified Hulth\'{e}n potential, let us first analyze the same boundary conditions (\ref{hp6}) as before.
We have%
\begin{equation}
k=\beta ^{2}\mp 2\alpha \kappa ,\qquad
\pi(\xi) = - \dfrac{\xi}{2} \pm \left[ (\alpha \pm \kappa) \xi - \alpha \right] \/, \label{mhp5}
\end{equation}
where
\begin{equation}
\kappa=\sqrt{\dfrac{1}{4}+(b-1){\beta}^2} \, > \dfrac{1}{2} \/ , \label{mhp6}
\end{equation}
for all four possible sign combinations.
The solutions can be found by putting:
\begin{equation}
u= \varphi(\xi) y(\xi) = \xi ^{\alpha }(1-\xi)^{\kappa+1/2} y(\xi) ,\label{mhp7}
\end{equation}%
which results in the hypergeometric equation (\ref{gauss}):
\begin{equation}
\xi(1-\xi) y^{\prime \prime }+\left[ 2\alpha +1-2(\alpha+\kappa + 1) \xi \right] y^{\prime }
-\left[ (2\alpha + 1)\left(\kappa + \dfrac{1}{2}\right)-\beta ^{2}\right] y=0 \/,%
\label{mhp8}
\end{equation}
with
\begin{equation}
A=\alpha + \kappa + \dfrac{1}{2} + \gamma , \quad
B=\alpha + \kappa + \dfrac{1}{2} - \gamma ,  \quad
C=2\alpha + 1 , \quad
\gamma = \sqrt{\alpha^2 + b\beta^2}
\label{mhp9}
\end{equation}
(details of calculations are presented in Table~\ref{tab:ModHulthen} and in a complementary
Mathematica file).
The required solution, that is bounded at $\xi =0,$ has the form%
\begin{equation}
y= {}_{2}F_{1}\biggl(
\begin{array}{@{}c@{}}
1/2 + \kappa + \alpha +\gamma ,\ 1/2 + \kappa +\alpha -\gamma  \\
2\alpha +1%
\end{array}%
;\xi \biggr) , \label{gaussolu}
\end{equation}%
up to a constant, and the first boundary condition is satisfied $u(0)=0\/,$ when $\alpha > 0 \/.$
In view of (\ref{gausslim}), one gets
\begin{align}
u(1) &= \lim_{\xi \rightarrow 1^{-}} {(1-\xi)^{-(\kappa-1/2)}} \left[(1-\xi)^{2\kappa} {}_{2}F_{1}\biggl(
\begin{array}{@{}c@{}}
1/2 +\kappa +\alpha +\gamma ,  1/2 +\kappa +\alpha -\gamma  \\
2\alpha +1%
\end{array}%
;\xi \biggr) \right] \label{mhp10} \\
&= \frac{\Gamma(2\alpha+1)\,\Gamma(2\kappa)}{\Gamma(1/2 +\kappa +\alpha+\gamma)\,\Gamma(1/2 +\kappa +\alpha-\gamma)}
\, \times \lim_{\xi \rightarrow 1^{-}} {(1-\xi)^{-(\kappa-1/2)}}=\infty \/.
\notag
\end{align}%
Therefore, one has to look for square-integrable solutions that are not bounded at the origin.
By terminating the hypergeometric series in (\ref{gaussolu}), or by using the Nikiforov--Uvarov condition (\ref{A7}), we arrive at the following quantization rule:
\begin{equation}
\alpha +\kappa +\dfrac{1}{2} -\gamma = -n \quad (n=0, 1, 2, \  \dots \ \le n_{\rm{max}}) \/,
\label{mhp11}
\end{equation}
or
\begin{equation}
\alpha =\dfrac{b\beta^2-(n+ \kappa +1/2)^2}{2(n+ \kappa +1/2)} > 0 \/ .
\label{mhp12}
\end{equation}
The corresponding energy levels are given by
\begin{equation}
E_n = -V_0 \left( \dfrac{b\beta^2-(n+ \kappa +1/2)^2}{2\beta(n+ \kappa +1/2)} \right)^2 \/ ,
\label{mhpEnergy}
\end{equation}
where
\begin{equation}
{b {\beta}^{2}-\left(\kappa +n+\dfrac{1}{2} \right)^{2}} = {\beta}^{2} - n^2 -(2\kappa +1)\left(n+\dfrac{1}{2}\right) > 0 \/.
{\label{mhpEnergyCom}}
\end{equation}
Once again, there exists a minimum size of potential hole before any energy eigenvalue at all can be obtained, viz. $b\beta^2=(\kappa +1/2)^2\/.$
Equation $(n+ \kappa +1/2)^2 \le b\beta^2 $ determines the finite number of eigenvalues, $n \le n_{\rm{max}}\/,$ in a potential hole of a given size.

The radial wave functions take the form%
\begin{equation}
R_{n}(r) =C_{n}\ \xi ^{\alpha }(1-\xi)^{\kappa +1/2} \ _{2}F_{1}\biggl(
\begin{array}{@{}c@{}}
-n,\ 2\alpha +2\kappa +n + 1 \\
2\alpha +1%
\end{array}%
;\xi \biggr) ,\quad \xi =e^{-r/a},\label{mhpFunctions}
\end{equation}%
where the hypergeometric series terminates and $C_n$ is a constant to be determined.
Once again, this result can also be obtained with the help of the Rodrigues-type formula (\ref{A8}):
\begin{equation}
\left[ \xi ^{2\alpha +n}(1-\xi) ^{2\kappa +n}\right]^{(n)} =
\frac{\Gamma(2\alpha+n +1)}{\Gamma(2\alpha +1)} \ \xi ^{2\alpha}(1-\xi)^{2\kappa}
{}_{2}F_{1}\biggl(
\begin{array}{@{}c@{}}
-n,\ 2\alpha +2\kappa +n+1 \\
2\alpha +1%
\end{array}%
;\xi \biggr) \/.
\label{mhpRodriguesFormula}
\end{equation}
(It can be thought of as a special case of (15.5.9) from \cite{NIST}.)

The normalization condition (\ref{KMP6}) for the radial wave function becomes
\begin{equation}
a C^2_{n}\ \int_{0}^{1}\xi ^{2\alpha -1}(1-\xi)^{2\kappa +1}y_{n}^{2}(\xi) \, d\xi =1\label{mhp13} \/ ,
\end{equation}
where%
\begin{equation}
\xi ^{2\alpha }(1-\xi)^{2\kappa} y_{n}(\xi) =\frac{\Gamma(2\alpha+1)}{\Gamma(2\alpha +n +1)}\left[ \xi
^{2\alpha +n}(1-\xi)^{2\kappa +n}\right] ^{(n)}.
\label{mhp14}
\end{equation}
Integrating by parts $n$ times, one gets%
\begin{align}
&\int_{0}^{1}(\xi ^{-1}-1)\ _{2}F_{1}\biggl(
\begin{array}{@{}c@{}}
-n,\ 2\alpha +2\kappa +n +1\\
2\alpha +1%
\end{array}%
;\xi \biggr) \left[ \xi ^{2\alpha +n}(1-\xi)^{2\kappa +n}\right]^{(n)} d\xi \\
&=\left. \left( (\xi ^{-1} -1)\ _{2}F_{1}\biggl(
\begin{array}{@{}c@{}}
-n,\ 2\alpha +2\kappa +n +1 \\
2\alpha +1%
\end{array}%
;\xi \biggr) \left[ \xi ^{2\alpha +n}(1-\xi)^{2\kappa +n}\right]
^{(n-1)}\right) \right\vert _{\xi =0}^{1}  \notag \\
&\; \; \quad -\int_{0}^{1}\left[ (\xi ^{-1}-1) \ _{2}F_{1}\biggl(
\begin{array}{@{}c@{}}
-n,\ 2\alpha +2\kappa +n +1\\
2\alpha +1%
\end{array}%
;\xi \biggr) \right] ^{\prime }\left[ \xi ^{2\alpha +n}(1-\xi)^{2\kappa +n}\right]^{(n-1)} d\xi   \notag \\
&= \dots = \notag \\
& (-1) ^{k-1}\left. \left( \left[ (\xi ^{-1} -1)\
_{2}F_{1}\biggl(
\begin{array}{@{}c@{}}
-n,\ 2\alpha +2\kappa +n +1\\
2\alpha +1%
\end{array}%
;\xi \biggr) \right] ^{(k-1)}\hspace{-20pt}\left[ \xi ^{2\alpha +n}(1-\xi)^{2\kappa +n}\right] ^{(n-k)}\right) \right\vert _{\xi
=0}^{1}  \notag \\
&  +(-1)^{k}\!\int_{0}^{1}\left[ (\xi ^{-1}-1) \ _{2}F_{1}\biggl(
\begin{array}{@{}c@{}}
-n,\ 2\alpha +2\kappa +n +1 \\
2\alpha +1%
\end{array}%
;\xi \biggr) \right] ^{(k)}\hspace{-8pt}
\left[ \xi ^{2\alpha +n}(1-\xi)^{2\kappa +n}\right] ^{(n-k)} d\xi   \notag \\
&= \dots =(-1)^{n}\int_{0}^{1}\left[ (\xi ^{-1} -1)\ _{2}F_{1}\biggl(
\begin{array}{@{}c@{}}
-n,\ 2\alpha +2\kappa +n +1 \\
2\alpha +1%
\end{array}%
;\xi \biggr) \right] ^{(n)}\!\!\xi ^{2\alpha +n}(1-\xi) ^{2\kappa +n} d\xi \/ .  \notag
\label{mhp15}
\end{align}%
By the power series expansion,%
\begin{align*}
& \left[ (\xi ^{-1}-1) \ _{2}F_{1}\biggl(
\begin{array}{@{}c@{}}
-n,\ 2\alpha +2\kappa +n +1  \\
2\alpha +1%
\end{array}%
;\xi \biggr) \right] ^{(n)} \\
&\qquad\qquad=\left( \frac{1}{\xi }\right)
^{(n)} - \frac{(-n)_{n}(2\alpha +2\kappa +n +1)_{n}}{(n!) (2\alpha +1)_{n}}(\xi^{n})^{(n)} \notag \\
&\qquad\qquad=(-1)^{n}\frac{(n!)}{\xi ^{n+1}} - \frac{%
(-n)_{n}(2\alpha +2\kappa +n +1)_{n}}{(n!)(2\alpha+1)_{n}}(n!),  \notag
\end{align*}%
and the integral evaluation can be completed with the aid of the following Euler beta integrals (\ref{eqnAppB2}):
\begin{align*}
\int_{0}^{1}\xi^{2\alpha -1}(1-\xi)^{(2\kappa +n +1)-1} \ d\xi  &=\frac{\Gamma(2\alpha) \Gamma(2\kappa +n +1)}{\Gamma(2\alpha +2\kappa +n+1)}, \\
\int_{0}^{1}\xi^{2\alpha +n}(1-\xi)^{2\kappa +n} \ d\xi  &=\frac{\Gamma(2\alpha+n+1) \Gamma(2\kappa +n+1)}{\Gamma(2\alpha+2\kappa +2n+2)} \/ .
\end{align*}
The final result is given by
\begin{equation}
C_n^2 = \dfrac{\Gamma(2\alpha +n+1)\Gamma(2\alpha +2\kappa +n+1)(2\alpha +2\kappa +2n+1)}{a(2\alpha) \Gamma^2(2\alpha)\Gamma(2\kappa +n+1)(2\kappa +2n+1)n!} \/ .
\label{mhp16}
\end{equation}
Our modification of the Hulth\'{e}n potential can be used for study of vibrations for diatomic molecules, when $m$ becomes the reduced mass of two atoms.
Therefore, it is informative to compare the classical Morse potential and the modified
Hulth\'{e}n one.
Suppose that at the common point of the potential minimum $r=r_{\text{min}}$
the following conditions hold:%
\begin{align}
U_{\text{Morse}}(r_{\text{min}}) & = U_{\text{ModHulth\'{e}n}}(r_{\text{min}%
}), \\
\frac{d}{dr}U_{\text{Morse}}(r_{\text{min}}) & = \frac{d}{dr}U_{\text{ModHulth%
\'{e}n}}(r_{\text{min}})=0,  \notag \\
\frac{d^{2}}{dr^{2}}U_{\text{Morse}}(r_{\text{min}}) & = \frac{d^{2}}{dr^{2}}%
U_{\text{ModHulth\'{e}n}}(r_{\text{min}}). \notag
\label{mhp15a}
\end{align}
Then $r_{\text{min}}=r_{0}\/, \,$ exp($r_{0}/a)=2b-1\/$ and%
\begin{equation}
D=\frac{V_{0}}{4(b-1)},\qquad 2D\frac{\alpha ^{2}}{r_{0}^{2}}=\frac{%
(2b-1)^{2}V_{0}}{8a^{2}(b-1)^{3}}
\end{equation}
(see our Mathematica file for more details).
The Morse potential (\ref{mp1}) can be rewritten in an equivalent form:%
\begin{align}
U_{\text{Morse}}(r) &= \frac{V_{0}}{4(b-1)}\left( 2b-1\right) ^{\frac{2b-1}{%
2(b-1)}} \\
&\quad\times \left[ \left( 2b-1\right) ^{\dfrac{2b-1}{2(b-1)}}\exp \left( -%
\dfrac{(2b-1)r}{(b-1)a}\right) -2\exp \left( -\dfrac{(2b-1)r}{2(b-1)a}%
\right) \right] ,  \notag \label{mhp15b}
\end{align}%
where the parameters of the modified Hulth\'{e}n potential (\ref{mhp1}) have been
utilized.

\begin{figure}
\centering
\includegraphics[width=0.8\textwidth]{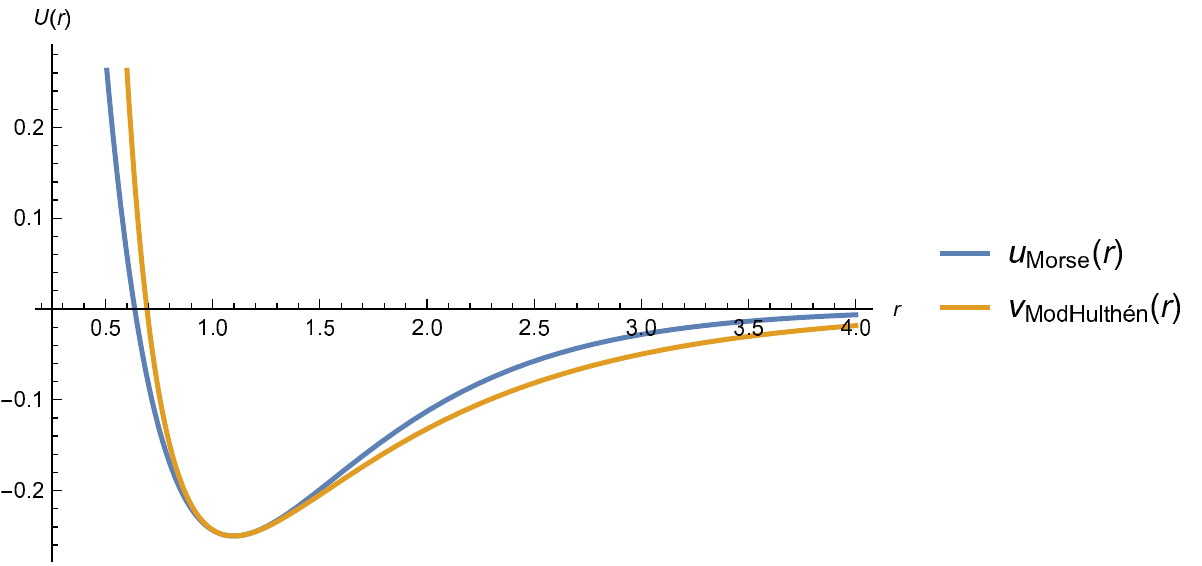}
\caption{An example, comparing the Morse potential (blue) and the modified Hulth\'{e}n
  potential (brown), when $V_{0}=a=1$, $b=2$, and $r_{\text{min}}=\ln 3\simeq 1.09861$.}
\label{Figure2}
\end{figure}

\begin{figure}
\centering
\includegraphics[width=0.8\textwidth]{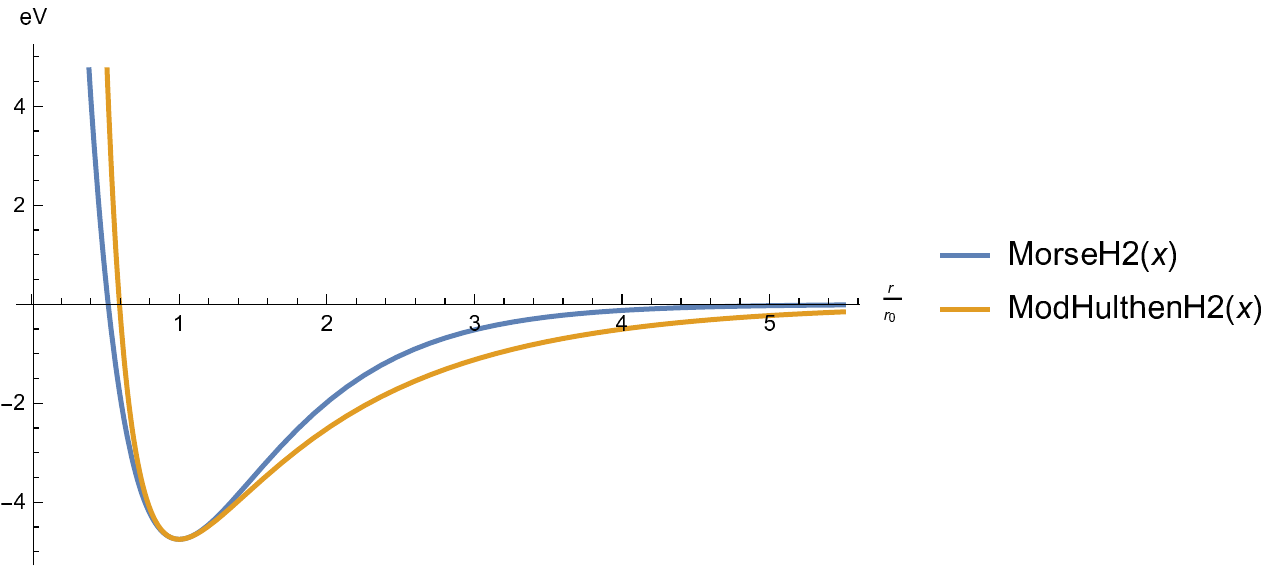}
\caption{The Morse potential (blue)~\cite{Flugge} and the modified Hulth\'{e}n
  potential (brown), for the molecule ${\text{H}_2}$.}
\label{Figure3}
\end{figure}

A graphical example, comparing both potentials,
when $V_{0}=a=1\/,$ $b=2\/,$ and $r_{\text{min}}=\ln 3\simeq 1.09861\/,$ is presented in Figure~\ref{Figure2} and further
discussed in our Mathematica notebook for the reader's convenience, as well as potentials for molecules $\text{H}_2\/$,
HCl, and $\text{I}_2\/$ with a completion of the data from~\cite{Flugge} in
Table~\ref{tab:MorseModHulthenParameters} (see also Figure~\ref{Figure3}).
%
%

\begin{table}
\caption[]{Comparison of the parameters for the Morse potential and the modified
  Hulth\'{e}n potential: $E($eV$)=E($cm$^{-1})\times 1.2398\times 10^{-4}$ according
  to~\cite{Flugge}.}
\label{tab:MorseModHulthenParameters}
\begin{center}
\begin{tabular}{|l|l|l|l|l|l|}
\hline
Molecule & \mybox{${\dfrac{\hbar ^{2}}{2mr_{0}^{2}}}\ ($cm$^{-1})$} & $D\ ($cm$^{-1})$
& $\alpha $ & $b$ & $V_{0}\ ($cm$^{-1})$ \\ \hline
H$_{2}$ & $60.8296$ & $38292$ & $1.440$ & $1.5904$ & $67394$ \\ \hline
HCl & $10.5930$ & $37244$ & $2.380$ & $4.51744$ & $5\allowbreak 24010$ \\
\hline
I$_{2}$ & $0.0374$ & $12550$ & $4.954$ & $68.848$ & $1984\,90$ \\ \hline
\end{tabular}
\end{center}
\end{table}
%

%
%

%
In order to find parameters of the modified Hulth\'{e}n potential in terms
of the Morse ones, we have solved numerically the following equation:%
\begin{equation}
\alpha =\varphi (b):=\frac{(2b-1)\ln (2b-1)}{2(b-1)},\quad b\geq 1
\label{mhp15c}
\end{equation}%
with the monotone function $\varphi (b).$ Then $V_{0}=4(b-1)D$ and $%
a=r_{0}/\ln (2b-1).$ (More details are provided in the Mathematica file.)

\textbf{Remark.} The so-called generalized Morse potential is usually introduced as
\begin{equation}
U(r)=D\left( 1-\frac{e^{\alpha r_{0}}-1}{e^{\alpha r}-1}\right) ^{2}
\end{equation}%
($D,$ $\alpha ,$ $r_{0}$ are parameters) with the following asymptotic
\begin{equation}
U(\infty )=\lim_{r\rightarrow \infty }U(r)=D >0
\end{equation}%
(see \cite{DelSol}, \cite{Rong2003}, and the references therein). The difference%
\begin{equation}
U(r)-U(\infty )=-2D\left( e^{\alpha r_{0}}-1\right) \ \frac{1-e^{-\alpha
r}\left( 1+e^{\alpha r_{0}}\right) /2}{\left( 1-e^{-\alpha r}\right) ^{2}}%
e^{-\alpha r}
\label{mhp17}
\end{equation}%
has the form of our modified Hulth\'{e}n potential (\ref{mhp1}) with the parameters
given by%
\begin{equation}
V_{0}=2D\left( e^{\alpha r_{0}}-1\right) ,\quad b=\frac{1}{2}\left(
1+e^{\alpha r_{0}}\right) ,\quad a=\frac{1}{\alpha }.
\label{mhp18}
\end{equation}
Therefore, this case has been studied here as well (see also section~19 for an independent computer algebra approach).

%
%

%
%

%
%

%
%
%

%
%

%

%
\begin{table}
\caption[]{The modified Hulth\'{e}n potential \eqref{mhp1}
in the spherically symmetric case $l=0$.
Substitution: $R(r) = u(\xi)$, $\xi=\exp(-r/a)$,
$\alpha^2 = -(2m E/\hbar^2)\ a^2 >0$, $\beta^2 = (2m V_0/\hbar^2)\ a^2 >0$.}
\label{tab:ModHulthen}
\begin{center}
\begin{tabular}{|l|l|}
\hline
$\sigma (\xi)$ & $\xi (1-\xi)  $ \\ \hline
$\widetilde{\sigma }(\xi)$ & $ - \left( {\alpha}^2 + b {\beta}^2 \right){\xi}^2 + \left(2{\alpha}^2 +{\beta}^2\right)\xi -{\alpha}^2 $ \\ \hline
$\widetilde{\tau }(\xi)$ & $ 1 - \xi$ \\ \hline
$k$ & \mybox{${\beta^2} - 2\alpha \kappa , \qquad \kappa=\sqrt{(1/4)+(b-1){\beta}^2}$}
\\ \hline
$\pi(\xi) $ & $ \alpha - (\alpha + \kappa + 1/2) \xi  $ \\
\hline
$\tau (\xi) =\widetilde{\tau }+2\pi $ & $ 2\alpha + 1 - 2(\alpha +\kappa +1) \xi  $ \\ \hline
$\lambda =k+\pi ^{\prime }$ & $ {\beta}^2 - (2\alpha +1)(\kappa+1/2)  $ \\ \hline
$\varphi (\xi) $ & $\xi^{\alpha} (1-\xi)^{\kappa+1/2} $ \\ \hline
$\rho (\xi) $ & $\xi^{2\alpha} (1-\xi)^{2\kappa} $ \\ \hline
$ E_n $ & \mybox{$ -V_0 \left( \dfrac{b\beta^2-(n+ \kappa +1/2)^2}{2\beta(n+ \kappa +1/2)} \right)^2 \/, \;  n=0,1,2,\ \dots \; \left(n \le n_{\rm{max}} \right)$} \\ \hline
$y_{n}(\xi)$ & ${}_2F_1(-n, \, 2\alpha +2\kappa  +n +1 ;\, 2\alpha +1 ;\,  \xi)$, \\
& \mybox{$\alpha ={\dfrac{b\beta^2-(n+ \kappa +1/2)^2}{2(n+ \kappa +1/2)} > 0}$} \\ \hline
$C_{n}^2$ & \mybox{$ \dfrac{\Gamma(2\alpha +n+1)\Gamma(2\alpha +2\kappa +n+1)(2\alpha +2\kappa +2n+1)}{a(2\alpha) \Gamma^2(2\alpha)\Gamma(2\kappa +n+1)(2\kappa +2n+1)n!}
$} \\ \hline
\end{tabular}
\end{center}
\end{table}

\section{Rotation Correction of Modified Hulth\'{e}n Potential}

In view of%
\begin{equation}
\frac{dU(r)}{dr}=V_{0}\ e^{r/a}\frac{1-2b+e^{r/a}}{a\left( e^{r/a}-1\right)
^{3}}=0,
\label{rmhp1}
\end{equation}%
the minimum of the modified Hulth\'{e}n potential \eqref{mhp1} occurs when%
\begin{equation}
e^{r_{\text{min}}/a}=2b-1>1.
\label{rmhp1a}
\end{equation}
By letting $r=a(1+x)x_{0}\/$ in the neighborhood of this minimum $r\approx r_{%
\text{min}}$ (or $x\approx 0$), with the aid of Mathematica, we derive the
following expansion:%
\begin{align}
&\frac{1}{(1+x)^{2}}-C_{0}-e^{-(1+x)x_{0}}\frac{C_{1}+C_{2}e^{-(1+x)x_{0}}}{%
\left( 1-e^{-(1+x)x_{0}}\right) ^{2}} \label{rmhp2} \\
& \quad=\left( -4+\frac{3bx_{0}}{b-1}-\frac{1-2b+4b^{2}}{6(b-1)^{2}}%
x_{0}^{2}\right) x^{3}+\text{O}(x^{4}),\quad x\rightarrow 0,  \notag
\end{align}%
where%
\begin{align}
C_{0} &= 1+4(b-1)\frac{3(b-1)-(3b-1)x_{0}}{(2b-1)^{2}x_{0}^{2}}, \label{rmhp3coeffs} \\
C_{1} &= 8(b-1)^{2}\frac{6(1-b)+(4b-1)x_{0}}{(2b-1)^{2}x_{0}^{2}},  \notag \\
C_{2} &= 8(b-1)^{2}\frac{6b(b-1)+(1-2b(b+1))x_{0}}{(2b-1)^{2}x_{0}^{2}} \notag
\end{align}%
and  $e^{x_{0}}=2b-1>1$ (see our complementary Mathematica file).
Therefore,%
\begin{equation}
\frac{l(l+1)}{r^{2}}\approx {\frac{l(l+1)}{r^2_{\text{min}}}}\left[ C_{0}+e^{-r/a}%
\frac{C_{1}+C_{2}e^{-r/a}}{\left( 1-e^{-r/a}\right) ^{2}}\right] \qquad (r_{\text{min}}=ax_0) \/,
{\label{rmhp4}}
\end{equation}
and our equation \eqref{mhp2} for the modified Hulth\'{e}n potential holds, once again, but with the following modified values of the parameters:
\begin{alignat}{3}
\alpha ^{2} &\rightarrow \alpha _{1}^{2}=\alpha ^{2}+\frac{l(l+1)}{x_{0}^2 }C_{0}, &\quad
\beta ^{2} &\rightarrow \beta _{1}^{2}=\beta ^{2}-\frac{l(l+1)}{x_{0}^{2}}C_{1}, \notag
\\
b &\rightarrow  b_{1}=\frac{\beta _{2}^{2}}{\beta _{1}^{2}}, & \quad
\beta_{2}^{2} &= b\beta ^{2}+{\frac{l(l+1)}{x_{0}^{2}}}C_{2}, \label{rmhp5}
\end{alignat}
namely,%
\begin{equation}
\xi ^{2}\frac{d^{2}u}{d\xi ^{2}}+\xi \frac{du}{d\xi }+\left[ -{\alpha_1^{2}}+{\beta_1^{2}}\
\frac{\xi \left(1-b_1 \xi\right)}{\left(1-\xi\right)^2 }\right] u=0 . \/
{\label{rmhp6}}
\end{equation}%
As a result, we have arrived at the generalized equation of hypergeometric type (Table~\ref{tab:rmhp}) and our analysis from the previous section is valid, say,
up to a proper change of parameters.

For example, the vibration and rotation energy levels are given by%
%
\begin{align}
E_{v,l} &= -V_{0}\left( \frac{\beta _{2}^{2}-\left( \kappa _{1}+v+1/2\right)
^{2}}{2\beta \left( \kappa _{1}+v+1/2\right) }\right) ^{2} \\
&\quad\;+\frac{\hbar ^{2}l(l+1)}{2mr_{\text{min}}^{2}}\left[ 1+4(b-1)\frac{%
3(b-1)-(3b-1)x_{0}}{(2b-1)^{2}x_{0}^{2}}\right] {\label{rmhp7}}  \notag
\end{align}
($v= 0, 1, 2,\ \dots \ \le v_{\text{max}}\/$). Here, %
\begin{equation}
{\beta _{2}^{2}-\left(\kappa _{1}+v+\dfrac{1}{2} \right)^{2}} = \beta _{1}^{2} - v^2 -(2\kappa_1 +1)\left(v+\dfrac{1}{2}\right) > 0 \/ .
{\label{rmhp8}}
\end{equation}
The corresponding normalized wave functions are the same as before in {\eqref{mhpFunctions}} and {\eqref{mhp16}} but
with the new values of parameters {\eqref{rmhp5}}
(we leave further details to the reader).

Our study of the rotation correction of the modified Hulth\'{e}n potential will be continued elsewhere (see also the Mathematica file).

\begin{table}
\caption[]{The rotation correction of the  modified Hulth\'{e}n potential (\ref{rmhp4}),
where the coefficients $C_0$, $C_1$, and $C_2$ are given by (\ref{rmhp3coeffs}).}
{\label{tab:rmhp}}
\begin{center}
\begin{tabular}{|l|l|}
\hline
$\sigma (\xi)$ & $\xi (1-\xi)$ \\ \hline
$\widetilde{\sigma }(\xi)$ & \mybox{$ - \left( {\alpha}^2_1 + b_1 {\beta}^2_1 \right){\xi}^2 + \left(2{\alpha}^2_1 +{\beta}^2_1\right)\xi -{\alpha}^2_1 $} \\ \hline
$\widetilde{\tau }(\xi)$ & $ 1-\xi $ \\ \hline
$k$ & \mybox{$ {\beta}_1^2 -2{\alpha}_1 {\kappa}_1 \/, \quad {\kappa}_1 =\sqrt{(1/4)+(b_1 -1){\beta}_1^2} $} \\ \hline
$\pi(\xi) $ & \mybox{${\alpha}_1 - ({\alpha}_1 +{\kappa}_1 +1/2) \xi$} \\ \hline
$\tau (\xi) =\widetilde{\tau }+2\pi $ & \mybox{$(2{\alpha}_1+1) -2({\alpha}_1 +{\kappa}_1 +1) \xi $} \\ \hline
$\lambda =k+\pi ^{\prime }$ & \mybox{$ {\beta}_1^2 -\left(2 {\alpha}_1 +1\right) \left({\kappa}_1 +1/2\right) $} \\ \hline
$\varphi (\xi) $ & \mybox{$\xi^{\alpha_1} (1-\xi)^{\kappa_1 +1/2}$} \\ \hline
$\rho (\xi) $ & \mybox{$\xi^{2\alpha_1} (1-\xi)^{2\kappa_1}$} \\ \hline
\end{tabular}
\end{center}
\end{table}
%
%

%

\section{Generalized Morse Potential}

The potential of the form:%
\begin{equation}
U(r)=D\left( 1-\frac{\gamma}{e^{a r}-1}\right) ^{2},\quad \gamma=e^{a
r_{0}}-1\quad \quad \left( 0\leq r<\infty \right) {\label{gMp1}}
\end{equation}%
by (\ref{mhp17}) is reduced to the modified Hulth\'{e}n potential (\ref{mhp1}%
) up to a proper change of parameters (\ref{mhp18}). This is why, we can use
solutions from section 17.

On the second thought, the standard change of the variables \cite{DelSol}:%
\begin{equation}
R(r)=\sqrt{a }u(\eta ),\qquad \eta =\left( e^{a r}-1\right) ^{-1}{%
\label{gMp2}}
\end{equation}%
results in the generalized equation of hypergeometric type with the
following coefficients:%
\begin{equation}
\sigma (\eta )=\eta (1+\eta ),\quad \widetilde{\sigma }(\eta )=\epsilon
-\kappa (1-\gamma \eta )^{2},\quad \widetilde{\tau }(\eta )=2\eta +1,\label{gMp3}
\end{equation}%
where%
\begin{equation}
\kappa=\frac{2mD}{a^{2}\hbar ^{2}},\quad \epsilon =\frac{2mE}{a^{2}\hbar ^{2}};
\qquad {\alpha}^2 =\kappa -\epsilon >0, \quad {\beta}^2 =\kappa(\gamma+1)^2 -\epsilon >0
\label{gMp4}
\end{equation}%
(see Table~{\ref{tab:gMp}} and the Mathematica file). The following substitution%
\begin{equation}
u=\varphi (\eta )y(\eta )=\frac{\eta ^{\alpha }}{(1+\eta )^{\beta }}y(\eta )%
\label{gMp5}
\end{equation}%
results in the hypergeometric equation of Gauss (\ref{gauss}) in the variable $-\eta $
with the following values of parameters:%
\begin{equation}
A=\alpha -\beta +\delta ,\quad B=\alpha -\beta -\delta +1,\quad C=2\alpha
+1;\quad \delta =\frac{1}{2}+\sqrt{\frac{1}{4}+\kappa \gamma ^{2}}.
\label{gMp6}
\end{equation}
Bound states correspond to the polynomial solutions, when $A=\alpha -\beta
+\delta =-n$ and $n$ are some nonnegative integers. Thus $\alpha $ and $%
\beta $ are $n$-dependent, once again, and satisfy the following equations:%
\begin{equation}
\beta _{n}-\alpha _{n}=n+\delta ,\qquad \beta _{n}^{2}-\alpha
_{n}^{2}=\kappa \gamma (\gamma +2).
\label{gMp7}
\end{equation}%
One finds that%
\begin{equation}
\alpha _{n}=\frac{1}{2}\left( \frac{\kappa \gamma (\gamma +2)}{n+\delta }%
-n-\delta \right) ,\quad \beta _{n}=\frac{1}{2}\left( \frac{\kappa \gamma
(\gamma +2)}{n+\delta }+n+\delta \right)
\label{gMp8}
\end{equation}%
and the energy levels have the form \cite{DelSol}:%
\begin{equation}
E_{n}=D-\frac{\alpha ^{2}\hbar ^{2}}{8m}\left( n+\delta -\frac{\kappa \gamma
(\gamma +2)}{n+\delta }\right) ^{2}.\label{gMp9}
\end{equation}
The corresponding normalized wave functions are given by%
\begin{align}
R_{n}(r) &=\sqrt{a}C_{n}e^{-{\beta_n} ar}\left( e^{ar}-1\right) ^{{\beta_n} -{\alpha_n}
}\ _{2}F_{1}\left(
\begin{array}{c}
-n,\ 1-n-2\delta  \\
2\alpha _{n}+1%
\end{array}%
;\frac{1}{1-e^{ar}}\right)   \label{gMp10} \\
&=\sqrt{a}C_{n}e^{- {\alpha_n} ar}\left(1- e^{-ar}\right) ^{\delta}\
_{2}F_{1}\left(
\begin{array}{c}
-n,\ 2{\alpha_{n}}+2\delta + n \\
2{\alpha_{n}}+1%
\end{array}%
;e^{-ar}\right)   \notag
\end{align}
[we have used (15.8.1) of \cite{NIST}) in order to obtain the standard form (\ref{mhpFunctions})], with the normalization coefficients
found in \cite{DelSol} as follows%
\begin{equation}
C_{n}^{2}=\frac{a(\alpha _{n}+n+\delta )\Gamma (2\alpha _{n}+n+1)\Gamma
(2\alpha _{n}+n+2\delta )}{n!(n+\delta )\Gamma (2\alpha _{n})\Gamma (2\alpha
_{n}+1)\Gamma (n+2\delta )}\/,
\label{gMp11}
\end{equation}%
due to the normalization condition:%
\begin{equation}
\int_{0}^{\infty }R_{n}^{2}(r)\ dr=1\/.
\label{gMp12}
\end{equation}%
There is only a finite set of the energy levels:%
\begin{equation}
n=0,1,2,\ \ldots \ n_{\text{max}} \leq \sqrt{\kappa
\gamma (\gamma +2)}-\delta .
\label{gMp12a}
\end{equation}

One can use our results from section~17 in order to verify all these formulas, originally presented in
\cite{DelSol} (see also \cite{AiOthman}, \cite{Rong2003}, and the references therein). We leave further
details to the reader.
%

%

\begin{table}
\caption[]{The generalized Morse potential (\ref{gMp1}).
The coefficients $\epsilon\/$ and $\kappa$ are given by (\ref{gMp4}).}
{\label{tab:gMp}}
\begin{center}
\begin{tabular}{|l|l|}
\hline
$\sigma (\eta)$ & $\eta (1+ \eta)$ \\ \hline
$\widetilde{\sigma }(\eta)$ & \mybox{$ \epsilon - \kappa \left( 1 -\gamma \eta \right)^2  $} \\ \hline
$\widetilde{\tau }(\eta)$ & $ 2\eta +1 $ \\ \hline
$k$ & \mybox{$ 2\left(\alpha^2 -\alpha \beta +\kappa \gamma \right)\/, \quad {{\alpha =\sqrt{\kappa -\epsilon }} \ }, \;  \beta = \sqrt{\kappa (\gamma +1)^2 - \epsilon} $} \\ \hline
$\pi(\eta) $ & \mybox{${\alpha} + ({\alpha} -{\beta}) \eta $} \\ \hline
$\tau (\eta) =\widetilde{\tau }+2\pi $ & \mybox{$ 2{\alpha}+1 + 2({\alpha} -{\beta}  +1) \eta $} \\ \hline
$\lambda =k+\pi ^{\prime }$ & \mybox{$ \alpha -\beta +2{\alpha}^2 - 2\alpha \beta +2\kappa \gamma  $} \\ \hline
$\varphi (\eta) $ & \mybox{$\eta^{\alpha} (1+\eta)^{-\beta}$} \\ \hline
$\rho (\eta) $ & \mybox{$\eta^{2\alpha} (1 +\eta)^{-2\beta}$} \\ \hline
\end{tabular}
\end{center}
\end{table}
%
%

%
\medskip
In a similar fashion, one can consider the Wood--Saxon potential and the motion of the electron in magnetic field \cite{AiOthman},
\cite{DelSol}, \cite{Flugge}, \cite{BBKK2007} \cite{La:Lif}, and \cite{Ni:Uv}.
%
\medskip

\textbf{Acknowledgments.\/}
{We are grateful to Dr.~Steven Baer, Dr.~Kamal Barley, Dr.~Sergey Kryuch\-kov, Dr.~Eugene Stepanov, Dr.~Jos\'e Vega-Guzm\'an, and Dr.~Alexei Zhedanov
for valuable discussions and help.}
\medskip

\begin{appendices}

%
\section{Data for the Classical Orthogonal Polynomials}
\label{app:A}

The basic information about classical orthogonal polynomials, namely, for the Jacobi ${P_n^{(\alpha, \beta)}(x)}\/,$ Laguerre ${L_n^{\alpha}(x)}\/,$
and Hermite $H_n(x)$ polynomials, is presented, for the reader's convenience, in Table~\ref{tab:cop}\/.
It contains the coefficients of the differential equation (\ref{A3}), the intervals of orthogonality $(a,b)$, the weight functions $\rho(x)$ and constants $B_n$
in the Rodrigues-type formula (\ref{A8}), the leading terms:
\begin{equation}
y_n(x)=a_nx^n+b_nx^{n-1}+\ldots \label{eqnApp1}
\end{equation}
for these polynomials, their squared norms:
\begin{equation}
d_n^2=\int_a^b y_n^2(x)\rho(x)\, dx, \label{eqnAA1}
\end{equation}
and the coefficients of the three-term recurrence relation:
\begin{equation}
x\,y_n(x)=\alpha_n\,y_{n+1}(x)+\beta_n\,y_n(x)+\gamma_n\,y_{n-1}(x), \label{eqnAA2}
\end{equation}
where
\begin{equation}
\alpha_n=\frac{a_n}{a_{n+1}},\quad\beta_n
=\frac{b_n}{a_n}-\frac{b_{n+1}}{a_{n+1}},
\quad\gamma_n=\alpha_{n-1}\,\frac{d_n^2}{d_{n-1}^2}. \label{eqnAA3}
\end{equation}
(More details can be found in \cite{AndAskRoy}, \cite{Ni:Su:Uv}, \cite{Ni:Uv},
\cite{NIST}, and \cite{Suslov2020}\/.)

%
\begin{table}
\caption{Data for the Jacobi ${P_n^{(\alpha, \beta)}(x)}\/,$ Laguerre
  ${L_n^{\alpha}(x)}\/,$ and Hermite $H_n(x)$ polynomials.}
\label{tab:cop}
\begin{center}
\begin{tabular}{|l|c|c|c|}
\hline
$y_n(x)$ & \mybox{$P_n^{(\alpha, \,\beta)}(x)\, (\alpha > -1, \beta > -1)$}
& $L_n^{\alpha}(x)\, (\alpha > -1)$ & $H_n(x)$ \\ \hline
$(a, b)$ & $(-1, 1)$ & $(0, \infty)$ & $(-\infty, \infty)$ \\ \hline
$\rho(x)$ & $(1-x)^{\alpha}(1+x)^{\beta}$
& $x^{\alpha}\,e^{-x}$ & $e^{-x^2}$ \\ \hline
$\sigma(x)$ & $1-x^2$ & $x$ & $1$ \\ \hline
$\tau(x)$ & $\beta - \alpha - (\alpha + \beta + 2)\,x$
& $1 + \alpha - x$ & $-2x$ \\ \hline
$\lambda_n$ & $n(\alpha + \beta + n + 1)$ & $n$ & $2n$ \\ \hline
$B_n$ & \mybox{$\dfrac{(-1)^n}{2^n n!}$} & $\dfrac{1}{n!}$ & $(-1)^n$\\ \hline
$a_n$ &
\mybox{$\dfrac{\Gamma(\alpha + \beta + 2n + 1)}
{2^n n! \Gamma(\alpha + \beta + n + 1)}$}
& $\dfrac{(-1)^n}{n!}$ & $2^n$ \\ \hline
$b_n$ &
\mybox{$\dfrac{(\alpha - \beta) \Gamma(\alpha + \beta +2n)}
{2^n (n-1)! \Gamma(\alpha + \beta + n + 1)}$}
& $(-1)^{n-1} \dfrac{\alpha + n}{(n-1)!}$ & $0$ \\ \hline
$d^2$ &
\mybox{$\dfrac{2^{\alpha + \beta + 1} \Gamma(\alpha + n + 1) \Gamma(\beta + n +1)}
{n! (\alpha + \beta +2n +1) \Gamma(\alpha + \beta + n +1)}$}
& $\dfrac{\Gamma(\alpha + n +1)}{n!}$
& $2^n n! \sqrt{\pi}$ \\ \hline
$\alpha_n$ &
\mybox{$\dfrac{2(n+1)(\alpha + \beta +n +1)}
{(\alpha +\beta +2n +1)(\alpha +\beta +2n +2)}$}
& $-(n+1)$ & $\dfrac{1}{2}$ \\ \hline
$\beta_n$ &
\mybox{$\dfrac{\beta^2 - \alpha^2}
{(\alpha +\beta +2n)(\alpha +\beta +2n +2)}$}
& $\alpha +2n +1$ & $0$ \\ \hline
$\gamma_n$ &
\mybox{$\dfrac{2(\alpha +n)(\beta +n)}
{(\alpha +\beta +2n)(\alpha +\beta +2n +1)}$}
& $-(\alpha +n)$ & $n$ \\ \hline
\end{tabular}
\end{center}
\end{table}
\section{An Integral Evaluation}
\label{app:B}

The following useful integral:%
\begin{align}
J_{nms}^{\alpha \beta }& =\int_{0}^{\infty }e^{-x}x^{\alpha +s}\
L_{n}^{\alpha }(x) L_{m}^{\beta }(x) \ dx
\\
& =(-1)^{n-m}\frac{\Gamma(\alpha+s+1)\,\Gamma(\beta+m+1)\,\Gamma(s+1)}{m!\,(n-m)!\,
  \Gamma(\beta+1)\,\Gamma(s-n+m+1)}   \notag  \\
& \quad \times \ _{3}F_{2}\biggl(
\begin{array}{@{}c@{}}
-m,\ s+1,\ \beta -\alpha -s \\
\beta +1,\quad n-m+1%
\end{array}%
;1\biggr) ,\quad n\geq m,  \notag
\label{eqnAppB1}
\end{align}%
where parameter $s$ may take some integer values and $_{3}F_{2}(1)$ is
the generalized hypergeometric series \cite{AndAskRoy}, \cite{NIST},
has been evaluated in \cite{Sus:Trey} and \cite{Suslov2020} (see also \cite{Fues1926}, \cite{SchrodingerParabolic}\/).
Special cases have been used above for the normalization of the wave functions; see (\ref{sol20}) and (\ref{mp20}).
We use also throughout the manuscript  the familiar Euler beta and gamma integrals:
\begin{equation}
B(\alpha, \ \beta)=\int_0^1 t^{\alpha-1} (1-t)^{\beta-1}\ dt= \dfrac{\Gamma(\alpha)\Gamma(\beta)}{\Gamma(\alpha +\beta)} \/,
\label{eqnAppB2}
\end{equation}
provided $\Re{(\alpha)}>0 $ and $\Re{(\alpha)}>0 \/,$
\begin{equation}
\Gamma(\alpha)=\int_0^{\infty} t^{\alpha-1} e^{-t}\ dt \/, \qquad \Re{(\alpha)}>0.
\label{eqnAppB3}
\end{equation}
(See \cite{AndAskRoy}, \cite{Ni:Uv}, and \cite{NIST} for more details.)

\section{Mathematica File}
\label{app:c}

We have discussed basic potentials of the nonrelativistic and relativistic quantum mechanics that can be integrated in the Nikiforov and Uvarov paradigm
with the aid of the Mathematica computer algebra system.
(The corresponding Mathematica notebook is available from the authors by a request.
It is also posted on Wolfram community \url{https://community.wolfram.com/groups/-/m/t/2897057}
and featured in the editorial columns \url{https://community.wolfram.com/content?curTag=staff+picks} .)
\smallskip

In section 2, the general formulas are derived. In the notebook, they are stored in global variables that will be used in all the subsequent sections.
For this purpose, allow Mathematica to evaluate all initialization cells.
After that one can run each case independently from the others.
The results, for the most integrable cases that are available in the literature, are presented
in the Tables~\ref{tab:StatSchr}--\ref{tab:gMp}\/.

\end{appendices}
%

%


\begin{thebibliography}{99}
\bibitem{AiOthman} A.~B.~Al-Othman and A.~S.~Sandouqa, \emph{Comparison study of bound states for diatomic molecules using
Kratzer, Morse, and modified Morse potentials\/,} Physica Scripta, \textbf{97}(3),  035401, 2022.
\url{https://doi.org/10.1088/1402-4896/ac514c}.

\bibitem{AndAskRoy} G.~E.~Andrews, R.~Askey, and R.~Roy, \textsl{Special
Functions}, Cambridge University Press, New York, 1999.

\bibitem{Barleyetal2021} K.~Barley, J.~Vega-Guzm\'{a}n, A.~Ruffing, and S.~K.~Suslov,
\emph{Discovery of the relativistic Schr\"{o}dinger equation\/,} Physics--Uspekhi,
\textbf{69}(1), 90--103, 2022 [in English];
\textbf{192}(1), 100--114, 2022 [in Russian];
\url{https://iopscience.iop.org/article/10.3367/UFNe.2021.06.039000}.

\bibitem{Bethe:Sal} H.~A.~Bethe and E.~Salpeter, \textsl{Quantum Mechanics
of One- and Two-Electron Atoms\/}, Dover Publications, Mineola, New York,
2008.

\bibitem{Berk2006}
C.~Berkdemir, A.~Berkdemir, and J.~Han, \emph{Bound state solutions of the Schr\"{o}dinger equation for modified Kratzer's potential\/},
\newblock Chemical Physics Letters \textbf{417}(4--6), 326--329, 2006.
  \url{https://www.sciencedirect.com/science/article/abs/pii/S0009261405015812}.

\bibitem{Blokh} D.~I.~Blokhintsev, \textsl{Quantum Mechanics\/}, D.~Reidel,
Dordrecht, 1964.

\bibitem{Dav}
A.~S.~Davydov, \textsl{Quantum Mechanics\/},
\newblock Pergamon Press, Oxford and New York, 1965.

\bibitem{Dar}
C.~G.~Darwin, \emph{The wave equations of the electron\/},
\newblock Proceedings of the Royal Society A: Mathematical, Physical and
  Engineering Sciences \textbf{118}(780), 654--680, 1928.
  {\url{https://doi.org/10.1098/rspa.1928.0076}}

\bibitem{DelSol} A.~Del Sol Mesa, C.~Quesne, and Yu.~F.~Smirnov,
\emph{Generalized Morse potential: Symmetry and satellite potentials\/,}
   Journal of Physics A: Mathematical and General \textbf{31}(1), 321--335, 1998.
  {\url{https://iopscience.iop.org/article/10.1088/0305-4470/31/1/028}}

\bibitem{Flugge}
S.~Fl\"{u}gge, \textsl{Practical Quantum Mechanics\/},
\newblock Springer-Verlag, Berlin, Hedelberg, New York, 1999.

\bibitem{Fluggetal67} S.~Fl\"{u}gge, P.~Walger, and A.~Weiguny,  \emph{A generalization of the Morse potential\/},
\newblock Journal of Molecular Spectroscopy \textbf{23}(3),
  243--257, 1967.
  {\url{https://doi.org/10.1016/S0022-2852(67)80013-4}}

\bibitem{Fock1978}
V.~A.~Fock, \textsl{Fundamentals of Quantum Mechanics\/},
  Moscow: Mir Publishers, 1978.
  \url{https://mirtitles.org/2013/01/01/fock-fundamentals-of-quantum-mechanics/}.

\bibitem{Fues1926} E.~Fues,  \emph{Das Eigenschwingungsspektrum zweiatomiger Molek\"{u}le in der Undulationsmechanik},\linebreak[1]
\newblock Annalen der Physik \textbf{385}(12),
  367--396, 1926.
  \newblock \url{https://onlinelibrary.wiley.com/doi/10.1002/andp.19263851204}
  \newblock [in German].

\bibitem{Gol:Kriv}
I.~I.~Gol'dman and V.~D.~Krivchenkov, \textsl{Problems in Quantum Mechanics\/},
\newblock Dover Publications, Inc., New York, 1993.

\bibitem{BBKK2007} B.~G\"{o}n\"{u}l and K.~K\"{o}ksal, \emph{Solutions for a
generalized Woods-Saxon potential}, Physica Scripta \textbf{76}(5), 565--570 (2007).
  {\url{http://dx.doi.org/10.1088/0031-8949/76/5/026}}

\bibitem{Gor}
W.~Gordon, \emph{Die Energieniveaus des Wasserstoffatoms nach der Diracschen
  Quantentheorie des Elektrons\/},
\newblock Zeit\-schrift f{\"{u}}r Physik \textbf{48}(1), 11--14, 1928.
 {\url{https://doi.org/10.1007/BF01351570}}
 \newblock [in German]

\bibitem{Greiner}
W.~Greiner, \textsl{Relativistic Quantum Mechanics: Wave Equations\/}, 2nd ed.,
\newblock Springer-Verlag, Berlin and Hedelberg, 1997.

\bibitem{Karp70} M.~Karplus and R.~N.~Porter, \textsl{Atoms \& Molecules:
An Introduction for Students of Physical Chemistry\/}, The Benjamin/Cummings Company, Menlo Park,
California, 1970.

\bibitem{Ko:Sua:Su}
C.~Koutschan, E.~Suazo, and S.~K.~Suslov, \emph{Fundamental laser modes in paraxial
  optics: from computer algebra and simulations to experimental observation\/}.
\newblock Applied Physics B \textbf{121}(3), 315--336, 2015.
 {\url{https://doi.org/10.1007/s00340-015-6231-9}}

\bibitem{KZ10} C.~Koutschan and D.~Zeilberger, {\emph{The 1958 Pekeris--Accad--WEIZAC ground-breaking collaboration that computed ground states
of two-electron atoms (and its 2010 redux)}},  {Mathematical Intelligencer\/}  {\textbf{33}}, 52--57, 2011.
   \url{https://doi.org/10.1007/s00283-010-9192-1}

\bibitem{Kryuch:Sus:Vega12}
S.~I.~Kryuchkov, S.~K.~Suslov, and  J.~M.~Vega-Guzm{\'{a}}n, \emph{The minimum-uncertainty
  squeezed states for atoms and photons in a cavity\/}.
\newblock Journal of Physics B: Atomic, Molecular and Optical Physics
  \textbf{46}(10), 104007 (2013).
   {\url{https://iopscience.iop.org/article/10.1088/0953-4075/46/10/104007}}

\bibitem{Langer1937}
R.~E. Langer, \emph{On the connection formulas and the
  solutions of the wave equation\/},  \emph{Physical Review} \textbf{51} no.~8
  (1937), 669--676.
  {\url{https://journals.aps.org/pr/abstract/10.1103/PhysRev.51.669}}

\bibitem{La:Lif}
L.~D.~Landau and E.~M.~Lifshitz, \textsl{Quantum Mechanics: Non-Relativistic Theory},
  3rd ed., Butter\-worth--Heine\-mann, Oxford, 1998.

\bibitem{Lop:Sus:VegaHarm}
R.~M.~L{\'o}pez, S.~K.~Suslov, and J.~M.~Vega-Guzm{\'a}n, \emph{On a hidden symmetry of
  quantum harmonic oscillators\/}.
\newblock Journal of Difference Equations and Applications \textbf{19}(4),
  543--554, 2013.
   {\url{https://www.tandfonline.com/doi/abs/10.1080/10236198.2012.658384}}

\bibitem{Morse29}
P.~M.~Morse, \emph{Diatomic molecules according to the wave mechanics\/. II\/}.
\newblock Physical Review \textbf{34}(1),
  57--64, 1929.
  {\url{https://journals.aps.org/pr/abstract/10.1103/PhysRev.34.57}}

\bibitem{Ni:Su:Uv}
 A.~F.~Nikiforov, S.~K.~Suslov, and V.~B.~Uvarov, \textsl{Classical Orthogonal Polynomials
  of a Discrete Variable\/},
\newblock Springer Series in Computational Physics. Springer Berlin Heidelberg,
  Berlin, Heidelberg, 1991.

\bibitem{Ni:Uv}
 A.~F.~Nikiforov and V.~B.~Uvarov, \textsl{Special Functions of Mathematical Physics: A
  Unified Introduction with Applications\/},
\newblock Birkh{\"a}user, Boston, MA, 1988.

\bibitem{NIST} \textsl{NIST Handbook of Mathematical Functions\/},
(F.~W.~J.~Olver and D.~W.~Lozier, \ eds.), Cambridge University Press, New
York, 2010. \url{https://dlmf.nist.gov/}.

\bibitem{Pek58} C.~L.~Pekeris. \emph{Ground state of two-electron atoms\/},  Physical Review
\textbf{112}(5), 1649--1658, 1958.
\url{https://doi.org/10.1103/PhysRev.112.1649}.

\bibitem{PT}
G.~P\"{o}schl and E.~Teller, \emph{Bemerkungen zur Quantenmechanik des anharmonischen Oszillators.}
\newblock Zeitschrift f\"{u}r Physik \textbf{83} March issue, 143--151 (1933),
\url{https://doi.org/10.1007/BF01331132}.

\bibitem{Rong2003}
Z.~Rong, H.~G.~Kjaergaard, and M.~L.~Sage, \emph{Comparison of the Morse and Deng--Fan potentials
  for X--H bonds in small molecules},
Molecular Physics \textbf{101}(14), 2285--2294, 2003.
\url{https://www.tandfonline.com/doi/abs/10.1080/0026897031000137706}

\bibitem{RM}
N.~Rosen and P.~M.~Morse, \emph{On the Vibrations of Polyatomic Molecules\/.}
Phys. Rev. \textbf{42} (2), 210--217 (1932),
\url{https://doi.org/10.1103/PhysRev.42.210}.

\bibitem{Schrodinger2010} E.~{Schr{\"o}dinger}, \emph{Quantisation as a problem of
  proper values\/ {(Part I)}},  in \textsl{Collected Papers on Wave Mechanics\/}, New
  York, Providence, Rhode Island: AMS Chelsea Publishing, 2010, (Original:
  Annalen der Physik (4), vol. 79(6), pp. 489--527, 1926 [in German]), pp.~1--12.
  {\url{https://onlinelibrary.wiley.com/doi/10.1002/andp.19263840404}}

\bibitem{SchroedingerOscillator} E.~{Schr{\"o}dinger}, \emph{Quantisation as a problem of
  proper values {(Part II)}\/},  in \textsl{Collected Papers on Wave Mechanics\/}, New
  York, Providence, Rhode Island: AMS Chelsea Publishing, 2010, (Original:
  Annalen der Physik (4), vol. 79(6), pp. 489--527, 1926 [in German]), pp.~13--40.
  {\url{https://onlinelibrary.wiley.com/doi/10.1002/andp.19263840602}}


\bibitem{SchrodingerParabolic} E.~{Schr{\"o}dinger}, \emph{Quantisation as a problem of
  proper values {(Part III)}\/},  in \textsl{Collected Papers on Wave Mechanics\/}, New
  York, Providence, Rhode Island: AMS Chelsea Publishing, 2010, (Original:
  Annalen der Physik (4), vol. 386~\# 18, pp. 109--139, 1926 [in German]), pp.~62--101\/.
  {\url{https://onlinelibrary.wiley.com/doi/abs/10.1002/andp.19263851302}}.

\bibitem{SchrodingerCohrent} E.~{Schr{\"o}dinger}, \emph{The continuous transition from
  micro-to macro-mechanics\/},  in \textsl{Collected Papers on Wave Mechanics\/},
  \textbf{28}, New York, Providence, Rhode Island: AMS Chelsea Publishing, 2010,
  (Original: Die Naturwissenschaften, vol. 28, pp. 664--666, 1926 [in German]),
  pp.~41--44.
  {\url{https://doi.org/10.1007/BF01507634}}

\bibitem{SchrodingerTimeDependent}   E.~Schr{\"o}dinger, \emph{Quantisation as a problem of
  proper values {(Part IV)\/}},  in {\textsl{Collected Papers on Wave Mechanics\/}}, New
  York, Providence, Rhode Island: AMS Chelsea Publishing, 2010, (Original:
  Annalen der Physik (4), vol. 81, pp. 109--139, 1926 [in German]), pp.~102--123.
   {\url{https://onlinelibrary.wiley.com/doi/10.1002/andp.19263861802}}.

\bibitem{SchrodingerReview}  E.~Schr{\"o}dinger, \emph{An undulatory theory of the mechanics of atoms and molecules\/},
   Physical Review \textbf{28}(6), 1049--1070, 1926.
   \url{https://journals.aps.org/pr/abstract/10.1103/PhysRev.28.1049}.

\bibitem{Schiff}
L.~I.~Schiff, \textsl{Quantum Mechanics\/}, 3rd edn.
\newblock International series in pure and applied physics. McGraw-Hill, Inc.,
  New York, 1968.

\bibitem{Sommerfeld1951}
A.~Sommerfeld, \textsl{Atombau und Spektrallinien\/}, 2
  ed., \textbf{1}, Friedrich Vieweg \& Sohn, Braunschweig, German, 1951.
  [in German]

\bibitem{Sus:Trey}
 S.~K.~Suslov and B.~Trey, \emph{The {H}ahn polynomials in the nonrelativistic and
  relativistic {C}oulomb problems\/}.
\newblock Journal of Mathematical Physics \textbf{49}(1), 012104 (2008).
  {\url{https://doi.org/10.1063/1.2830804}}

\bibitem{Suslov2020}
S.~K.~Suslov, J.~M.~Vega-Guzm{\'{a}}n, and K.~Barley,
\emph{An introduction to special functions with some applications to quantum
  mechanics\/,}  in  \textsl{Orthogonal Polynomials: 2nd AIMS-Volkswagen Stiftung
  Workshop, Douala, Cameroon, 5-12 October, 2018}
  (M.~Foupouagnigni and   W.~Koepf, eds.), \textsl{Tutorials, Schools, and
  Workshops in the Mathematical Sciences\/} no. AIMSVSW 2018, Springer Nature
  Switzerland AG, March 2020, pp.~517--628.
  {\url{https://link.springer.com/chapter/10.1007/978-3-030-36744-2_21}}

\bibitem{Varshalovich1988}
D.~A. Varshalovich, A.~N.~Moskalev, and V.~K.~Khersonskii,
  \textsl{Quantum Theory of Angular Momentum\/}, Singapore,
  New Jersey, Hong Kong:  World Scientific, 1988.

\bibitem{Watson}
G.~N.~Watson,  \textsl{A Treatise on the Theory of {B}essel Functions\/}, 2nd edn.
\newblock Cambridge Mathematical Library. Cambridge University Press,
  Cambridge, England, 1995.
\newblock [Reprint of the second (1944) edition]

\bibitem{Whi:Wat}
E.~T.~Whittaker and G.~N.~Watson, \textsl{A Course of Modern Analysis: An Introduction to
  the General Theory of Infinite Processes and of Analytic Functions; with an
  Account of the Principal Transcendental Functions\/}.
\newblock Cambridge Mathematical Library. Cambridge University Press,
  Cambridge, England, 1950.
\newblock [Reprint of the 4th (1927) edition]

\end{thebibliography}
\end{document}